# Solving large-scale MEG/EEG source localisation and functional connectivity problems simultaneously using state-space models

Jose Sanchez-Bornot[1*], Roberto C. Sotero[2], J. A. Scott Kelso[1,3], Özgür Şimşek[4], and Damien Coyle[1,4*]

[1] Intelligent Systems Research Centre, School of Computing, Engineering and Intelligent Systems, Ulster University, Magee campus, Derry~Londonderry, UK.

[2] Department of Radiology and Hotchkiss Brain Institute, University of Calgary, Calgary, AB, Canada.

[3] Human Brain & Behavior laboratory, Center for Complex Systems & Brain Sciences, Florida Atlantic University, Boca Raton, Florida, USA.

[4] Bath Institute for the Augmented Human, University of Bath, Bath, BA2 7AY, United Kingdom.

[*] Corresponding authors: JSB: jm.sanchez-bornot@ulster.ac.uk; DC: dhc30@bath.ac.uk

## Abstract

State-space models are widely employed across various research disciplines to study unobserved dynamics. Conventional estimation techniques, such as Kalman filtering and expectation maximisation, offer valuable insights but incur high computational costs in large-scale analyses. Sparse inverse covariance estimators can mitigate these costs, but at the expense of a trade-off between enforced sparsity and increased estimation bias, necessitating careful assessment in low signal-to-noise ratio (SNR) situations. To address these challenges, we propose a three-fold solution: 1) Introducing multiple penalised state-space (MPSS) models that leverage data-driven regularisation; 2) Developing novel algorithms derived from backpropagation, gradient descent, and alternating least squares to solve MPSS models; 3) Presenting a *K*-fold cross-validation extension for evaluating regularisation parameters. We validate this MPSS regularisation framework through lower and more complex simulations under varying SNR conditions, including a large-scale synthetic magneto- and electro-encephalography (MEG/EEG) data analysis. In addition, we apply MPSS models to concurrently solve brain source localisation and functional connectivity problems for real event-related MEG/EEG data, encompassing thousands of sources on the cortical surface. The proposed methodology overcomes the limitations of existing approaches, such as constraints to small-scale and region-of-interest analyses. Thus, it may enable a more accurate and detailed exploration of cognitive brain functions.

## Keywords:

state-space models, source localization, functional connectivity, large-scale analysis, MEG, EEG.

## Introduction

Solving magneto- and electro-encephalographic (MEG/EEG) source localisation and functional connectivity (FC) problems can significantly contribute to the identification and study of information-processing pathways, enhancing our understanding of spontaneous and evoked events in the human brain (Lopes da Silva, 2013). Localised brain region activations underpin information processing, with intraregional and interregional communication facilitating the creation and synchronisation of these activations in networks such as attentional and default-mode networks (Raichle, 2015). The understanding of these phenomena is vital for comprehending human cognitive function as an emerging property of the integration and segregation of information processing (Bressler and Kelso, 2001; Tognoli and Kelso, 2014), which may also be helpful for developing new artificial intelligence (AI) methods.

MEG/EEG and electrocorticogram (ECoG) are popular neuroimaging techniques for studying neuronal dynamics due to their millisecond-resolution tracking of brain activity (Lopes da Silva, 2013). While these techniques have limitations, such as ECoG's invasiveness or MEG/EEG's lack of spatial specificity, the primary shortcomings lie in the analytical tools used to derive information from the recorded signals. High computational costs often restrict established brain activity mapping methods to region-of-interest (ROI) analyses or small-scale simulations. This is evident in the use of the expectation maximisation (EM) algorithm for solving dynamic causal models (DCMs) (Friston et al., 2003) and the use of state-space models based on EM and Kalman filtering (Barton et al., 2009; Cheung et al., 2010; Galka et al., 2004; Long et al., 2011; Shumway and Stoffer, 1982; Van de Steen et al., 2019; Yamashita et al., 2004). However, neglecting high-dimensional brain dynamics can yield misleading results due to overlooked cause-effect relationships (Bastos and Schoffelen, 2016). Moreover, brain source localization and FC methods often address these problems separately (Gross et al., 2001; Haufe et al., 2013; Haufe and Ewald, 2016; Nolte et al., 2004; Pascual-Marqui et al., 1994; Pascual-Marqui, 1999; Sekihara et al., 2001; Sekihara and Nagarajan, 2015; Stam et al., 2007; Stam and van Dijk, 2002; Valdes-Sosa et al., 2006; Van Veen et al., 1997; Vega-Hernández et al., 2008), despite their interdependence (Hincapié et al., 2017; Mahjoory et al., 2017; Manomaisaowapak et al., 2021; Pirondini et al., 2018). Although state-space models tackle this issue (Cheung et al., 2010; Galka et al., 2004; Yamashita et al., 2004), only a few studies have dealt with large-scale analysis (Long et al., 2011).

A further shortcoming of existing methods for FC assessment is that bias may occur when the analysis is performed directly on MEG/EEG sensor data (Cao et al., 2022; Nolte et al., 2004; Sanchez-Bornot et al., 2018; Van de Steen et al., 2019), e.g., due to volume conduction. Bias may also arise in FC analysis based on fMRI measurements, such as the case of resting-state FC or rs-fMRI studies (Frässle et al., 2021; Greicius et al., 2009; Power et al., 2012). Since fMRI favours spatial localisation at the expense of low temporal resolution (Brookes et al., 2011; Tewarie et al., 2019), rs-fMRI can measure primarily a signals temporal correlation while ignoring lagged interactions; thus, giving a false sense of quantification of actual brain networks. Moreover, FC analysis based directly on the information extracted from brain sources can be biased in methods that reduce dimensions, such as ROI analyses (Bastos and Schoffelen, 2016; Hillebrand et al., 2012; Sanchez-Bornot et al., 2021). With the emergence of Big Data Analytics (Raghupathi and Raghupathi, 2014) and high-performance computing (HPC) tools (Bouchard et al., 2016), it may be significantly advantageous to avoid such limited approaches by examining the neuronal dynamics in their natural high-dimensional manifolds. So far, however, few neuroimaging studies are exploiting the HPC tools critical for large-scale analysis of the brain as a complex dynamical system (Long et al., 2011; Sanchez-Bornot et al., 2021).

To address the aforementioned limitations, this study has two main objectives: 1) To develop a novel methodology for solving state-space models in different signal-to-noise ratio (SNR) scenarios (**Figs. 1A-E**), and 2) To solve MEG/EEG source localisation and FC problems simultaneously (**Figs. 1F-G**). For the first objective, we propose a simple backpropagation algorithm for solving state-space models (**Fig. 1B**), enabled by the weight decay technique applied in deep learning to handle noisy signals, as justified from a Bayesian perspective (Yang and Wang, 2020). We then introduce more general multiple penalised state-space (MPSS) models, allowing for more robust data fitting across various SNR scenarios. Moreover, we propose a state-space gradient descent (SSGD) algorithm (**Fig. 1C**) for potentially more stable solutions than backpropagation for MPSS models. Given the quadratic nature of the optimisation problem, we also present a state-space alternating least-squares (SSALS) algorithm

(**Fig. 1D**) for faster convergence in large-scale analysis. An additional essential technique of our proposed framework is a novel extension of *K*-fold cross-validation for estimating the regularisation parameters in MPSS models (**Fig. 1E**).

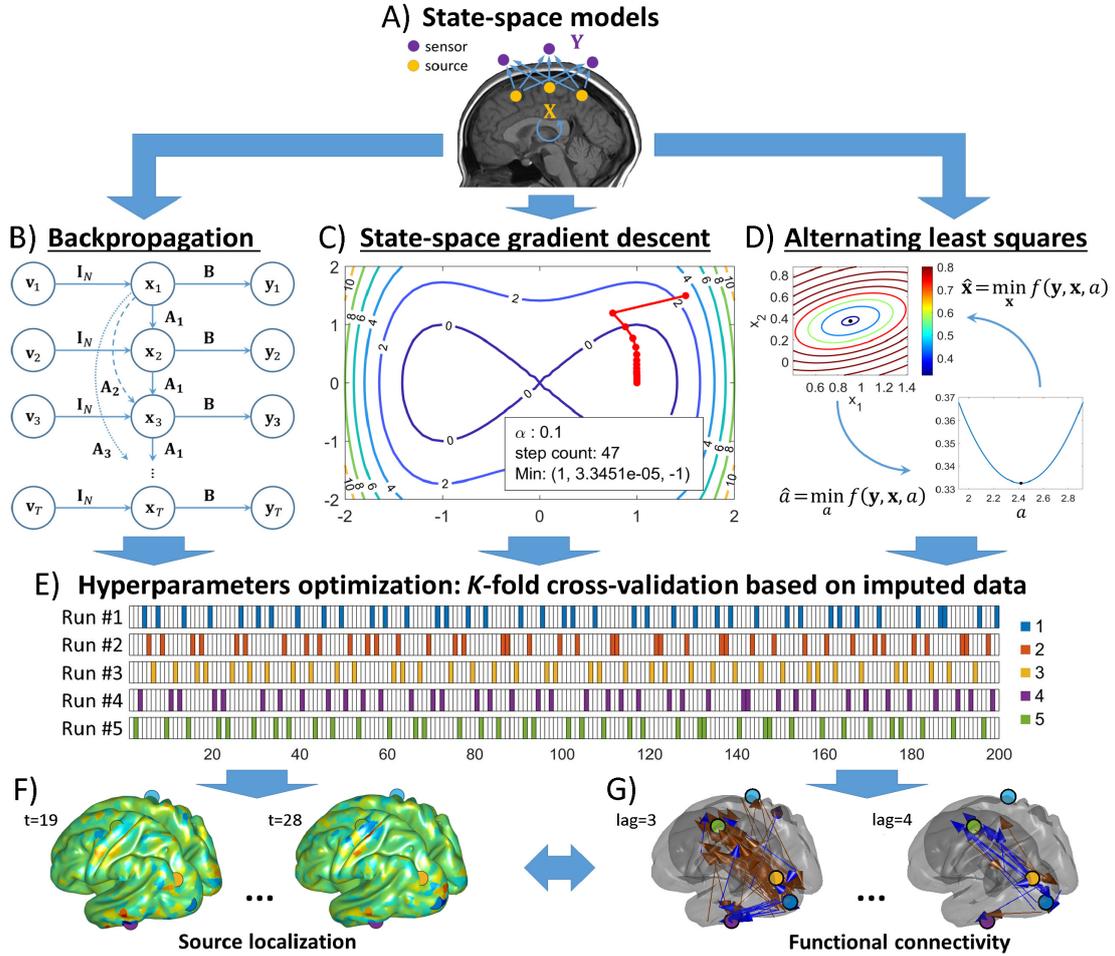

**Figure 1:** Solving state-space models with data-driven regularization techniques. **A)** State-space model example: MEG/EEG data generated by dynamic changes in neuronal activity and communication, as depicted by **Eqs. (1, 2)**. Only sensor measurements $y_1, \ldots, y_T$ are observable, while hidden state variables $x_1, \ldots, x_T$ represent neuronal dynamics. Directed connections (straight arrows) signify the influence of the electromagnetic field on MEG/EEG sensor recordings. The recursive connection (circular arrow) emphasizes the recurrent nature of brain communication, as expressed in **Eq. (2)** via multiplication of autoregressive matrices $A_1, \ldots, A_P$ and corresponding past activations. **B)** Graphical model of data simulation using **Eqs. (1, 2)**, also representing the forward propagation stage of the proposed backpropagation algorithm. Primarily, interactions for lag=1,2,3 time delays are shown according to the multivariate autoregressive (MVAR) model. Observation noise vectors $w_1, \ldots, w_T$ are omitted for simplicity, while state noise vectors $v_1, \ldots, v_T$ are depicted as input vectors multiplied by the identity operator $I_N \in \mathcal{R}^{N \times N}$. **C)** Local optimization of the model using the proposed state-space gradient descent (SSGD) algorithm, illustrated for a two-dimensional nonlinear cost function. SSGD and backpropagation rely on choosing the gradient descent step size ($\alpha$) and momentum (Yang and Wang, 2020), affecting the computation of iterative solutions (red piecewise linear path) and essential for ensuring convergence to local minima. **D)** The proposed state-space alternating least squares (SSALS) algorithm provides more efficient iteratively closed-form solutions for large-scale analysis, primarily by alternating the estimation of parameters $x_1, \ldots, x_T$ and $A_1, \ldots, A_P$. **E)** Proposed extension of classic *K*-fold cross-validation based on imputed data from the state-space model, enabling hyperparameter evaluation during model learning with proposed algorithms, illustrated for $K = 5$ fold runs. Each row displays a partition of randomly selected time series samples (coloured items) excluded from the training data in each corresponding cross-validation run. X-axis: time series samples. **F-G)** This methodology allows simultaneous estimation of brain source localization and FC problems by estimating $\hat{x}_1, \ldots, \hat{x}_T$ (shown at t=19 and t=28 time instants) and $\hat{A}_1, \ldots, \hat{A}_P$ (shown for lag=3 and lag=4 time delays), represented by colourmaps and arrows superimposed on the brain cortical surface, respectively, as demonstrated for a simulated example. **F)** Coloured spheres are placed on the cortical surface along with estimated source activities to emphasize the location of five simulated dipoles. **G)** Identical coloured spheres are overlaid on the (transparent) cortical surfaces as in **F** with the same viewpoint orientation. Connections are also shown among 96 of the 100 most prominent estimated dipoles, represented by arrows, using a 15% threshold relative to the maximum connectivity value for improved visibility. Brown-coloured arrows indicate estimated positive connections, while blue-coloured arrows denote negative connections.

To address the second objective, we apply our methodology to estimate large-scale brain sources and their underlying FC for simulated MEG/EEG data in resting state and event related experimental conditions, assessing the accuracy of estimated spatiotemporal components and FC maps with ground-truth information. We then evaluate the methodology using single-subject MEG/EEG data for a familiar face recognition task as a proof of concept. In conclusion, our proposed methods can improve the performance of state-of-the-art techniques for solving state-space models and contribute to advancing the research on source localisation and FC problems. This methodology can also accelerate data-intensive neuroscience and neuroimaging studies to improve our understanding of cognitive brain functioning.

## Materials and Methods

Instead of using probabilistic approaches such as the Kalman filter and EM for solving state-space models, we developed and evaluated alternative algorithms such as those based on backpropagation, gradient descent and alternating least square approaches. The state-space model is represented using multivariate autoregressive (MVAR) modelling as follows (**Eq. (2)**):

$$\mathbf{y}_t = \mathbf{B}\mathbf{x}_t + \mathbf{w}_t; \text{ with } \mathbf{w}_t \sim N(0, \sigma_o^2 \mathbf{I}_M); t = 1, 2, \dots, T, \quad (1)$$

$$\mathbf{x}_t = \sum_{p=1}^{P} \mathbf{A}_p \mathbf{x}_{t-p} + \mathbf{v}_t; \text{ with } \mathbf{v}_t \sim N(0, \sigma_s^2 \mathbf{I}_N); t = P+1, \dots, T, \quad (2)$$

where $\mathbf{x}_t \in \mathcal{R}^{N \times 1}$ (column vector) represents the state dynamics, for $t = 1,2, \dots, T$ time instants. We use the MVAR generative model to produce surrogate data for the modelled neuronal activity in $N$ spatial locations of the brain, while the MVAR's coefficients $\mathbf{A}_p \in \mathcal{R}^{N \times N}$, $p = 1, \dots, P$, represent the surrogate neuronal communication for lagged (time-delayed, lag=1, …, $P$) interactions. The observed dynamics $\mathbf{y}_t \in \mathcal{R}^{M \times 1}$ represent $M$-variate measurements (time series) obtained from the MEG/EEG sensors (**Eq. (1)**), and $\mathbf{B} \in \mathcal{R}^{M \times N}$ is the mixing (lead field) matrix. The state-space model also includes noise terms $\mathbf{v}_t \in \mathcal{R}^{N \times 1} \sim N(0, \sigma_s^2 \mathbf{I}_N)$ and $\mathbf{w}_t \in \mathcal{R}^{M \times 1} \sim N(0, \sigma_o^2 \mathbf{I}_M)$ to account for the perturbation in state and observation dynamics, respectively. For simplicity, we assume that noise terms are independent and follow multivariate normal distributions with variances $\sigma_o^2$ and $\sigma_s^2$.

### Backpropagation algorithm for state-space models

To develop a backpropagation implementation of the state-space model (**Eqs. 1, 2**), we first created a graphical representation of its algebraic operations (**Fig. 1B**). This graphical model represents the general case of one or more delayed influences and resembles the unrolled graphical representation of interactions in recurrent neural networks (RNNs) (Lillicrap and Santoro, 2019; Werbos, 1990), in which the MVAR's coefficients $\mathbf{A}_p, p = 1, \dots, P$, play a similar role to the shared weights in RNNs.

**Table 1** shows a backpropagation algorithm to solve the state-space model. In order to improve convergence, we implement backpropagation using gradient descent (GD) with the momentum modification, and using the weight decay technique (Yang and Wang, 2020). The only known data to solve this problem are the observations $\mathbf{y}_t$, while $\mathbf{x}_t$ and $\mathbf{A}_p$ are the model parameters. Backpropagation enables the estimation of $\mathbf{A}_p$ and $\mathbf{v}_t$ directly and uses the MVAR's equation to generate the time series $\hat{\mathbf{x}}_t$ from these estimates during the forward propagation phase, i.e., $\hat{\mathbf{x}}_t = \sum_{p=1}^{P} \hat{\mathbf{A}}_p \hat{\mathbf{x}}_{t-p} + \hat{\mathbf{v}}_t$, for $t = P+1, \dots, T$. Also, notice that we are penalizing the error term, correspondingly to the weight decay implementation, which results in the regularised optimisation problem $F = \frac{1}{2T} \sum_{t=1}^{T} \{(\mathbf{z}_t - \mathbf{y}_t)^2 + \lambda \|\mathbf{v}_t\|_2^2\}$ instead of the original optimisation function shown in the table, which is similar to the expression in Ridge regression (Hastie et al., 2001).

**Table 1:** Backpropagation algorithm for the state-space model represented in **Eqs. (1, 2)** and **Fig. 1B**.

| **Error function** |
|---|
| $F = \frac{1}{2T}\sum_{t=1}^{T}(\mathbf{z}_t - \mathbf{y}_t)^2,$  where $\mathbf{z}_t$ and $\mathbf{y}_t$ represent the measured data and predicted outcome, respectively, from the backpropagation forward pass step. |
| **Forward propagation** (for $i = 1, \ldots, I$ iterations) |
| $\mathbf{x}_t(i) = \mathbf{v}_t(i);$ for $t = 1, \ldots, P,$  $\mathbf{x}_t(i) = \sum_{p=1}^{P}\mathbf{A}_p(i)\mathbf{x}_{t-p}(i) + \mathbf{v}_t(i);$ for $t = P+1, \ldots, T,$  $\mathbf{y}_t(i) = \mathbf{B}\mathbf{x}_t(i);$ for $t = 1, \ldots, T.$ |
| **Partial derivatives using chain rule for the $i$-th iteration** (iteration index removed for simplicity) |
| $\boldsymbol{\delta}_t^y = \partial F/\partial \mathbf{y}_t = -(\mathbf{z}_t - \mathbf{y}_t)/T;$ for $t = 1, \ldots, T,$  $\boldsymbol{\delta}_T^x = \partial F/\partial \mathbf{x}_T = \mathbf{B}^T\boldsymbol{\delta}_T^y,$  $\boldsymbol{\delta}_t^x = \partial F/\partial \mathbf{x}_t = \mathbf{B}^T\boldsymbol{\delta}_t^y + \sum_{p=max(P+1-t,1)}^{min(P,T-t)}\mathbf{A}_p^T\boldsymbol{\delta}_{t+p}^x;$ calculate backwards for $t = T-1, \ldots, 2, 1.$   Note that application of the chain rule is backwards for correct calculations/updates. Also note that, although dynamics in $\mathbf{x}_t$ affect $\mathbf{x}_{t+1}, \ldots, \mathbf{x}_{t+p}$, in the extreme cases $\mathbf{x}_{T-1}$ only affects $\mathbf{x}_T$, and $\mathbf{x}_1$ only affects $\mathbf{x}_{P+1}$, in agreement with the forward propagation phase (previous step). |
| **Parameter updates using gradient descent** |
| $\mathbf{v}_t(i+1) = \mathbf{v}_t(i) - \alpha\bigl(\boldsymbol{\delta}_t^x + (\lambda/T)\mathbf{v}_t(i)\bigr);$ for $t = 1, \ldots, T,$  $\mathbf{A}_p(i+1) = \mathbf{A}_p(i) - \alpha\sum_{t=P+1}^{T}\boldsymbol{\delta}_t^x\mathbf{x}_{t-p}(i)^T;$ for $p = 1, \ldots, P,$   where $\alpha$ is the step size adopted along the gradient descent direction. Here, $\lambda \geq 0$ is a regularisation parameter often used to implement weight decay in the training of artificial neural networks (ANNs), which can be justified using a Bayesian interpretation and assuming that the noise terms $\mathbf{v}_t$ follow a normal distribution. Similarly, we can rewrite the second expression as $\mathbf{A}_p(i+1) = \mathbf{A}_p(i) - \alpha\bigl(\sum_{t=P+1}^{T}\boldsymbol{\delta}_t^x\mathbf{x}_{t-p}(i)^T + (\lambda_2/T)\mathbf{A}_p(i)\bigr)$, with $\lambda_2 \geq 0$; however, for simplicity, here we ignored the application of weight decay on the estimation of $\mathbf{A}_p$, for $p = 1, \ldots, P$. |

## Multiple penalised state-space (MPSS) models

State-space models can also be examined from a Bayesian perspective by using probability distributions to represent the state and observation dynamics and incorporate *a priori* information that may enhance the estimators' stability. For example, conditional and *a priori* distributions justify the implementation of weight decay in backpropagation (see **Eqs. (4, 5)** below). These statistical designs are also widely considered for solving the brain inverse problem, as shown by MUSIC, LORETA, and other methods (Grech et al., 2008; Vega-Hernández et al., 2008).

In the present case, we consider the following assumptions:

- Conditional distribution for the observation variable:

$$\mathbf{y}_t|\mathbf{B}, \mathbf{x}_t \sim N(\mathbf{B}\mathbf{x}_t, \sigma_o^2 \mathbf{I}_M). \qquad (3)$$

- Conditional distribution for the state variable:

$$\mathbf{x}_{t>P}|\{\mathbf{A}_1, \ldots, \mathbf{A}_P\}, \{\mathbf{x}_1, \ldots, \mathbf{x}_{t-1}\} \sim N\bigl(\sum_{p=1}^{P}\mathbf{A}_p\mathbf{x}_{t-p}, \sigma_s^2 \mathbf{I}_N\bigr). \qquad (4)$$

- A *priori* distribution for the state variable:

$$\mathbf{x}_{t \leq P} \sim N(\mathbf{0}_N, \sigma_s^2 \mathbf{I}_N), \tag{5}$$

where $\mathbf{0}_N$ and $\mathbf{I}_N$ are the zero vector and the identity matrix with $N$ elements and $N \times N$ dimension, respectively.

- A *priori* distribution for the autoregressive coefficients:

$$vec(\mathbf{A}) \sim N(\mathbf{0}_{N^2 P}, \sigma_{2,a}^2 \mathbf{I}_{N^2 P}), \tag{6}$$

where $\mathbf{A} = [\mathbf{A}_1, \ldots, \mathbf{A}_P] \in \mathcal{R}^{N \times NP}$ is the matrix that contains all the autoregressive coefficients, shown by using Matlab notation for horizontal concatenation for a better understanding; $vec(\mathbf{A})$ is the vectorized representation following Matlab's column-major notation.

Additionally, we can also use sparse priors:

$$vec(\mathbf{x}) \sim Laplace(\mathbf{0}, \sigma_{1,x}^2 \mathbf{I}_{NT}), \tag{7}$$

where $\mathbf{x}$ contains all the state's variables for all time instants, similar to the notation used for $\mathbf{A}$, and/or

$$vec(\mathbf{A}) \sim Laplace(\mathbf{0}, \sigma_{1,a}^2 \mathbf{I}_{N^2 P}); \tag{8}$$

although, in practice, we may only adopt sparse priors in a large-scale scenario to control dimensionality through numerical optimization.

In general, using the Bayesian approach enables the proposition of a multiple penalised state-space (MPSS) framework, where the maximum a posteriori estimate for the parameters $\mathbf{x} \in \mathbb{R}^{N \times T}$ and $\mathbf{A} \in \mathcal{R}^{N \times NP}$ are obtained from solving the MPSS optimisation problem:

$$F = \frac{1}{2T} \begin{pmatrix} \sum_{t=1}^T \|\mathbf{y}_t - \mathbf{B}\mathbf{x}_t\|_2^2 + \lambda \sum_{t=P+1}^T \|\mathbf{x}_t - \sum_{p=1}^P \mathbf{A}_p \mathbf{x}_{t-p}\|_2^2 + \lambda \sum_{t=1}^P \|\mathbf{x}_t\|_2^2 + \lambda_2^{(a)} \|\mathbf{A}\|_F^2 \\ + 2\lambda_1^{(x)} \|\mathbf{x}\|_1 + 2\lambda_1^{(a)} \|\mathbf{A}\|_1 \end{pmatrix}, \tag{9}$$

$$\hat{\mathbf{x}}, \hat{\mathbf{A}} = \underset{\mathbf{x}, \mathbf{A}}{argmin}\, F\left(\mathbf{x}, \mathbf{y}, \mathbf{A}, \mathbf{B}, \lambda, \lambda_2^{(a)}, \lambda_1^{(x)}, \lambda_1^{(a)}\right). \tag{10}$$

### State-space gradient descent (SSGD) algorithm to solve MPSS models

For estimating MPSS models, we initially propose a state-space gradient descent (SSGD) approach as an alternative to the aforementioned backpropagation algorithm. SSGD is implemented based on the following partial derivatives:

$$T \frac{\partial F}{\partial \mathbf{x}_t} = \begin{matrix} -\mathbf{B}^T(\mathbf{y}_t - \mathbf{B}\mathbf{x}_t) + \lambda\left(\mathbf{x}_t - \sum_{p=1}^P \mathbf{A}_p \mathbf{x}_{t-p}\right) \\ -\lambda \sum_{q=1}^{min(P, T-t)} \mathbf{A}_q^T\left(\mathbf{x}_{t+q} - \sum_{p=1}^P \mathbf{A}_p \mathbf{x}_{t+q-p}\right) + \lambda_1^{(x)} sgn(\mathbf{x}_t) \end{matrix}; \text{ for } t = P+1, \ldots, T, \tag{11}$$

$$T \frac{\partial F}{\partial \mathbf{x}_t} = \begin{matrix} -\mathbf{B}^T(\mathbf{y}_t - \mathbf{B}\mathbf{x}_t) - \lambda \sum_{q=max(P+1-t,1)}^{min(P,T-t)} \mathbf{A}_q^T\left(\mathbf{x}_{t+q} - \sum_{p=1}^P \mathbf{A}_p \mathbf{x}_{t+q-p}\right) \\ + \lambda \mathbf{x}_t + \lambda_1^{(x)} sgn(\mathbf{x}_t) \end{matrix}; \text{ for } t = 1, \ldots, P, \tag{12}$$

$$T \frac{\partial F}{\partial \mathbf{A}_p} = -\lambda \sum_{t=P+1}^T \left(\mathbf{x}_t - \sum_{q=1}^P \mathbf{A}_q \mathbf{x}_{t-q}\right) \mathbf{x}_{t-p}^T + \lambda_1^{(a)} sgn(\mathbf{A}) + \lambda_2^{(a)} \mathbf{A}; \text{ for } p = 1, \ldots, P. \tag{13}$$

Realistically, to estimate the parameters $\mathbf{x}_t$, $t = 1, 2, \ldots, T$, and $\mathbf{A}_p$, $p = 1, \ldots, P$, we must first find suitable values for the hyperparameters $\lambda$, $\lambda_2^{(a)}$, $\lambda_1^{(x)}$, and $\lambda_1^{(a)}$. Next we propose an extension of $K$-fold cross-validation to select these values. Otherwise, assuming the conditional and *a priori* distributions are known, we can estimate "naïve" solutions corresponding to $\lambda = \sigma_o^2/\sigma_s^2$, $\lambda_2^{(a)} = \sigma_o^2/\sigma_{2,a}^2$, $\lambda_1^{(x)} = 0.5\,\sigma_o^2/\sigma_{1,x}^2$ and $\lambda_1^{(a)} = 0.5\,\sigma_o^2/\sigma_{1,a}^2$.

### *K*-fold cross-validation based on imputed data for state-space models

We propose an extension of the $K$-fold cross-validation method (Hastie et al., 2001) to estimate the prediction error in state-space models as represented in **Eqs. (1, 2)**. As in the original cross-validation method, we randomly separate the data samples (time point measurements in our case) into $K$-fold subsets. One critical difference, in our case, is that the data correspond to the temporal sequence $\mathbf{y}_t$, $t = 1, \dots, T$, with an underlying autoregressive (generative) model. Leaving out a patch of adjacent samples is highly detrimental to model estimation due to the temporal dependency. Thus, we implement the $K$-fold partition by dividing the samples into nonoverlapping adjacent time windows of length $K$ and subsequently randomly assigning each time-window sample to one of the $K$-fold subsets. This procedure guarantees a balanced partition of the samples into $K$ subsets $\left\{y_{\tau_1}^{(1)}, \dots, y_{\tau_{|S_1|}}^{(1)}\right\} \in S_1, \dots, \left\{y_{\tau_1}^{(K)}, \dots, y_{\tau_{|S_K|}}^{(K)}\right\} \in S_K$, where $|S|$ represents the subset's cardinality. At the same time, it guarantees that no more than two adjacent time-instant samples are assigned to the same subset (see **Fig. 1E** for the case of $K = 5$ and $T = 200$).

The other essential difference with the classical cross-validation approach is that usually, the data consist of pairs $(\mathbf{x}_t, \mathbf{y}_t)$, which are assigned randomly to each $K$-fold subset. However, in our state-space model, we only know $\mathbf{y}_t$ and $\mathbf{x}_t$ must be estimated together with the other model parameters. Therefore, for each $K$-fold run, to estimate the prediction error for the hold-out samples $\{\mathbf{y}_\tau\} \in S_k$, we first estimate their corresponding "missing" part $\{\mathbf{x}_\tau\}$. Reflecting on this feature, we refer to our procedure as $K$-fold cross-validation based on imputed data ($K$-fold CVI) as we first need to impute $\{\mathbf{x}_\tau\}$ by treating $\{\mathbf{y}_\tau\} \in S_k$ as missing values, for each run $k = 1, \dots, K$, before evaluating the prediction error.

Corresponding to this modification, we change the original MPSS optimisation function (**Eqs. (9, 10)**) to account for the $K$-fold hold-out subset at each iteration, as follows:

$$F_k = \frac{1}{2T}\left(\begin{array}{c}\sum_{t=1}^T L_t^{(k)}\|\mathbf{y}_t - \mathbf{B}\mathbf{x}_t\|_2^2 + \lambda \sum_{t=P+1}^T \left\|\mathbf{x}_t - \sum_{p=1}^P \mathbf{A}_p \mathbf{x}_{t-p}\right\|_2^2 + \lambda \sum_{t=1}^P \|\mathbf{x}_t\|_2^2 \\ + \lambda_2^{(a)}\|\mathbf{A}\|_F^2 + 2\lambda_1^{(x)}\|\mathbf{x}\|_1 + 2\lambda_1^{(a)}\|\mathbf{A}\|_1\end{array}\right), \tag{14}$$

$$\hat{\mathbf{x}}^{(k)}, \widehat{\mathbf{A}}^{(k)} = \underset{\mathbf{x},\mathbf{A}}{argmin}\, F_k\left(\mathbf{x}, \mathbf{y}, \mathbf{A}, \mathbf{B}, \lambda, \lambda_2^{(a)}, \lambda_1^{(x)}, \lambda_1^{(a)}\right); \text{for } k = 1, \dots, K, \tag{15}$$

where $L_t^{(k)} = \begin{cases} 0, & \text{if } \mathbf{y}_t \in S_k \\ 1, & \text{otherwise} \end{cases}$ is a cost function, set up to remove the contribution of the hold-out data samples $\{\mathbf{y}_\tau\} \in S_k$ for each $K$-fold run.

After that, we calculate the prediction error by averaging the errors for the predicted observation for each hold-out sample, for all the $K$-fold runs:

$$PE = \frac{1}{(T-P)MK}\sum_{k=1}^K \sum_{\mathbf{y}_t \in S_k}\left\|\mathbf{y}_t^{(k)} - \mathbf{B}\hat{\mathbf{x}}_t^{(k)}\right\|_2^2. \tag{16}$$

### Alternating least squares (SSALS) and hybrid algorithms for state-space models

Here, we develop an SSALS algorithm for large-scale analysis. As the optimisation problem above is quadratic, the solutions have a closed form obtained by solving the optimisation problem separately for the parameter subsets $\{\mathbf{x}_t\}$, $t = 1, \dots, T$, and $\{\mathbf{A}_p\}$, $p = 1, \dots, P$. For example, conditioning on the previous solution for the autoregressive coefficients, $\{\widehat{\mathbf{A}}_p^{(i)}\}$, at iteration $i = 1, \ldots, I$, the closed-form solution for $\{\mathbf{x}_t\}$ is

$$\widehat{\mathbf{X}}_V^{(i+1)} = \left(\mathbf{D}_L^T \mathbf{D}_L \otimes \mathbf{B}^T \mathbf{B} + \lambda (\mathbf{I}_{TN} - \mathbf{W})^T (\mathbf{I}_{TN} - \mathbf{W})\right)^{-1} (\mathbf{D}_L^T \mathbf{D}_L \otimes \mathbf{I}_N) vec(\mathbf{B}^T \mathbf{Y}) \qquad (17)$$

where $\widehat{\mathbf{X}}_V^{(i+1)} = vec(\widehat{\mathbf{X}}^{(i+1)})$ is the vectorized representation ($\widehat{\mathbf{X}}^{(i+1)} \in \mathcal{R}^{N \times T}$), $\otimes$ is the Kronecker's product, $\mathbf{D}_L$ is a diagonal matrix with the diagonal entries set to the values of the cost function $L_t^{(k)}$, and $\mathbf{W}$ is a matrix encoding the contribution of the autoregressive term as shown below.

Notice that in **Eq. (14)**

$$\sum_{t=P+1}^{T} \left\| \mathbf{x}_t - \sum_{p=1}^{P} \mathbf{A}_p \mathbf{x}_{t-p} \right\|_2^2 + \sum_{t=1}^{P} \|\mathbf{x}_t\|_2^2 = \|(\mathbf{I}_{TN} - \mathbf{W}) vec(\mathbf{X})\|_2^2, \qquad (18)$$

where

$$\mathbf{W} = \begin{bmatrix} \mathbf{0}_N & \mathbf{A}_1 & \cdots & \mathbf{A}_P & \mathbf{0}_N & \ddots & & & \\ \mathbf{0}_N & \mathbf{0}_N & \mathbf{A}_1 & \cdots & \mathbf{A}_P & \mathbf{0}_N & \ddots & & \\ \ddots & \mathbf{0}_N & \mathbf{0}_N & \mathbf{A}_1 & \cdots & \mathbf{A}_P & & & \\ & \ddots & & & \cdots & & \ddots & & \\ & & & \mathbf{0}_N & \mathbf{0}_N & \mathbf{A}_1 & \cdots & \mathbf{A}_P & \\ & & & & \ddots & \mathbf{0}_N & \mathbf{0}_N & \mathbf{0}_N & \mathbf{0}_N \\ & \ddots & & & & \ddots & \mathbf{0}_N & \mathbf{0}_N & \mathbf{0}_N \\ & & & & \ddots & \ddots & \mathbf{0}_N & \mathbf{0}_N & \mathbf{0}_N \end{bmatrix}, \qquad (19)$$

is a diagonal-block matrix with the autoregressive matrices in the corresponding super diagonal positions for each lag=1,…,P. Also, notice that the involved matrix dimension is $NT \times NT$, but these are very highly-sparse matrices with a convenient block structure. That is, they can be treated as banded-block matrices for compact representation, a fact that may be very convenient to speed calculations. Notice also that $\{\mathbf{x}_t\}$ must be considered in time-reversal order, so $vec(\mathbf{X})$ can be correctly multiplied by $\mathbf{W}$ in the right-hand side of **Eq. (18)**.

Furthermore, given the numerical values $\widehat{\mathbf{X}}^{(i+1)}$ obtained from **Eq. (17)**, then

$$\widehat{\mathbf{A}}^{(i+1)} = \mathbf{X}_{P+1:T} \mathbf{Z}^T (\mathbf{Z}\mathbf{Z}^T)^{-1}, \qquad (20)$$

is the classical least-squares solution for MVAR equations, where $\widehat{\mathbf{A}} = [\widehat{\mathbf{A}}_1, \ldots, \widehat{\mathbf{A}}_P]$ and $\mathbf{Z} = [\mathbf{X}_{P:T-1}, \ldots, \mathbf{X}_{1:T-P}]$ are both defined by following Matlab's horizontal concatenation notation. Besides, $\mathbf{X}_{i:j}$ is a shortcut for $\mathbf{X}(:, i:j)$ that represents the subblock matrix with elements between columns $i$ and $j$, including both subindices. Solving the Yule-Walker equations can be more efficient (Barnett and Seth, 2014), but the least squares notation is simpler for our presentation purposes.

More general than the alternating least squares (ALS) algorithm, we introduce a hybrid method between gradient descent (GD) and ALS, called HGDALS, based on interleaving GD and ALS iterations. While ALS has a much faster convergence rate than GD, the latter is less computationally expensive. The advantage of this combination is not clear in the quadratic case, as ALS alone produces very fast convergence to the solution. However, combining GD and ALS could render better results for a general nonlinear optimisation problem. Succinctly, HGDALS relies on applying one iteration of ALS with as many iterations, $I_{\text{GD}}$, of GD as calculated by the formula

$$\frac{\Delta F_{\text{GD}}}{\Delta F_{\text{ALS}}} \approx \frac{I_{\text{GD}} T_{\text{GD}}}{T_{\text{ALS}}}, \qquad (21)$$

which balances the ratio between the optimisation function evaluation changes by GD ($\Delta F_{\text{GD}}$) and ALS ($\Delta F_{\text{ALS}}$), with the ratio of their computational cost or calculation time for a single (previous) iteration. Therefore, if ALS is not producing faster local convergence than GD, HGDALS will automatically increase the number of GD's iterations. Conveniently, we simultaneously limit the number of GD's iterations, so the GD's total time is not more than, for example, 20% of ALS computational time, i.e., $I_{\text{GD}} T_{\text{GD}} \leq 0.2 T_{\text{ALS}}$. A summary of the HGDALS follows in **Table 2**.

**Table 2:** HGDALS algorithm.

```
I_GD = 1;
for i = 1, ..., I
    F = function_evaluation();
    time = clock();
    <model optimisation with ALS>
    T_ALS = clock() - time;
    ΔF_ALS = F - function_evaluation();
    if (i > 1)
        I_GD = min(0.2 T_ALS/T_GD, ceil((ΔF_GD T_ALS)/(ΔF_ALS T_GD)));
    end
    F = function_evaluation();
    time = clock();
    for j = 1, ..., I_GD
        <model optimisation with GD>
    end
    T_GD = (clock() - time)/I_GD;
    ΔF_GD = F - function_evaluation();
end
```

## Simulations with state-space models

We validate the algorithms discussed in this study using synthetic signals randomly generated with the state-space model in **Eqs. (1, 2)** for different scenarios. These models use fixed (ground truth) autoregressive matrices $\mathbf{A}_p$, $p = 1, \ldots, P$, from which time series $\mathbf{x}_t$ and $\mathbf{y}_t$ are randomly generated. Particularly, we create three different simulations: the more straightforward of those generate bivariate and three-variate time series with $P = 1$ and $P = 3$ time-lagged influences, respectively, while the two more complex cases involve the simulation of two and five state variables, with $P = 5$ time-lagged influences, corresponding to simulated sources with random locations in a template brain cortical surface. However, the estimation in the latter cases will involve estimating the source dynamics for thousands of sources (brain dipoles) and their connections, as the ground-truth actual number of sources is ignored during the estimation step.

We use a numerical tolerance of $10^{-6}$ on the partial derivatives (cost function) norms as convergence criteria for backpropagation and the other algorithms, unless otherwise stated. Only the measurement data ($\mathbf{y}_t$) will be used during the model estimation step for each of the tested algorithms, whereas $\mathbf{x}_t$ and $\mathbf{A}_p$ will serve as ground truth for the evaluation of the estimated coefficients. To evaluate the accuracy of estimates $\hat{\mathbf{x}}_t$ and $\hat{\mathbf{A}}_p$, we can use the relative squared error (RSE) formula. For example, for evaluating the accuracy of each individual estimated time series, the error can be calculated as $RSE = 100 * \frac{\sum_{t=1}^{T}(x_t - \hat{x}_t)^2}{\sum_{t=1}^{T}(x_t - \bar{x})^2}$, where $\bar{x} = \frac{1}{T}\sum_{t=1}^{T} x_t$, represented in percentage (%) with respect to the base error for enhanced interpretation.

*Small-scale simulation of time series data*

The bivariate simulations are created using $T = 200$, $M = 5$, $N = 2$, and $P = 1$ (lag=1), by setting $\sigma_o = 0.1$, $\sigma_o = 0.5$, or $\sigma_o = 1$, while always setting $\sigma_s = 1$ for each case, to simulate three different SNR scenarios, with fixed $\mathbf{A}_1 = \begin{bmatrix} -0.5 & 0 \\ 0.7 & -0.5 \end{bmatrix}$.

Similarly, the three-variate simulations are created using $T = 240$, $M = 5$, $N = 3$, and $P = 3$ (lag=1,2,3), and setting the noise parameters as in the bivariate case for three different SNR scenarios. In contrast to the bivariate simulation, the complexity increases because the three-variate model involves three latent variables ($N = 3$) and three time-lagged interactions ($P = 3$). For the three-variate simulation, the ground-truth autoregressive coefficients are fixed as follows (Stokes and Purdon, 2018):

$$\mathbf{A} = \left\{ \begin{bmatrix} -0.9000 & 0 & 0 \\ -0.3560 & 1.2124 & 0 \\ 0 & -0.3098 & -1.3856 \end{bmatrix}, \begin{bmatrix} -0.8100 & 0 & 0 \\ 0.7136 & -0.4900 & 0 \\ 0 & 0.5000 & -0.6400 \end{bmatrix}, \begin{bmatrix} 0 & 0 & 0 \\ -0.3560 & 0 & 0 \\ 0 & -0.3098 & 0 \end{bmatrix} \right\},$$

where the coefficient matrices corresponding to the time-lagged interactions are horizontally stacked inside curly brackets from left to right in increasing delay order.

*Large-scale brain simulations for resting state and event related MEG/EEG data*

The two large-scale simulations discussed in this section are based on Haufe and Ewald's MEG/EEG toolbox (Haufe and Ewald, 2016) and simulate resting state and event-related experimental conditions. Haufe and Ewald's toolbox provides high- and low-resolution mesh data (surface's triangle and vertex points) for the so-called "New York" brain, with 74,382 and 2,004 vertices, respectively, as well as the vertices for standard and inflated cortical surfaces. The high-resolution meshes enables plotting of the results on the brain cortical surface. At the same time, the low-resolution vertices provide the locations to calculate the dynamic state variables in our procedure; therefore, for these simulations we will estimate $N = 2004$ sources. The toolbox also provides lead field matrices calculated for MEG and EEG forward problems for dipoles perpendicularly oriented to the cortical surface. In our case, for the first large-scale simulation we simulate MEG data, and thereby M=298 sensors; whereas, for the second large-scale simulation, we generate synthetic EEG data, and thus M=108. In all cases, as the dimension of the largest matrix in **Eq. (17)** is $NT \times NT$, we could not use a value of $T > 50$ because of limited RAM memory resources. Thereby, in these simulations, we set $T = 30$ while considering the sampling frequency $F_S = 250$ Hz (equivalent to 4 ms time resolution) and $T = 18$ for $F_S = 125$ Hz (8 ms time resolution), respectively, and simulated $P = 5$ time-lagged influences (lag=1,…,5) in both cases, which is equivalent to considering influences as long as 20 or 40 ms from the past.

To be as realistic as possible, we tested values for $P$ consistently with estimates of neurophysiological data, where the evidence indicates that communication delays can be as long as 40-50 ms (Ringo et al., 1994). However, using a larger value of $P$ could make challenging the tuning of these simulations for appropriate values of the autoregressive coefficients $\mathbf{A}_p$ due to numerical instabilities, and it will force to use very small values that are difficult to estimate. Correspondingly to the above simulation settings, here we used the autoregressive matrices as presented below for each case (shown from left to right inside curly brackets, for lag=1,…,5). The synthetic MEG/EEG signals were generated in a similar approach to that described in the literature (Liang et al., 2022; Liu et al., 2019; Sanchez-Bornot et al., 2018) but using $SNR = 20$ and $SNR = 5$ decibels (dB), respectively, for these two cases.

1) Resting state simulation:

$$\mathbf{A} = \left\{ \begin{bmatrix} 1.356 & 0 & 0 & 0 & 0 \\ 0.8 & 1.356 & 0 & 0 & 0 \\ 0 & 0 & 1.356 & 0 & 0 \\ 0 & 0 & 0 & 1.5 & 0 \\ 0 & 0 & 0 & 0 & 1.5 \end{bmatrix}, \begin{bmatrix} -0.49 & 0 & 0 & 0 & 0 \\ -0.8 & -0.49 & 0 & 0 & 0 \\ 0.8 & 0 & -0.49 & 0 & 0 \\ 0 & 0 & 0 & -0.75 & 0 \\ 0 & 0 & 0 & 0 & -0.75 \end{bmatrix}, \right.$$

$$\left. \begin{bmatrix} 0 & 0 & 0 & 0 & 0 \\ 0 & 0 & 0 & 0 & 0 \\ -0.8 & 0 & 0 & 0 & 0 \\ 0.8 & 0 & 0 & 0 & 0 \\ 0 & 0 & 0 & 0 & 0 \end{bmatrix}, \begin{bmatrix} 0 & 0 & 0 & 0 & 0 \\ 0 & 0 & 0 & 0 & 0 \\ 0 & 0 & 0 & 0 & 0 \\ -0.8 & 0 & 0 & 0 & 0 \\ 0 & 0 & 0 & 0 & 0 \end{bmatrix}, \begin{bmatrix} 0 & 0 & 0 & 0 & 0 \\ 0 & 0 & 0 & 0 & 0 \\ 0 & 0 & 0 & 0 & 0 \\ 0 & 0 & 0 & 0 & -0.1 \\ 0 & 0 & 0 & 0.1 & 0 \end{bmatrix} \right\},$$

2) Event related simulation:

$$\mathbf{A} = \left\{ \begin{bmatrix} 0.8 & 0 & 0 & 0 & 0 \\ 0 & 0.8 & 0 & 0 & 0 \\ 0 & 0 & 0.8 & 0 & 0 \\ 0 & 0 & 0 & 0.8 & 0 \\ 0 & 0 & 0 & 0 & 0.8 \end{bmatrix}, \begin{bmatrix} -0.5 & 0 & 0 & 0 & 0 \\ 0 & -0.5 & 0 & 0 & 0 \\ 0 & 0 & -0.5 & 0 & 0 \\ 0 & 0 & 0 & -0.5 & 0 \\ 0 & 0 & 0 & 0 & -0.5 \end{bmatrix}, \right.$$

$$\left. \begin{bmatrix} 0 & 0 & 0 & 0 & 0 \\ 1.0 & 0 & 0 & 0 & 0 \\ 0 & 0 & 0 & 0 & 0 \\ 0 & 0 & 0 & 0 & 0 \\ 0 & 0 & 0 & 0 & 0 \end{bmatrix}, \begin{bmatrix} 0 & 0 & 0 & 0 & 0 \\ 0 & 0 & 0 & 0 & 0 \\ 0 & 0 & 0 & 0 & 0 \\ 0 & 0 & 0 & 0 & -1.0 \\ 0 & 0 & 0 & 1.0 & 0 \end{bmatrix}, \begin{bmatrix} 0 & 0 & 0 & 0 & 0 \\ 0 & 0 & 0 & 0 & 0 \\ 0 & 0 & 0 & 0 & 0 \\ 0 & 0 & 0 & 0 & 0 \\ 0 & 0 & 0 & 0 & 0 \end{bmatrix} \right\}.$$

More precisely, for the event related simulation, we generated 80 synthetic datasets by running 20 Monte Carlo replications based on different random locations of the brain sources in four different conditions, which involved the 2x2 combinations of simulations considering either two or five sources with an activity patch extension of 6 or 15 cm². For clarity, each dataset contains two 3D matrices of dimension equals to $\#Channels \times \#Samples \times \#Epochs$, for the synthetic generated latent and observed time series. The simulation based on five dipoles uses exactly the same autoregressive matrix above, whereas the two source simulation uses the same matrix but reduced to the first two regions (rows). Notice that for the autoregressive matrices, there is not really interaction for the 5-th lag as the entries are zero in the matrix. However, the zero-entries were set there explicitly to highlight that all $p = 1, ..., 5$ will be estimated.

Finally, to resemble better the actual conditions in which human experiments are conducted, the event-related data was generated as a continuous EEG signal where as many continuous segments as trials were generated. Each segment duration was set to 750 ms, in which a random stimulus onset occurred between 200 and 500 ms and the different connections among regions were programmed to become active after each stimulus onset, while modulated by a Hanning's window of 120 ms length. Moreover, a random jitter was represented using the normal distribution with mean ± std equals to 20 ± 10 ms to represent the dynamics build up mediating between stimulating a region with an input pulse and the activation of its outgoing connections. The onset signal was modelled as a single square pulse with height equals to 0.3 and duration of 24 ms, which was added to the model-generated inner dynamics. Whereas in the two-ROIs simulation only the first region is stimulated externally, in the five-ROIs case the regions 1 and 4 are stimulated with inter-pulse interval of 48 ms. For the analysis of these datasets, the continuously generated EEG signals were epoched using the time onset values, in the interval $-150 \leq t \leq 282$ ms around the stimulus onset, and the rest of the analysis was performed as it is done with real data. In the comparison study discussed in the results section, due to the aforementioned RAM limitations, our proposed method used only the epoched data in the interval (50; 200] ms ($T = 18$ samples), which included the simulated intraregional communication, whereas the other evaluated inverse solution methods used the complete epoched data.

### Algorithm optimization to speed calculations in large-scale analysis

Using $T < 50$ in the large-scale simulations above can be seen as a critical limitation given the vast number of parameters. However, we can exploit a computational trick to run our calculations with a much larger number of samples. As usual in event-related studies, many trials or epochs are recorded for the same experimental conditions. Therefore, we adapt to this situation by running many replications (epochs) with the settings defined above. Accordingly, we implemented the MPSS optimisation function (see **Eqs. (14, 15)**) for multiple epochs as follows:

$$F_k = \frac{1}{2TE}\left(\begin{array}{c}\sum_{e=1}^{E}\sum_{t=1}^{T}L_t^{(k)}\left\|\mathbf{y}_t^{(e)} - \mathbf{B}\mathbf{x}_t^{(e)}\right\|_2^2 + \lambda\sum_{e=1}^{E}\sum_{t=P+1}^{T}\left\|\mathbf{x}_t^{(e)} - \sum_{p=1}^{P}\mathbf{A}_p\mathbf{x}_{t-p}^{(e)}\right\|_2^2 \\ +\lambda\sum_{e=1}^{E}\sum_{t=1}^{P}\|\mathbf{x}_t\|_2^2 + \lambda_2^{(x)}\|\mathbf{X}\|_F^2 + \lambda_2^{(a)}\|\mathbf{A}\|_F^2\end{array}\right), \tag{22}$$

$$\hat{\mathbf{x}}^{(k)}, \hat{\mathbf{A}}^{(k)} = \underset{\mathbf{x},\mathbf{A}}{argmin}\, F_k\left(\mathbf{x}, \mathbf{y}, \mathbf{A}, \mathbf{B}, \lambda, \lambda_2^{(x)}, \lambda_2^{(a)}\right); \text{for } k = 1, \ldots, K, \tag{23}$$

where $E$ is the number of epochs ($E = 200$ in simulations). In this case, we state the optimisation problem based only on the hyperparameters $\lambda$, $\lambda_2^{(x)}$, and $\lambda_2^{(a)}$. The use of sparse penalty functions (corresponding to using $\lambda_1^{(a)}$ or $\lambda_1^{(x)}$) is presently ignored in our study due to the lack of an efficient implementation for large-scale analysis. It should be noted that despite the MVAR matrices being fixed during each simulated condition, the generated time series $\mathbf{x}_t^{(e)}$, $e = 1, \ldots, E$, change randomly and independently across the epochs.

Interestingly, the last optimisation problem can be solved very efficiently if we run the *K*-fold partition only for the time dimension, i.e., a unique cross-validation partition of the temporal indices for all epochs (coded by the loss function $L_t^{(k)}$ in **Eq. (22)**). Notice that the dynamic variables $\left\{\mathbf{x}_t^{(e)}\right\}$, $e = 1, \ldots, E$, can be estimated separately for each epoch as these are independent measurements when conditioning on the estimated autoregressive matrices. However, more conveniently, we can modify **Eq. (17)** to estimate the time series in closed form, as follows:

$$\hat{\mathbf{X}}_{1:E}^{(i+1)} = \left(\mathbf{D}_L^T\mathbf{D}_L \otimes \mathbf{B}^T\mathbf{B} + \lambda(\mathbf{I}_{TN} - \mathbf{W})^T(\mathbf{I}_{TN} - \mathbf{W})\right)^{-1}(\mathbf{D}_L^T\mathbf{D}_L \otimes \mathbf{I}_N)\mathbf{Z}_{1:E} \tag{24}$$

where $\hat{\mathbf{X}}_{1:E}^{(i+1)} = \left[\hat{\mathbf{X}}_V^{(i+1,e=1)}, \ldots, \hat{\mathbf{X}}_V^{(i+1,e=E)}\right]$ and $\mathbf{Z}_{1:E} = \left[vec(\mathbf{B}^T\mathbf{Y}^{(e=1)}), \ldots, vec(\mathbf{B}^T\mathbf{Y}^{(e=E)})\right]$ (matrices $\hat{\mathbf{X}}_{1:E}^{(i+1)}$ and $\mathbf{Z}_{1:E}$ are both of $NT \times E$ dimensions). This trick has tremendous advantages as the inverse of the $NT \times NT$ matrix is the most computationally expensive operation. While **Eq. (24)** keeps the same inverse matrix order as in **Eq. (17)**, on the other hand, the number of samples can be increased considerably by generating a higher number of epochs without significantly increasing the computational cost.

### Performance metrics to evaluate the quality of inverse solutions

Here, we present the quantitative measures used to evaluate the different tested inverse solution methods in comparison to the solution estimated by our proposed MPSS models. For comparison purposes, we used the relative root squared error (RRSE) to quantify the temporal accuracy of estimated signals, and the receiver operating characteristic (ROC) curve to quantify their spatial accuracy (Grova et al., 2006; Liang et al., 2022; Liu et al., 2019). Regarding the latter, we also proposed a modified version for the ROC calculation which is introduced below. The RRSE stat can be calculated as follows:

$$RRSE = \sqrt{\frac{\sum_{i=1}^{N}\sum_{t=1}^{T}(\tilde{x}_{it}^{ERP}-\tilde{y}_{it}^{ERP})^2}{\sum_{i=1}^{N}\sum_{t=1}^{T}(\tilde{x}_{it}^{ERP})^2}}, \text{ or } RRSE = \sqrt{\frac{\sum_{i=1}^{N}\sum_{t=1}^{T}\sum_{e=1}^{E}(\tilde{x}_{ite}^{TRL}-\tilde{y}_{ite}^{TRL})^2}{\sum_{i=1}^{N}\sum_{t=1}^{T}\sum_{e=1}^{E}(\tilde{x}_{ite}^{TRL})^2}} \quad (26)$$

where $\mathbf{X}^{TRL}, \mathbf{Y}^{TRL} \in \mathbb{R}^{N \times T \times E}$ and $\mathbf{X}^{ERP}, \mathbf{Y}^{ERP} \in \mathbb{R}^{N \times T}$ are the 3D and 2D matrices containing the simulated and estimated event related data, for the original trial-basis signals ($\mathbf{X}^{TRL}$) and after averaging to obtain the ERP components ($\mathbf{X}^{ERP}$). However, these signals have been normalized before performing the RRSE calculation (e.g., $\tilde{x}_{it}^{ERP} = x_{it}^{ERP}/\max_{it}\{x_{it}^{ERP}\}$) to remove scale differences between the ground truth and estimated signals. These two versions for 2D and 3D matrices are considered as the SSALS/HGDALS solutions are 3D matrices; whereas, other methods produce ERP component estimates, such as the methods included in the comparison from the SPM toolbox (Friston et al., 2008; López et al., 2014; Penny et al., 2011). We also define here $\hat{q}_i = \left(\sum_{t=1}^{T} \hat{x}_{it}^2\right)^{1/2}$ and $\tilde{q}_i = \hat{q}_i/\max\{\hat{q}_1, \dots, \hat{q}_N\}$ as each source estimated and normalized energy, where the latter normalized measure will be used next.

About the ROC calculation, we agree with Grova et al. (2006) that a modification to the calculation of the ROC curve is necessary to address the case of brain inverse solutions. Grova et al. (2006) pointed out the issue of the imbalance between the number of active sources in simulations (known ground truth), in the order of dozens or hundreds, and inactive sources, in the order of thousands. However, we observed empirically with a Monte Carlo analysis that this does not have a clear negative impact in the calculation of ROC statistics for random solutions with different smooth/sparse characteristics (**Supp. Material Fig 1**). To deal with this issue, Grova et al. (2006) proposed to randomly drawn the same amount of inactive as active sources, first sampling only from nearby inactive sources to the ground truth and then only from distant sources, for subsequently calculating the ROC and its area (AROC) stats for the active and selected nearby/faraway sources, separately, which they called $AROC_{close}$ and $AROC_{far}$. These values are finally averaged to produce the single statistic $AROC = \frac{1}{2}(AROC_{close} + AROC_{far})$ (Grova et al. (2006)'s Eq. (11)). In general, this modification considers that a false positive identification for a nearby estimated source is not as critical as those committed for a faraway estimate, given that it gives both errors the same weight and there is a much higher number of distant sources than there are close sources. Below, we consider a different modification of the ROC statistic that implements this idea from a more robust mathematical perspective.

Let $\mathcal{U} = \{1, \dots, N\}$ be the universe set containing all the possible source indices and $\mathcal{A}$ the subset of the ground truth active indices; thus, its complement $\mathcal{A}^C$ contains the ground truth inactive source indices. Moreover, for a threshold value $0 \leq \theta \leq 1$, the estimated active subset can be defined as $\hat{\mathcal{A}}(\theta) = \{i: \tilde{q}_i \geq \theta\}$. Using these definitions, the true/false positives (TP/FP) and true/false negatives (TN/FN) can be defined as in **Table 3**.

**Table 3:** Contingency table for classical ROC analysis (∩ is the intersection symbol and |·| represents the subset cardinality).

| $TP(\theta) = \|\mathcal{A} \cap \hat{\mathcal{A}}(\theta)\|$ | $FN(\theta) = \|\mathcal{A} \cap \hat{\mathcal{A}}^C(\theta)\|$ |
|---|---|
| $FP(\theta) = \|\mathcal{A}^C \cap \hat{\mathcal{A}}(\theta)\|$ | $TN(\theta) = \|\mathcal{A}^C \cap \hat{\mathcal{A}}^C(\theta)\|$ |

In contrast to the representation in **Table 3**, for convenience, we can represent these formulas using equivalent expressions. For example, corresponding to the false positive representation, we have $FP(\theta) = \sum_{i=1}^{N} \mathfrak{I}\left(i \in \mathcal{A}^C \,\&\, i \in \hat{\mathcal{A}}(\theta)\right)$, where "∈" and "&" are the set theory "belong to" and logical AND operators, and $\mathfrak{I}(\cdot)$ is an indicator function valued to 1 if the argument is true and 0 otherwise. Finally, the proposed modification of the AROC statistics, called here area under weighted ROC

(AWROC) statistics, can be calculated as in the standard case, but using a weighted measure to calculate the false positive instead, as follows:

$$FP(\theta) = \frac{1}{\mu} \sum_{i=1}^{N} \mathfrak{I}\left(i \in \mathcal{A}^C \& i \in \hat{\mathcal{A}}^C(\theta)\right) \xi_I(i), \tag{25}$$

where $\mu = \sum_{i=1}^{N} \xi_I(i)/|\mathcal{A}^C|$ is a normalization constant, and $\xi_I(i)$ is a function that evaluates the criticality of the false positive errors (we use explicitly the symbol $\xi_I$ in the notation as it may resemble the definition of Type I error in classical statistical analysis) by taking into consideration how far the source that is falsely identified is from the ground truth. That is, a false positive committed for a nearby source must be a much less serious statistical offense than the ones committed for distant sources. The normalization constant $\mu$ guarantees that the AWROC is an unbiased statistics for random solutions, as shown empirically (**Supp. Material Fig 1**). **Fig. 2** shows an example for the five simulated sources shown in **Fig. 1**, where the Minimum Euclidean Distance (MED) of each $i$-th point of the brain cortical surface, with respect to the five ground truth sources, is shown together with the corresponding cortical distribution of $\xi_I(i)$. The latter is calculated using the straightforward normalized formula:

$$\xi_I(i) = \log_{10}\left(1 + 9\, MED(i)\Big/ \max_{j=1,\ldots,N}\{MED(j)\}\right), \tag{26}$$

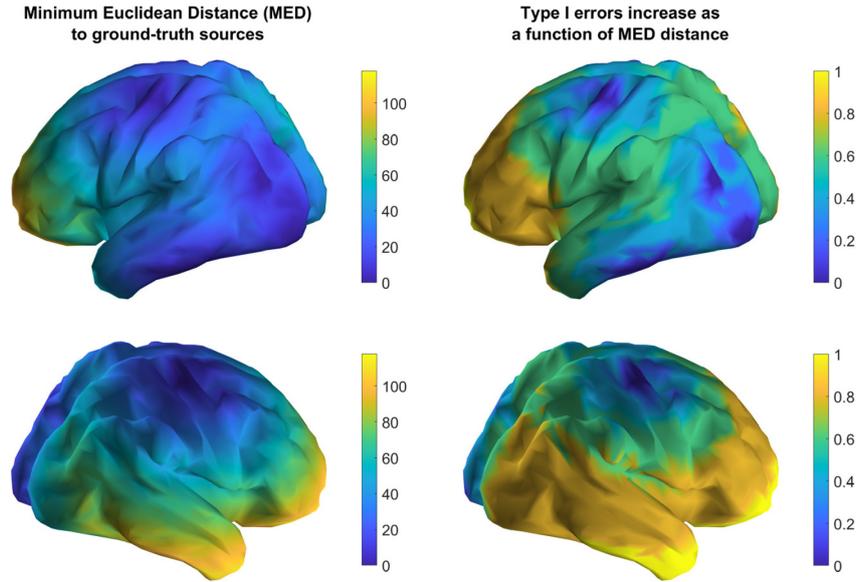

**Figure 2:** Minimum Euclidean Distance (MED) and measure of false positive errors according to the distance to the ground truth sources shows as distributed across the brain cortical surface. Top view: left hemisphere. Bottom view: right hemisphere.

## Results

### Solving straightforward state-space models with proposed algorithms

We evaluated the accuracy of proposed algorithms exhaustively using the small-scale simulations and different SNR conditions (see **Material Methods**) using 100 Monte Carlo replications, with the same simulated data used for all the algorithms. The comparison is based on the RSE statistic (**Material Methods**), as well as the evaluation of the autoregressive coefficient estimation and the algorithmic computational time.

## Analysis for simulated bivariate time series

First, we evaluate the proposed methods with the simplest small-scale model, i.e., the bivariate model introduced above. For this simulation, backpropagation converges and always recovers solutions nearby the ground truth for the close to noiseless ($\sigma_o = 0.1$) scenario, as evaluated for the Monte Carlo replications. Particularly, using the RSE statistics, the error percentages of estimated bivariate time series $\hat{\mathbf{x}}_t$ were 2.20±0.35% (mean ± mean's standard error bars) and 1.10±0.19%, respectively for the first and second latent variable, for the $\sigma_o = 0.1$ scenario (algorithm converged after 121,907.51±312.23 iterations, average computational time of 53 seconds/replica, with GD step size $\alpha = 1$, momentum $\mu = 0.99$); 27.85±6.76% and 14.54±3.03% for the $\sigma_o = 0.5$ scenario (13,468.16±348.92 iterations, 5.58 sec/replica, $\alpha = 1$, $\mu = 0.99$); and 87.41±6.22% and 55.54±6.86% for the $\sigma_o = 1.0$ scenario (202,956.8±8,808.83 iterations, 82.66 sec/replica, $\alpha = 10^{-2}$, $\mu = 0.99$, $tol = 10^{-5}$). As observed, convergence was faster for an intermediate level of noise and the GD step size must be reduced to $\alpha = 10^{-2}$ for the noisier case due to numerical instability. For the noisier case, the convergence tolerance parameter must be increased from the standard $tol = 10^{-6}$ to $10^{-5}$ due to the large number of iterations needed to achieve convergence. We also observed a large variability in the number of iterations (error bars = ±8,808.83) due to the initial random conditions which may have a more dramatic impact on convergence in this case.

In contrast, over the same Monte Carlo replications, the SSGD algorithm's RSE statistic for the estimation of the two latent variables were 2.20±0.35% and 1.10±0.19% for the $\sigma_o = 0.1$ scenario (10,461.00±9.19 iterations, 0.50 sec/replica); 27.85±6.76% and 14.54±3.03% for the $\sigma_o = 0.5$ scenario (16,679.00±297.13 iterations, 0.79 sec/replica); 89.32±5.19% and 57.40±7.25% for the $\sigma_o = 1.0$ scenario (133,066.14±20,493.18 iterations, 6.27 sec/replica). For all the SSGD analysis, we used $\alpha = 1$, $\mu = 0.99$, and $tol = 10^{-6}$. Whereas, for the SSALS algorithm, we obtained an RSE statistic of 2.20±0.35% and 1.10±0.19% for the $\sigma_o = 0.1$ scenario (3.86±0.04 iterations, 1.36 msec/replica); 27.76±6.66% and 14.51±3.00% for the $\sigma_o = 0.5$ scenario (16.06±0.59 iterations, 5.09 msec/replica); 90.12±5.42% and 56.18±6.79% for the $\sigma_o = 1.0$ scenario (1,155.70±75.00 iterations, 0.33 sec/replica).

For all the simulations above, the models were evaluated using $\lambda = \sigma_o^2/\sigma_s^2$ as in the naïve case. For the SSGD algorithm, the other hyperparameters were set to zero, i.e., $\lambda_2^{(a)} = \lambda_1^{(x)} = \lambda_1^{(a)} = 0$, and similarly for the SSALS algorithm $\lambda_2^{(x)} = \lambda_2^{(a)} = 0$. Unsurprisingly, both SSGD and SSALS algorithms converged to the same solutions as backpropagation for the same replications in the less noisier scenarios ($\sigma_o = 0.1$ and $\sigma_o = 0.5$), as corroborated by examining the RSE statistic (see also **Table 4**), but with more steady and faster convergence for SSGD, as compared to backpropagation, and with overall lightning (quadratic) convergence for SSALS in comparison to the other algorithms. Interestingly, in comparison with backpropagation, which also relies on the GD technique, the SSGD algorithm showed much better numerical stability. For example, without momentum ($\mu = 0$), SSGD also converged steadily although a bit slower, whereas backpropagation convergence was much slower as we have to set $\alpha = 10^{-3}$ or lower to avoid numerical instability.

In general, it should be noted that the RSE statistic increases dramatically for low SNR for all the algorithms, which may be a sign of overfitting. This trend occurs despite using $\lambda = \sigma_o^2/\sigma_s^2$ (naïve estimator) as acknowledging the model noise exactly for known variances (Hastie et al., 2001). Particularly, using this parameter in the implementation of our backpropagation algorithm (see **Table 1**) is equivalent to implementing the weight decay technique (Yang and Wang, 2020). Although we did not consider a similar regularization term for the estimation of the autoregressive matrices in this algorithm, the above comparison is fair as the corresponding hyperparameters were turned to zero in the other algorithms. Complementarily to these analyses, **Table 4** shows the outcome for the estimated autoregressive coefficients for each SNR scenario and algorithms, showing clearly that the parameter bias increases as the SNR decreases. Here, we consider that the more faithful estimates are provided by the SSALS algorithm due to its superior numerical stability and quadratic convergence. Therefore, this outcome may also suggest that the SSGD algorithm is superior to backpropagation in these simulations, as it converges closer to the SSALS solutions.

**Table 4**: Mean ± error bars' solutions for 100 Monte Carlo replications of the bivariate (lag=1) simulation for each SNR scenario using **Eqs. (1, 2)**. The error bars are calculated by estimating the standard deviation of the Monte Carlo calculated samples and then dividing this value by the squared root of the number of Monte Carlo simulations. The values of simulation parameters are $T = 200$, $M = 5$, $N = 2$, and $P = 1$. From left to right the ground-truth values of the autoregressive matrix are shown, together with the corresponding solutions for $\sigma_o = 0.1$, $\sigma_o = 0.5$, and $\sigma_o = 1$. We set $\sigma_s = 1$ in all simulations. The solutions are provided for the calculations based on the backpropagation, SSGD and SSALS algorithms, as presented from top-to-bottom rows in this order.

| Ground Truth | $\sigma_o = 0.1$ | $\sigma_o = 0.5$ | $\sigma_o = 1$ |
|---|---|---|---|
| | $\begin{bmatrix} -0.51 \pm 0.06 & 0.00 \pm 0.04 \\ 0.72 \pm 0.06 & -0.50 \pm 0.05 \end{bmatrix}$ | $\begin{bmatrix} -0.48 \pm 0.21 & -0.03 \pm 0.06 \\ 1.24 \pm 0.33 & -0.53 \pm 0.12 \end{bmatrix}$ | $\begin{bmatrix} 1.73 \pm 1.76 & -0.78 \pm 1.14 \\ 6.66 \pm 5.08 & -2.04 \pm 1.75 \end{bmatrix}$ |
| $\begin{bmatrix} -0.5 & 0 \\ 0.7 & -0.5 \end{bmatrix}$ | $\begin{bmatrix} -0.51 \pm 0.06 & 0.00 \pm 0.04 \\ 0.72 \pm 0.06 & -0.50 \pm 0.05 \end{bmatrix}$ | $\begin{bmatrix} -0.47 \pm 0.21 & -0.03 \pm 0.06 \\ 1.24 \pm 0.33 & -0.53 \pm 0.12 \end{bmatrix}$ | $\begin{bmatrix} 3.22 \pm 2.53 & -1.45 \pm 1.79 \\ 12.41 \pm 4.87 & -3.52 \pm 2.49 \end{bmatrix}$ |
| | $\begin{bmatrix} -0.51 \pm 0.06 & 0.00 \pm 0.04 \\ 0.72 \pm 0.06 & -0.50 \pm 0.05 \end{bmatrix}$ | $\begin{bmatrix} -0.48 \pm 0.21 & -0.03 \pm 0.06 \\ 1.23 \pm 0.33 & -0.53 \pm 0.11 \end{bmatrix}$ | $\begin{bmatrix} 3.56 \pm 2.23 & -1.64 \pm 1.52 \\ 11.58 \pm 4.03 & -3.85 \pm 2.19 \end{bmatrix}$ |

## Analysis for simulated three-variate time series

To further compare these algorithms, in this section we used the three-variate simulation and set the hyperparameter values as in the naïve estimators for each SNR scenario. Only SSGD and SSALS solutions are presented here as backpropagation convergence was significantly slower and less stable for the Monte Carlo replications. Moreover, we ignored the use of momentum for the SSGD algorithm ($\mu = 0$) due to numerical instability. For this method, the corresponding RSE statistics for the three latent variables were 2.50±0.06%, 1.99±0.06%, and 1.97±0.06% for the $\sigma_o = 0.1$ scenario (716,605.92±185,016.49 iterations, 88.82 sec/replica, $\alpha = 1$ and $tol = 10^{-6}$); 67.35±1.38%, 59.14±0.95%, and 52.38±0.35% for the $\sigma_o = 0.5$ scenario (686,235.88±33,749.76 iterations, 84.01 sec/replica, $\alpha = 10^{-3}$ and $tol = 10^{-5}$); 74.08±1.39%, 66.28±1.07%, and 54.38±0.36% for the $\sigma_o = 1.0$ scenario (617,378.55±51,054.21 iterations, 73.86 sec/replica, $\alpha = 10^{-3}$ and $tol = 10^{-5}$). Notice that we increased the tolerance due to slow convergence and reduced $\alpha$ to $10^{-3}$ in the noisier scenarios due to numerical instability. In contrast, for the SSALS algorithm, the RSE statistics were 2.42±0.06%, 1.88±0.05%, and 1.86±0.05 for the $\sigma_o = 0.1$ scenario (24.55±2.63 iterations, 0.04 sec/replica); 55.11±1.34%, 42.42±0.76%, and 39.61±0.62% for the $\sigma_o = 0.5$ scenario (3,757.89±337.23 iterations, 6.68 sec/replica); 150.54±4.18%, 83.3±1.69%, and 62.14±1.21% for the $\sigma_o = 1.0$ scenario (8,052.59±579.40 iterations, 16.80 sec/replica).

Consistent with the previous analysis, we observe that $\hat{\mathbf{x}}_t$ gradually diverged as the SNR decreased, as reflected by the RSE statistic. Moreover, as shown in **Table 5**, the same issue occurred with the estimated autoregressive coefficients. For the second simulation, in comparison to the first simulation, we had to deal with the worst numerical problems for the SSGD algorithm that forced a reduction of the GD step size and an increase of the convergence tolerance with increasingly more noise. Interestingly, for the $\sigma_o = 0.5$ scenario we rerun the SSGD algorithm for the same Monte replication but with increased tolerance $tol = 10^{-4}$ and $\alpha = 1$, and obtained different RSE values for the three latent variables: 44.74±1.14%, 30.1±0.64%, and 25.25±0.53%. These values are better than those obtained previously with more strict optimization parameters, which may have resulted in overfitting. Notice that increasing the SSGD algorithm tolerance value is equivalent to inducing early stopping, which is an optimization technique used for training ANNs to avoid overfitting. Also observe in **Table 5** that the solutions are more disparate for the SSALS algorithm in the noisier scenarios. If the SSALS algorithm more accurately captures the numerical solution due to its apparent superior convergence, then it is expected also that it will be more affected by overfitting.

**Table 5:** SSGD and SSALS mean ± error bars' solutions for 100 Monte Carlo replications of the three-variate (lag=1,2,3) simulation for each SNR scenario, using **Eqs. (1, 2)**. The values of simulation parameters are $T = 240$, $M = 5$, $N = 3$, and $P = 3$. (**I**) The simulation's ground-truth autoregressive coefficients were obtained according to Stokes and Purdon(Stokes and Purdon, 2018), presented between brackets (from the left to right) correspondingly to lag=1,2,3, in this order. (**II**) From top-to-bottom rows are shown the corresponding solutions for the SSGD algorithm, for $\sigma_o = 0.1$, $\sigma_o = 0.5$, and $\sigma_o = 1$ in this order. We set $\sigma_s = 1$ in all simulations. (**III**) Similarly, but for the solutions estimated by the SSALS algorithm.

| I) Ground-truth autoregressive coefficients |
|---|
| $\mathbf{A} = \left\{ \begin{bmatrix} -0.9000 & 0 & 0 \\ -0.3560 & 1.2124 & 0 \\ 0 & -0.3098 & -1.3856 \end{bmatrix}, \begin{bmatrix} -0.8100 & 0 & 0 \\ 0.7136 & -0.4900 & 0 \\ 0 & 0.5000 & -0.6400 \end{bmatrix}, \begin{bmatrix} 0 & 0 & 0 \\ -0.3560 & 0 & 0 \\ 0 & -0.3098 & 0 \end{bmatrix} \right\}$ |
| II) SSGD estimated autoregressive coefficients |
| $\left\{ \begin{bmatrix} -1.00 \pm 0.06 & -0.32 \pm 0.12 & 0.05 \pm 0.09 \\ -0.22 \pm 0.02 & 1.57 \pm 0.05 & -0.21 \pm 0.04 \\ 0.57 \pm 0.09 & -0.47 \pm 0.13 & -2.05 \pm 0.09 \end{bmatrix}, \begin{bmatrix} -0.77 \pm 0.17 & 0.60 \pm 0.28 & -0.01 \pm 0.21 \\ 0.91 \pm 0.05 & -1.01 \pm 0.10 & -0.24 \pm 0.09 \\ 1.17 \pm 0.22 & 1.19 \pm 0.29 & -1.81 \pm 0.21 \end{bmatrix}, \begin{bmatrix} 0.55 \pm 0.39 & -0.38 \pm 0.16 & -0.04 \pm 0.13 \\ -0.65 \pm 0.13 & 0.26 \pm 0.06 & -0.07 \pm 0.05 \\ 1.60 \pm 0.44 & -0.80 \pm 0.19 & -0.65 \pm 0.13 \end{bmatrix} \right\}$ |
| $\left\{ \begin{bmatrix} -0.39 \pm 0.03 & 0.64 \pm 0.03 & -0.97 \pm 0.02 \\ -0.53 \pm 0.05 & 1.42 \pm 0.03 & -1.27 \pm 0.06 \\ -0.24 \pm 0.04 & 0.92 \pm 0.03 & -1.38 \pm 0.04 \end{bmatrix}, \begin{bmatrix} -1.36 \pm 0.03 & 1.14 \pm 0.03 & 0.33 \pm 0.03 \\ -0.65 \pm 0.05 & 1.11 \pm 0.04 & -0.06 \pm 0.06 \\ -1.18 \pm 0.05 & 1.42 \pm 0.04 & 0.08 \pm 0.04 \end{bmatrix}, \begin{bmatrix} 0.87 \pm 0.03 & -0.85 \pm 0.02 & 0.15 \pm 0.03 \\ 0.41 \pm 0.06 & -1.22 \pm 0.04 & 0.84 \pm 0.07 \\ 0.64 \pm 0.05 & -1.15 \pm 0.04 & 0.62 \pm 0.06 \end{bmatrix} \right\}$ |
| $\left\{ \begin{bmatrix} -0.44 \pm 0.06 & 1.03 \pm 0.05 & -1.16 \pm 0.06 \\ -0.54 \pm 0.10 & 1.51 \pm 0.08 & -1.41 \pm 0.13 \\ -0.25 \pm 0.06 & 1.15 \pm 0.06 & -1.51 \pm 0.08 \end{bmatrix}, \begin{bmatrix} -1.70 \pm 0.06 & 1.07 \pm 0.07 & 0.66 \pm 0.08 \\ -0.78 \pm 0.11 & 1.34 \pm 0.08 & -0.16 \pm 0.13 \\ -1.50 \pm 0.07 & 1.44 \pm 0.07 & 0.30 \pm 0.08 \end{bmatrix}, \begin{bmatrix} 1.27 \pm 0.06 & -0.73 \pm 0.06 & -0.39 \pm 0.08 \\ 0.43 \pm 0.11 & -1.27 \pm 0.08 & 0.88 \pm 0.14 \\ 0.85 \pm 0.07 & -1.10 \pm 0.07 & 0.35 \pm 0.09 \end{bmatrix} \right\}$ |
| III) SSALS estimated autoregressive coefficients |
| $\left\{ \begin{bmatrix} -1.07 \pm 0.04 & -0.14 \pm 0.08 & 0.16 \pm 0.06 \\ -0.23 \pm 0.02 & 1.49 \pm 0.03 & -0.22 \pm 0.04 \\ 0.40 \pm 0.05 & -0.28 \pm 0.08 & -1.92 \pm 0.05 \end{bmatrix}, \begin{bmatrix} -1.02 \pm 0.09 & 0.18 \pm 0.15 & 0.24 \pm 0.13 \\ 0.94 \pm 0.03 & -0.87 \pm 0.06 & -0.29 \pm 0.07 \\ 0.74 \pm 0.11 & 0.70 \pm 0.16 & -1.50 \pm 0.12 \end{bmatrix}, \begin{bmatrix} -0.03 \pm 0.22 & -0.13 \pm 0.08 & 0.12 \pm 0.08 \\ -0.50 \pm 0.08 & 0.18 \pm 0.03 & -0.11 \pm 0.04 \\ 0.82 \pm 0.23 & -0.43 \pm 0.09 & -0.45 \pm 0.08 \end{bmatrix} \right\}$ |
| $\left\{ \begin{bmatrix} -4.14 \pm 1.75 & 8.17 \pm 2.76 & -1.64 \pm 3.14 \\ -5.94 \pm 2.17 & 2.91 \pm 1.25 & 6.14 \pm 1.71 \\ -1.16 \pm 1.99 & 1.33 \pm 1.03 & 0.36 \pm 1.24 \end{bmatrix}, \begin{bmatrix} -14.08 \pm 2.20 & -10.19 \pm 7.05 & 12.11 \pm 11.22 \\ 2.06 \pm 1.60 & 2.44 \pm 3.45 & 27.26 \pm 6.23 \\ 0.64 \pm 1.33 & 8.58 \pm 4.78 & 5.65 \pm 4.15 \end{bmatrix}, \begin{bmatrix} -9.73 \pm 3.47 & 9.43 \pm 11.84 & -2.63 \pm 15.44 \\ -14.11 \pm 3.09 & -36.49 \pm 6.78 & 50.76 \pm 9.51 \\ -6.58 \pm 3.21 & -18.76 \pm 8.13 & 25.07 \pm 10.89 \end{bmatrix} \right\}$ |
| $\left\{ \begin{bmatrix} -2.03 \pm 3.44 & 18.79 \pm 5.72 & 9.52 \pm 5.26 \\ 3.76 \pm 4.12 & 1.14 \pm 2.52 & 10.00 \pm 4.73 \\ -2.16 \pm 2.88 & 1.84 \pm 2.10 & 0.16 \pm 2.09 \end{bmatrix}, \begin{bmatrix} -14.66 \pm 5.14 & 3.30 \pm 16.54 & 58.31 \pm 21.84 \\ -16.54 \pm 5.40 & 1.51 \pm 9.07 & 58.00 \pm 19.03 \\ -8.16 \pm 3.61 & -9.14 \pm 7.38 & 16.74 \pm 9.22 \end{bmatrix}, \begin{bmatrix} 1.18 \pm 8.16 & -50.23 \pm 27.39 & 55.59 \pm 32.61 \\ -1.40 \pm 6.68 & -57.34 \pm 14.40 & 74.81 \pm 16.95 \\ -12.70 \pm 4.36 & -7.21 \pm 10.71 & 20.56 \pm 13.86 \end{bmatrix} \right\}$ |

## The problem of overfitting in state-space models

In previous analyses, the effects of overfitting were more noticeable for the noisier cases of $\sigma_o = 0.5$ and $\sigma_o = 1$, mainly for estimating the autoregressive coefficients as shown in **Tables 4-5**. In contrast, the corresponding estimated time series $\hat{\mathbf{x}}_t$ were very stable in most cases as reflected by the RSE statistic, and as demonstrated in **Fig. 3** for a particular replication of the two small-scale simulations, where solutions were obtained using the SSGD algorithm. For this example, for the three-variate simulation, it was remarkable that the curves for estimated time series were very close to the ground truth even for low SNR (**Fig. 3B**). Nonetheless, the accuracy for the estimation of the time series can also be severely deteriorated for low SNR, as shown for the first simulation (**Fig. 3A**, last row plot).

Therefore, an obvious observation is that the curse of overfitting increases for low SNR. In these cases, knowing the generative ground-truth model does not ensure a robust estimation (e.g., for the naïve estimator), as overfitting can occur either when estimating the state or space equation or balanced in between. For example, evaluating a model with a high degree of freedom (DOF) for the estimator $\hat{\mathbf{x}}_t$ and a low DOF for the estimator $\hat{\mathbf{A}}_p$ (e.g., a sparse estimator of the autoregressive coefficients) can be as suboptimal as estimating a model with a low DOF for $\hat{\mathbf{x}}_t$ (e.g., a sparse spatiotemporal solution) and a high DOF for $\hat{\mathbf{A}}_p$, despite apparently adequate data fitting. Therefore, a pertinent question concerns how we select the correct model with real data in large-scale analysis. As the next section demonstrates, a sensible answer is to use data-driven regularisation approach based on $K$-fold cross-validation.

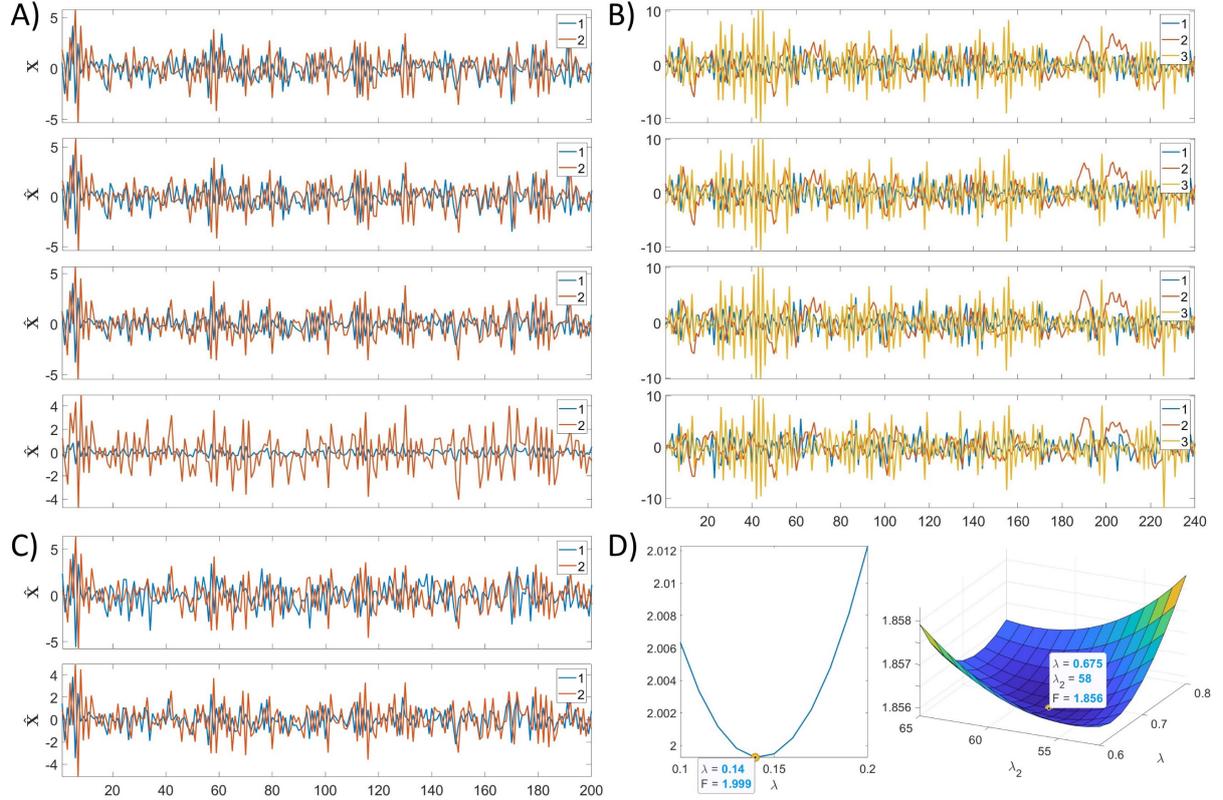

**Figure 3:** Solving state-space models using the SSGD algorithm for bivariate (lag=1) and three-variate (lag=1,2,3) simulations for a single replication in different signal-to-noise ratio (SNR) scenarios: $\sigma_o = 0.1$, $\sigma_o = 0.5$, and $\sigma_o = 1$, corresponding to decreasing SNR levels, with fixed $\sigma_s = 1$ for all cases. Ground-truth autoregressive coefficients are shown in **Tables 4-5** and dynamics are generated using **Eqs. (1, 2)**. **A**) Bivariate simulation: from top to bottom are plotted the ground-truth dynamics $x_t$ and estimated solutions $\hat{x}_t$ for decreasing SNR levels. **B**) Three-variate simulation: similar to **A**. In both **A** and **B**, distinct colours differentiate the two- or three-variate time series curves in the legend for easier visual comparison between ground-truth and estimated time series. **C**) Two additional solutions are calculated for the bivariate model for the same synthetic replication in the highest noise level ($\sigma_o = 1$) but using MPSS models described by **Eqs. (9, 10)**. Hyperparameters selection is based on the $K = 5$ fold cross-validation with imputed data ($K$-fold CVI) procedure. The first regularized solution (top row) is obtained by adjusting only $\lambda$ (line search), while the second solution (bottom row) is obtained by assessing both $\lambda$ and $\lambda_2^{(a)}$ (plane search), while the other hyperparameter values are set to zero. **D**) Cross-validation predicted error curves (**Eq. (16)** in **Materials and Methods**) are displayed for both line and plane searches on the left and right, respectively. The "optimal" hyperparameter values and corresponding minimum value of the curve and surface are highlighted with a filled circle marker (orange colour) and within the inset legends. The x-axis and x-y axes in these plots represent the hyperparameters' evaluated domains. The coloured surface employs the (Matlab) jet colormap for improved 3D visualization.

## Hyperparameter estimation using $K$-fold cross-validation based on imputed data

Here, we demonstrate an extension of the $K$-fold cross-validation procedure, based on imputed data ($K$-fold CVI, see **Materials and Methods**) to deal with the estimation of regularisation parameters in MPSS models, and thus illustrate how $K$-fold CVI can help to control overfitting. As a preliminary demonstration of our solution to the hyperparameter selection problem, **Table 6** shows the SSGD and SSALS solutions for the bivariate simulation in the noisiest scenario ($\sigma_o = 1$), for the more general MPSS framework. Solving MPSS models was demonstrated separately with the assessment of a single regularization parameter (i.e., estimating $\lambda$), called here as line search, or additionally, also considering a penalty term over the estimated autoregressive coefficients (i.e., estimating $\lambda$ and $\lambda_2^{(a)}$), called here as plane search. For this case, for the SSGD algorithm, the RSE statistics for the two latent variables were 85.75±2.27% and 46.72±1.01% for the line search (14.79 sec/replica; line-grid dimension equals to $21 \times 1$); and 57.96±1.10% and 32.96±0.64% for the plane search (54.13 sec/replica; plane-grid dimension equals to $31 \times 21$). Whereas, for the SSALS algorithm, the RSE statistics were 84.19±2.13% and 46.18±1.03% (0.32 sec/replica; line-grid dimension equals to $21 \times 1$) for the line search, and 58.32± 1.08% and 33.14±0.63 (2.60 sec/replica; plane-grid dimension equals $31 \times 10$) for the plane search analysis.

In comparison, recall from above sections that for the same noise scenario and using the same Monte Carlo replications, the RSE statistics for the naïve estimators were 89.32±5.19% and 57.40±7.25%, and 90.12±5.42% and 56.18±6.79%, respectively for the SSGD and SSALS algorithms (see **Figs 3C,D** as demonstration for a single replication using the SSGD algorithm). Therefore, we found that both line and plane search regularization solutions improved the naïve estimation, which is also evident from the comparison between the estimated autoregressive coefficients (compare outcomes in **Tables 4 and 6**). The other clear observation is that regularization improved the algorithms' computational performance (as shown by the measurements of seconds per replica or sec/replica above), as execution time remained more or less the same despite of having to estimate the solutions with a grid-search procedure, for $k = 5$ folds, for each Monte Carlo replication. Here, we used the same SSALS' Matlab code for calculating the naïve and the regularized solutions, but for the SSGD implementation we used its Matlab mex-function optimized implementation for the regularized case instead of the straightforward Matlab implementation, which was used before when comparing backpropagation vs SSGD, due to significantly faster computation with the mex function. Moreover, as expected from the regularization advantages, numerical stability of the SSGD algorithm also improved as their solutions were very close to SSALS, which was reflected in both the RSE values and estimated autoregressive coefficients (**Table 6**).

**Table 6**: SSGD and SSALS mean ± error bars' solutions for 100 Monte Carlo replications of the bivariate (lag=1) simulation for the noisiest scenario ($\sigma_o = 1$), The solutions were obtained from solving MPSS models using a line search *K*-fold cross-validation procedure to evaluate $\lambda$ or, similarly, a plane search to evaluate both $\lambda$ and $\lambda_2^{(a)}$.

| Method | Line search | Plane search |
|---|---|---|
| SSGD | $\begin{bmatrix} -0.46 \pm 0.01 & 0.03 \pm 0.01 \\ 0.56 \pm 0.02 & -0.44 \pm 0.01 \end{bmatrix}$ | $\begin{bmatrix} -0.18 \pm 0.02 & -0.02 \pm 0.01 \\ 0.51 \pm 0.01 & -0.51 \pm 0.01 \end{bmatrix}$ |
| SSALS | $\begin{bmatrix} -0.46 \pm 0.01 & 0.03 \pm 0.01 \\ 0.57 \pm 0.01 & -0.44 \pm 0.01 \end{bmatrix}$ | $\begin{bmatrix} -0.20 \pm 0.02 & -0.02 \pm 0.01 \\ 0.50 \pm 0.01 & -0.51 \pm 0.01 \end{bmatrix}$ |

For the three-variate simulation, in contrast, we ran two- and three-dimensional subspaces search for the lower SNR scenarios for a single replication as demonstration, and plotted the corresponding prediction error curves (**Fig. 4**). Here, the analysis was performed using only the SSGD algorithm because the SSALS algorithm does not implement L1-norm based sparse regularization. For clarity, for the naïve estimator for this replication, the corresponding values of the RSE statistic were 6.7%, 4.2%, and 1.8%, for $\sigma_o = 0.5$ (3,671,869 iterations, $\alpha = 10^{-2}$), and 32.7%, 18.3%, and 7.5% for $\sigma_o = 1$ (18,715,396 iterations, $\alpha = 10^{-3}$). In comparison, for the regularised solutions, for the same replica in the $\sigma_o = 0.5$ scenario, the RSE statistics were 5.2%, 3.7%, and 1.5% (7,532 iterations), 5.5%, 3.8%, and 1.6% (7,912 iterations), and 5.4%, 3.7%, and 1.5% (7,559 iterations), for the solutions corresponding to the search over subspaces $\left(\lambda, \lambda_2^{(a)}, \lambda_1^{(a)}\right)$, $\left(\lambda, \lambda_2^{(a)}\right)$, and $\left(\lambda, \lambda_1^{(a)}\right)$, in this order. For the same replica in the $\sigma_o = 1$ scenario, the RSE statistics were 11.8%, 10.0%, and 3.8% (6,440 iterations), 12.0%, 9.9%, and 3.8% (8,200 iterations), and 11.6%, 9.8%, and 3.7% (9,164 iterations), correspondingly. **Table 7** shows the estimated autoregressive coefficients for each case.

**Table 7:** SSGD regularised solutions for the same replication of the three-variate simulation for the noisier scenarios ($\sigma_o = 0.5$ and $\sigma_o = 1$). **I)** For $\sigma_o = 0.5$, from top to bottom, are shown the estimated autoregressive coefficients obtained by exploring the hyperparameters for the subspaces $\{(\lambda, \lambda_2^{(a)}, \lambda_1^{(a)}) | \lambda \geq 0, \lambda_2^{(a)} \geq 0, \lambda_1^{(a)} \geq 0\}$, $\{(\lambda, \lambda_2^{(a)}) | \lambda \geq 0, \lambda_2^{(a)} \geq 0\}$, and $\{(\lambda, \lambda_1^{(a)}) | \lambda \geq 0, \lambda_1^{(a)} \geq 0\}$, in this order. We used the hyperparameter values shown within the **Fig. 4** inset legends. **II)** Likewise for the $\sigma_o = 1$ scenario.

| | I) Autoregressive coefficients estimated for the $\sigma_o = 0.5$ case |
|---|---|
| $\lambda, \lambda_2^{(a)}, \lambda_1^{(a)}$ | $\mathbf{A} = \left\{ \begin{bmatrix} -0.94 & 0.06 & 0 \\ -0.23 & 1.07 & 0 \\ 0 & -0.07 & -1.49 \end{bmatrix}, \begin{bmatrix} -0.7 & 0 & 0 \\ 0.79 & -0.15 & 0 \\ 0 & 0 & -0.71 \end{bmatrix}, \begin{bmatrix} 0 & -0.07 & 0 \\ 0 & -0.24 & -0.02 \\ -0.39 & -0.04 & 0.01 \end{bmatrix} \right\}$ |
| $\lambda, \lambda_2^{(a)}$ | $\mathbf{A} = \left\{ \begin{bmatrix} -0.84 & 0.07 & -0.04 \\ -0.38 & 1.03 & 0.08 \\ 0 & -0.29 & -1.24 \end{bmatrix}, \begin{bmatrix} -0.57 & 0.04 & -0.07 \\ 0.56 & -0.12 & 0.11 \\ -0.04 & 0.35 & -0.34 \end{bmatrix}, \begin{bmatrix} 0.12 & -0.13 & -0.04 \\ -0.21 & -0.23 & 0.02 \\ -0.20 & -0.20 & 0.20 \end{bmatrix} \right\}$ |
| $\lambda, \lambda_1^{(a)}$ | $\mathbf{A} = \left\{ \begin{bmatrix} -0.93 & 0.03 & 0 \\ -0.20 & 1.07 & 0 \\ 0 & -0.06 & -1.50 \end{bmatrix}, \begin{bmatrix} -0.71 & 0 & 0 \\ 0.81 & -0.15 & 0 \\ 0 & 0 & -0.73 \end{bmatrix}, \begin{bmatrix} 0 & -0.06 & 0 \\ 0 & -0.23 & -0.02 \\ -0.39 & -0.02 & 0 \end{bmatrix} \right\}$ |
| | II) Autoregressive coefficients estimated for the $\sigma_o = 1$ case |
| $\lambda, \lambda_2^{(a)}, \lambda_1^{(a)}$ | $\mathbf{A} = \left\{ \begin{bmatrix} -0.89 & 0 & -0.03 \\ -0.29 & 1.02 & 0 \\ 0 & 0 & -1.18 \end{bmatrix}, \begin{bmatrix} -0.71 & 0 & 0 \\ 0.61 & 0 & 0 \\ 0 & 0 & -0.13 \end{bmatrix}, \begin{bmatrix} 0 & 0 & 0 \\ 0 & -0.33 & 0 \\ -0.59 & -0.09 & 0.35 \end{bmatrix} \right\}$ |
| $\lambda, \lambda_2^{(a)}$ | $\mathbf{A} = \left\{ \begin{bmatrix} -0.67 & 0.05 & -0.03 \\ -0.40 & 0.89 & 0.03 \\ 0.08 & -0.26 & -1.06 \end{bmatrix}, \begin{bmatrix} -0.41 & 0.09 & -0.01 \\ 0.46 & 0.09 & -0.03 \\ 0.01 & 0.36 & -0.02 \end{bmatrix}, \begin{bmatrix} 0.27 & -0.16 & -0.01 \\ -0.11 & -0.34 & -0.06 \\ -0.27 & -0.23 & 0.38 \end{bmatrix} \right\}$ |
| $\lambda, \lambda_1^{(a)}$ | $\mathbf{A} = \left\{ \begin{bmatrix} -0.96 & 0 & -0.02 \\ 0 & 1.17 & 0 \\ 0 & 0 & -1.56 \end{bmatrix}, \begin{bmatrix} -0.75 & 0 & 0 \\ 0.97 & -0.24 & 0 \\ 0 & 0 & -0.78 \end{bmatrix}, \begin{bmatrix} 0 & -0.06 & 0 \\ 0 & -0.20 & 0 \\ -0.45 & 0 & 0 \end{bmatrix} \right\}$ |

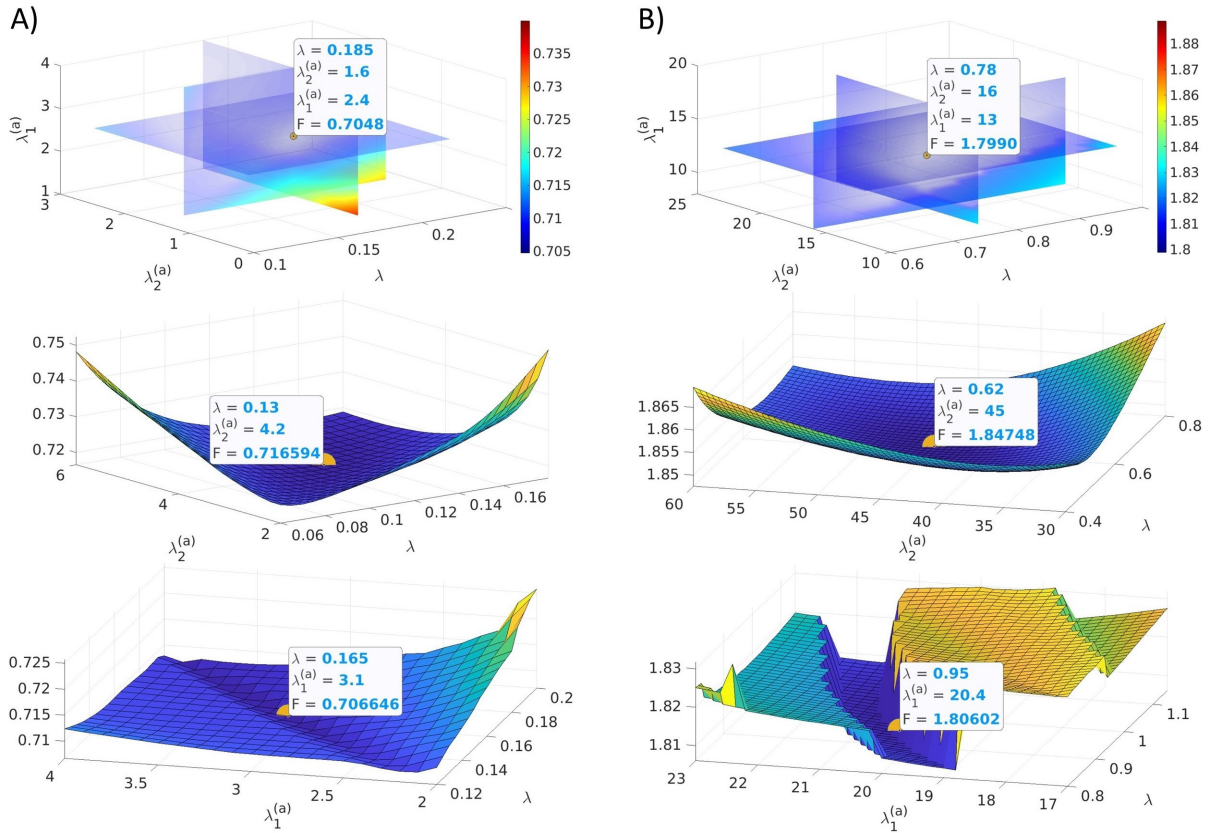

**Figure 4:** Prediction error curves obtained with the extension of $K$-fold cross-validation, based on imputed data ($K$-fold CVI), technique as demonstrated for the three-variate simulation for the noisier scenarios: $\sigma_o = 0.5$ (**A**) and $\sigma_o = 1$ (**B**). From top to bottom, the plots show the curves for separated multigrid search analyses corresponding to the subspaces

$\left\{\left(\lambda, \lambda_2^{(a)}, \lambda_1^{(a)}\right) \mid \lambda \geq 0, \lambda_2^{(a)} \geq 0, \lambda_1^{(a)} \geq 0\right\}$, $\left\{\left(\lambda, \lambda_2^{(a)}\right) \mid \lambda \geq 0, \lambda_2^{(a)} \geq 0\right\}$, and $\left\{\left(\lambda, \lambda_1^{(a)}\right) \mid \lambda \geq 0, \lambda_1^{(a)} \geq 0\right\}$, in this order. In the plots in 2nd and 3rd rows, the x-y axes show the hyperparameter search domains, while the z-axis show the corresponding values of the prediction error function (**Eq. (16)** in **Materials and Methods**), which are also represented by the colour-coded surfaces using (Matlab) jet colourmap. Similarly, for the 1st row plots, the search domain is represented by all x-y-z axes for the three used hyperparameters. Here, the prediction error is a 4D hypersurface or volume, where values are shown for the cutting (transparent) orthogonal planes converging on the hypersurface minima. In all the plots, the minimum points are represented with an orange-filled circle marker attached to an inset legend, where it is shown the optimal hyperparameter values with the corresponding minimum value for the estimated cross-validation prediction error.

Summarising our observations, firstly, regularisation has a noticeable positive effect on the estimation of better models, where improved solutions may be achieved by using more regularisation parameters. Additionally, the $\ell_1$-norm-based regularisation produced sparse estimators as expected (**Table 7**), which may enhance interpretation but at higher computational cost. Secondly, despite the increased computational time for hyperparameters search, regularisation improves the numerical condition for optimisation problems (Tikhonov and Arsenin, 1977), adding stability to the solutions and accelerating the convergence speed. However, searching the hyperparameter subspaces needs to be exercised with caution because the values can change abruptly, as shown by the ridged valley in **Fig. 4B** (bottom row). Mainly, they can change dramatically with low SNR level and high number of hyperparameters. Altogether, these results revealed the advantages of fitting MPSS models with the proposed data-driven regularisation approach.

Finally, to compare against current methodologies, we also calculated the solutions for the 100 Monte Carlo replications of the two small-scale simulations using the standard and Bayesian implementation of state-space models provided in Matlab R2022a's Econometric toolbox: functions "ssm.estimate" and "bssm.estimate" (James Durbin and Koopman, 2012). For the standard approach, our code used constrained (Matlab "fmincon" function) and interior-point search optimization and the covariance "sandwich" calculation method, as other settings failed to produce acceptable solutions (check scripts provided in **Supp. Materials Tables 1-6**). These constraints are not needed with the Bayesian solver as it assumes a *priori* Gaussian distributions for the parameters and the initial state variables and observation errors, for which covariances can be estimated using Monte Carlo simulations or importance sampling methods (James Durbin and Koopman, 2012). Note that our regularised methods produced comparable results to these approaches, although they may introduce unnecessary estimation bias if hyperparameters values are not appropriately selected (see **Supp. Materials Tables 7-8** and discussion therein). Clearly, our methods are significantly more computationally efficient than standard approaches, particularly evident for the three-variate simulations. However, this comparison depend on several factors such as the number of *K* folds, the granularity resolution of the search subspaces and the number of hyperparameters. We also did not consider the coding differences among these algorithms. Critically, an unquestionable advantage of our methods is that both $\hat{\mathbf{x}}_t$ and $\widehat{\mathbf{A}}_p$ estimates are produced. In contrast, state-of-the-art methods often calculate only estimators for $\widehat{\mathbf{A}}_p$, the initial state variables and observation errors.

## Solving large-scale MPSS models for synthetic MEG/EEG data

For the large-scale simulations, we solved the brain source localization and functional connectivity problems simultaneously using MPSS models and SSALS/HGDALS algorithm for synthetic MEG/EEG signals, for conditions resembling resting state and event-related experiments (see **Materials and Methods**). In the first section below, a preliminary analysis is conducted, involving five simulated sources as ground truth, to show in great detail the apparent moderate accuracy of estimated sources and their estimated dynamics for a single replication. In this case, synthetic MEG signals were generated with $SNR = 20$ dB. Whereas, in the second section, we conducted a Monte Carlo simulation to compare against state-of-the-art source localization methods. This latter analysis was performed for synthetic EEG signals, generated with $SNR = 5$ dB, and included 20 replications for random locations of either the two- or five-source scenarios combined with the simulation of activity patches of extension 6 or 15 cm$^2$ to represent different scenarios. This total amount of 80 random replications were created for a more in depth validation analysis in a comparison against state-of-the-art methods.

*Analysis for simulated resting state conditions*

**Fig. 5** shows several outcomes for the first of these analysis using synthetic MEG signals. The data was generated for conditions resembling resting state with epoched data. Assuming weak stationary conditions as represented by the generative (MPSS) models, as usually done also for real data, estimating the MPSS model in this noisy and highly underdetermined scenario should allow us to estimate the hidden dynamics, including the FC networks. Particularly, we estimate minimum norm solutions together with the SSALS/HGDALS estimator as shown on the cortical surface for comparison purposes in **Fig. 5A**. This solution reveals the estimated source activity that more closely matched the ground-truth sources at the left-hemisphere occipital, temporo-parieto-occipital junction, and inferior temporal regions, and bilateral (symmetric) somatosensory cortices (top row). In contrast, using the minimum norm inverse solution (bottom row), which is approximately equivalent to the SSALS/HGDALS solution for a single iteration, it is more difficult to guess where the relevant brain activity takes place with so many salient spots in the solution. A less experienced observer can be confused by these solutions because of the high- and low-intensity patches that lie adjacent to each other, but an expert can realise that some of these patches with opposed polarity may correspond to the same underlying source. In general, as result of signal leakage in estimated sources, dipoles with opposite polarities tend to lie on the opposite walls surrounding a brain sulcus. Thus, they will have similar waveforms but opposite signs (see **Supp. Materials Fig. 2**).

To better identify source locations, from the total amount of $N = 2004$, we selected the 100 most salient estimated dipoles according to the average spectral energy (**Supp. Materials Fig. 3**). As shown in **Fig. 5B**, using four views of the cortical surface, we corroborated that these dipoles are in clusters around the five simulated ground-truth sources (see also **Supp. Materials Fig. 4**). Additionally, **Fig. 5C** shows that the estimated time series dynamics for these dipoles are also very close to the ground-truth simulated epoched waveforms for each of the five simulated sources, after manually correcting for the polarity sign (**Supp. Materials Fig. 2**). As expected, the time series estimation may appear more accurate for the posterior than anterior regions by visual inspection because the posterior regions' activity changes can be explained by the activity changes in the anterior regions for the lag=1,…5 time delays. Here, the anterior/posterior description corresponds to the precedency order according to the simulated connections (**Fig. 5F**). Notice also the scale difference between ground-truth and estimated waveforms which is expected because we recovered between 11 and 31 dipoles for each simulated source (**Fig. 5C**).

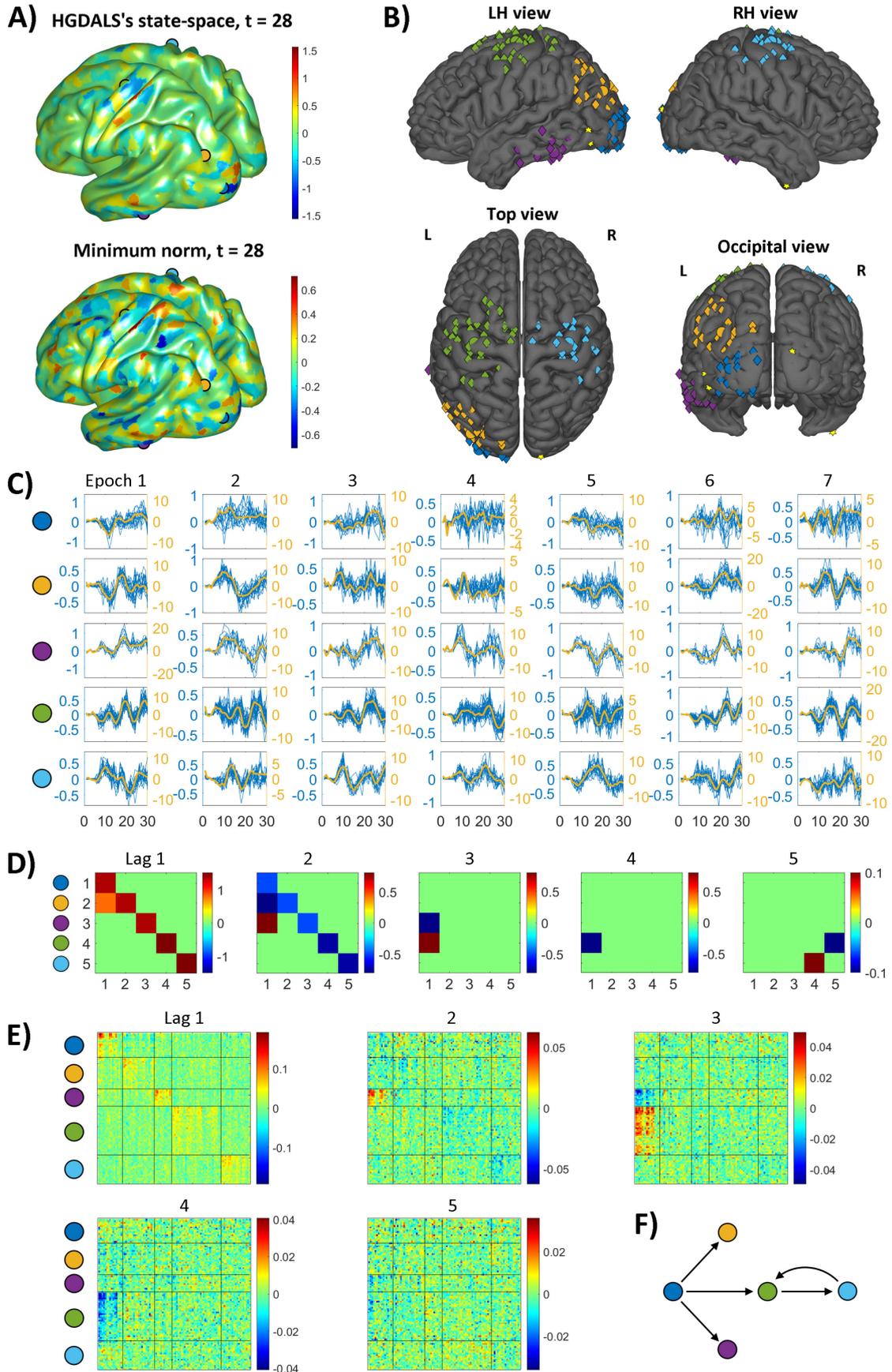

**Figure 5:** Large-scale MEG data simulation involving five brain sources. Ground-truth connectivity (autoregressive matrices) for lag=1,…,5 is provided in **Materials and Methods**, and ground-truth temporal dynamics are generated using the state-space model in **Eqs. (1, 2)**. Coloured spheres indicate simulated sources located in various brain regions in this sequence: occipital lobe (blue), temporo-parieto-occipital junction (orange), and inferior temporal gyrus (purple) in the left hemisphere,

and primary motor cortices in the left (green) and right (cyan) hemispheres. These spheres are superimposed on the cortical surfaces in **A** and **B** to emphasize source locations. **A**) Top and bottom rows display HGDALS and minimum norm solutions, respectively, for MEG simulated data at the same time point ($t = 28$, first epoch) for 2004 evenly distributed dipoles on the cortical surface. Colourmaps illustrate signal intensity and scale differences between solutions. **B**) Four standard brain views show the cortical locations of the most-salient 100 estimated dipoles, with 96/100 accurately identified (coloured diamond marker) as grouped into the five clusters corresponding to each ground-truth sources. Star marker (yellow colour) indicate location of false recovered dipoles. True-recovered dipole counts for each region are 16, 20, 11, 31, and 18, in the same order as mentioned above. **C**) Temporal dynamics for the 96 true-recovered (blue curves) and ground-truth dipoles (overlaid orange curves) are shown separately for each region (rows) and selected epochs (columns). Scale differences between estimated and ground-truth dynamics are emphasized by the coloured left and right y-axes (matching plotted curve colors). **D**) Ground-truth simulated connectivity maps for each lag, with regions arranged in the same order as mentioned above, as indicated by coloured circles. **E**) Estimated connectivity maps for the 96 true-recovered dipoles, grouped by assigned clusters as denoted by coloured circles. Horizontal and vertical lines separate (highlight) the clusters. In **D** and **E**, colourmaps depict connectivity strength, with rows and columns representing incoming and outgoing connections for each source, respectively. **F**) A graph of ground-truth connections demonstrates information flow (alternative view to **D**), with nodes representing the ground-truth sources, and directed edges indicating connections. Connection delays are omitted for simplicity (refer to **Materials and Methods**). All connections are forward, except for recursion between the last two regions.

Finally, **Figs. 5D, E** show the ground-truth and estimated connectivity matrices for the five sources and the most salient 100 dipoles, respectively, which also revealed outstanding results despite the higher dimension. For example, notice the very accurate recovery of the negative and positive interregional connections, particularly for the selected dipoles corresponding to simulated regions actively communicating for lag=2,3,4 in the ground-truth maps. That is, connections 1 → 3 (positive) for lag 2, 1 → 3 (negative) and 1 → 4 (positive) for lag 3, and 1 → 4 (negative) for lag 4. Additionally, notice that the estimated interactions in the block diagonals for lag=1,2 (i.e., intraregional communication among the salient dipoles clustered around the same ground-truth simulated source) are mostly positive for lag=1 and negative for lag=2, which agrees with the ground truth for each lagged interaction. Finally, note that particularly the estimated connections for lag=3,4 were overlaid in the template brain (transparent) cortical surface in **Fig. 1G** using a 15% threshold to plot only the most relevant connections. Therein, it is evident that our algorithm correctly captured the simulated negative and positive connections 1 → 3 and 1 → 4, respectively, for lag=3 (arrows between the regions denoted by blue and purple, and blue and green spheres), and the negative connection 1 → 4 for lag=4.

### Analysis for simulated event related conditions

The state-of-the-art inverse solution methods explored in this section include four methods provided in the Statistical Parametric Mapping (SPM12) toolbox (Penny et al., 2011) and two recently introduced Bayesian approaches to estimating spatiotemporal components with basis functions without (STBF) or with smoothness (SST) constraints (Liang et al., 2022; Liu et al., 2019). The four SPM12 methods are Empirical Bayesian Beamformer (EBB), Bayesian minimum norm without (MMN) and with smoothness priors in a way similar to LORETA (COH), and multiple sparse priors (MSP). All of these methods estimated the hyperparameter values using the Bayesian approach (Friston et al., 2008; Liang et al., 2022; Liu et al., 2019; López et al., 2014; Penny et al., 2011). Whereas, for estimating the MPSS models using the SSALS/HGDALS algorithms, we used the extension of the $K$-fold cross-validation approach to assess the model hyperparameters as previously discussed.

**Fig. 6** show the results based on the AWROC and RRSE statistics (**Materials and Methods**) to evaluate the spatial and temporal accuracy of the proposed SSALS approach vs state-of-the-art methods. The AWROC results reveal that in more realistic conditions (random source locations and 5 dB noise) our method was not clearly superior to the current methods (**Fig. 6A**). Although there are not overall significant differences among the different approaches, the Bayesian version of LORETA as implemented in the SPM toolbox (COH) seems to have a slightly superior spatial accuracy advantage, whereas EBB seems to have the worst outcome. On the other hand, for the RRSE stat, results reveal that MSP and EBB showed the best temporal accuracy with significant differences with respect to the other methods, whereas our approach showed the better results among the remaining approaches (see **Supp. Materials Figs. 5-12** as complement).

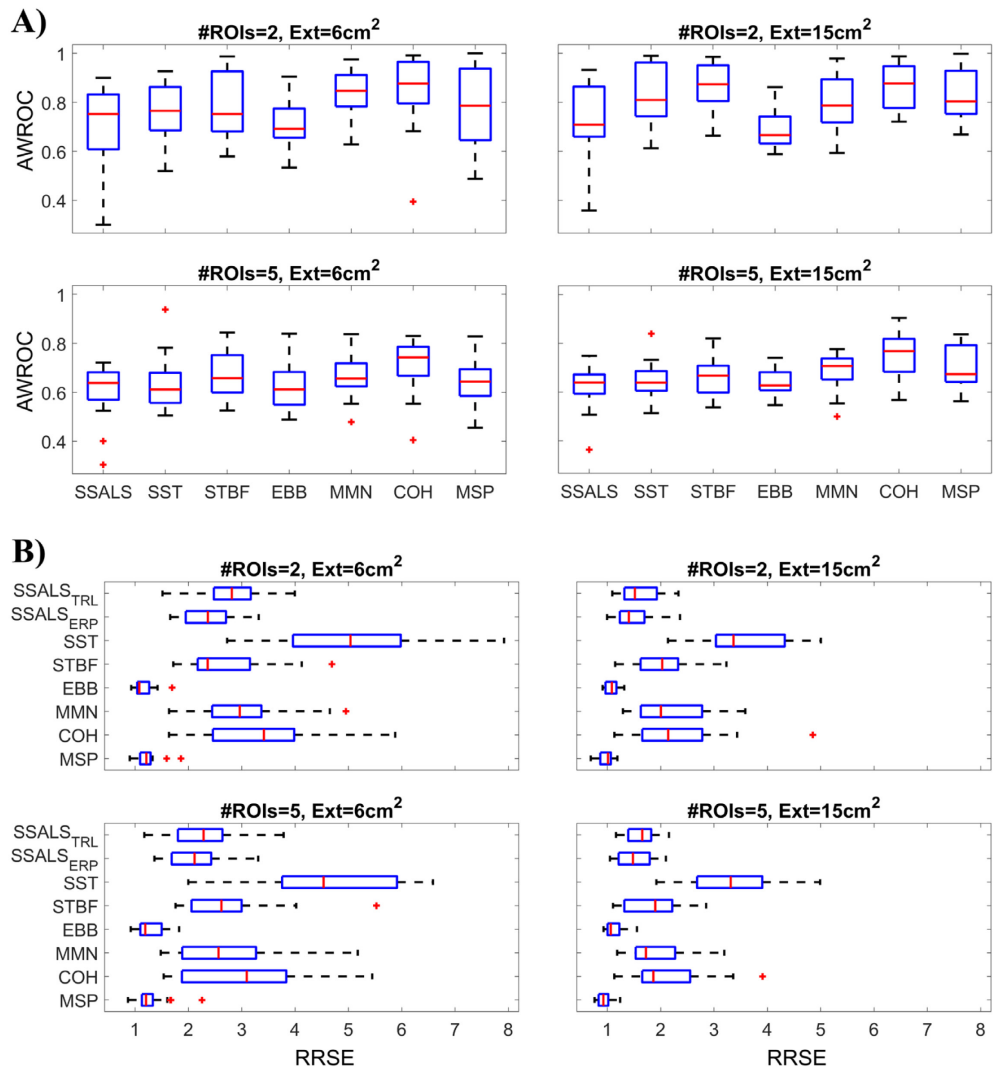

**Figure 6:** Illustration of spatial (**A**) and temporal (**B**) accuracy of different source localization methods, as evaluated using the area under weighted ROC (AWROC) and relative root squared error (RRSE) statistics, respectively. **A**) Boxplots of AWROC for the different tested methods (vertical orientation) for 80 Monte Carlo simulations: 20 simulation of each of four scenarios consisting of random locations of two or five simulated sources, combined with simulated activity patches of extension 6 or 15 cm$^2$. **B**) Similarly, but the boxplots are created for the RRSE stat (horizontal orientation) for the same simulations.

### Solving large-scale MPSS models for real event-related MEG/EEG data

The MEG/EEG data analysed in this section corresponds to a single subject from the Wakeman and Henson study (Wakeman and Henson, 2015). As a proof of concept, we conducted only a single-subject data analysis due to its substantial computational cost. The Wakeman and Henson study also provides individual anatomical MRI images and cortical surface segmentation. We used Fieldtrip (Oostenveld et al., 2011) to calculate the MEG/EEG lead fields with dipoles perpendicularly oriented to the cortical surface and used the scripts provided by the study to run preprocessing analysis (Wakeman and Henson, 2015). After controlling for eye-movement artefacts, we extracted the same 203 epochs for MEG/EEG analyses for the familiar face recognition task. All signals were filtered using a Butterworth low-pass filter for 25 Hz and downsampled to $F_S = 250$ Hz. We used $P = 5$ (lag=1,…,5) to estimate the past influences. As in the simulation above for large-scale data, we solved the modified optimisation problem for epoched data (**Materials and Methods**). Due to the computational cost of the algorithm, instead of calculating the entire time-varying FC, we conducted the analysis only for the time window of [100, 200] milliseconds after stimulus onset, according to the latency of face processing (Lin et al., 2018) and because it showed significant activity for this dataset as can be appreciated Wakeman and Henson's Fig. 1a (Wakeman and Henson, 2015).

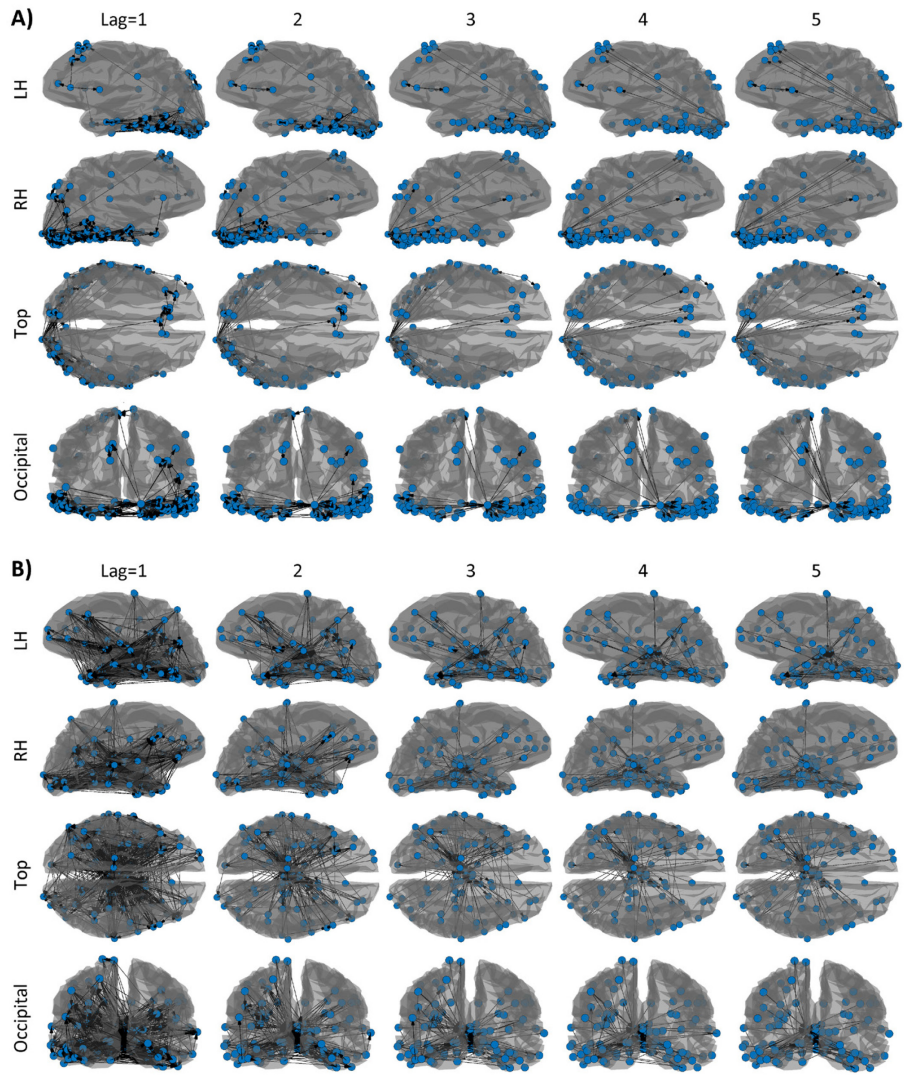

**Figure 7:** Estimated functional connectivity maps for most salient estimated dipoles (overlaid blue colour-filled spheres) using the SSALS/HGDALS algorithm for (**A**) EEG and (**B**) MEG data, separately. Estimated connections among the dipoles are represented as arrows, using a 15% threshold regarding the maximum value of connectivity for better visibility. Columns show the connectivity maps for each lagged interaction (lag=1,…,5). Rows show different views for each map.

**Fig. 7A** shows the results of the EEG analysis revealing the flow of information from the occipital to the inferior temporal lobe. This process corresponds to the activation of the ventral stream of the visual cortex as expected in a face processing task (Goodale and Milner, 1992; Lin et al., 2018; Miller et al., 2017), a remarkable finding also because of the consistency of flow direction. To a lesser degree, estimated visual cortex connections also target parietal and frontal regions. A further exciting and consistent outcome is that the interacting regions are mainly close to each other for the minor lags (lag=1,2, equivalent to influences from 4 to 8 milliseconds in the past). In contrast, the number of long-range interactions increases for the higher lags (lag=4,5, corresponding to influences from 16 to 20 milliseconds in the past), mainly between the occipital and frontal regions.

On the other hand, **Fig. 7B** may reveal a contradictory result from the MEG-based analysis. In this case, using only the MEG signals (from both magnetometer and gradiometer channels), we can capture subcortical structures that create a hub of incoming information mainly from visual and temporal areas. This finding is also tenable as the processing of face information involves subcortical brain regions such as the amygdala and hippocampus, which is also evident in MEG/EEG studies (Dubarry et al., 2014; Dumas et al., 2013) and concurrent intracortical recordings (Dubarry et al., 2014). Overall, the EEG data analysis outcome may be more consistent as it shows the activation of the ventral visual pathway related to face processing.

# Discussion

In this study, we presented a data-driven regularization method for addressing large-scale state-space models, particularly interesting for solving simultaneously the brain source localization and functional connectivity problems. Unlike previous research (Barton et al., 2009; Cheung et al., 2010; Friston et al., 2003; Galka et al., 2004; Shumway and Stoffer, 1982; Van de Steen et al., 2019; Yamashita et al., 2004), our approach directly resolves these issues in the high-dimensional manifold of brain activations and functional connectivity (FC), without dimensionality reduction as done in earlier studies (Cheung et al., 2010; Galka et al., 2004; Long et al., 2011). Traditional state-space model solutions rely on Kalman filtering, expectation maximization, or Bayesian/particle filtering (Barton et al., 2009; Cheung et al., 2010; Friston et al., 2003; Galka et al., 2004; Long et al., 2011; Puthanmadam et al., 2020; Shumway and Stoffer, 1982; Van de Steen et al., 2019; Yamashita et al., 2004), which may face difficulties when dealing with extensive data or more intricate situations, as shown through simulations. However, we demonstrated that even with moderately noisy data, simple algorithms like backpropagation can find stable solutions(**Fig. 1B** and **Table 1**). As supported by research on artificial neural networks (ANNs), backpropagation is effective in solving high-dimensional problems, being the preferred algorithm (Bengio, 2013; Yang and Wang, 2020), thus providing motivation for the development of suitable algorithms for large-scale state-space model as in this study.

By examining large-scale state-space models, we highlighted the frequently overlooked issue of overfitting in model estimation within this field (Barton et al., 2009; Cheung et al., 2010; Friston et al., 2003; Galka et al., 2004; Long et al., 2011; Shumway and Stoffer, 1982). We introduced multiple penalized state-space (MPSS) models and showcased that a data-driven regularization method, based on a novel *K*-fold cross-validation extension, can manage overfitting. To solve MPSS models, we proposed a state-space gradient descent (SSGD) algorithm, which may incorporate techniques used in training ANNs with backpropagation since both algorithms share a gradient descent (GD) optimization-based iterative approach. For instance, we demonstrated that SSGD can use momentum modification, and future implementations could also include stochastic GD, mini-batch GD, and other strategies to enhance learning and reduce the computational cost (Yang and Wang, 2020). Notably, SSGD outperformed backpropagation in our simulations, demonstrating more stable and quicker convergence.

We then introduced a state-space alternating least squares (SSALS) algorithm and its hybrid combination with SSGD (HGDALS) to solve MPSS models with thousands of dynamic state variables and their corresponding connectivity matrices. Particularly, as demonstrated with Monte Carlo simulations, although the SSGD and SSALS algorithms can effectively solve small-scale state-space models, SSALS is superior from both a numerical and computational perspective (see **Tables 4, 5**). Remarkably, the SSALS algorithm provides iterative closed-form expressions, which mathematically demonstrate that brain source localization and FC problems complement and constrain each other. Ignoring this connection may lead to biased results, hindering progress in understanding cognitive brain functions or developing early biomarkers for neurodegeneration, such as for studying Alzheimer's disease (Sanchez-Bornot et al., 2021).

It is important to note that our adaptation of the classical *K*-fold cross-validation approach for state-space models is crucial for implementing the MPSS framework (**Fig. 1E**). Our situation is comparable to time series forecasting, where applying cross-validation directly is not possible due to temporally correlated signals (Bergmeir et al., 2018). Nevertheless, we present a solution based on a practical data imputation procedure made possible by the MPSS framework (see **Eqs. (14, 15)** in **Materials and Methods**). Future research should further evaluate this and other cross-validation methods, such as nested cross-validation (Varoquaux et al., 2017), to ensure robustness to overfitting and proper model generalization. However, many existing cross-validation techniques assume uncorrelated noise, which is not accurate (Bergmeir et al., 2018; Chen et al., 1997), and might result in increased computational time without substantial benefits (Wainer and Cawley, 2021). Our MPSS regularization framework is thoroughly demonstrated using 100 Monte Carlo replications of the small-scale simulations, for both the proposed regularized and state-of-the-art methods (see **Figs. 3,4** and **Tables 6,7**; see also **Supp. Materials Tables 7,8** and discussion therein for complementary information).

Our ultimate goal is to simultaneously address two significant unresolved challenges in MEG/EEG neuroimaging: brain source localization and functional connectivity (FC) problems. The SSALS algorithm is well-suited for this task, as it relies on quadratic optimization and is capable of addressing large-scale problems, particularly when modified for analysing epoched data (see **Eqs. (22, 23)** and more details in **Materials and Methods**). The optimization problem considers both the linearity of the source localization problem, due to the electromagnetic quasi-static assumptions (Nolte, 2003), and MVAR models, frequently employed to address the FC problem (Bastos and Schoffelen, 2016; Valdes-Sosa et al., 2006). A potential criticism is that neural mass (Friston et al., 2003; Jansen and Rit, 1995; Sotero et al., 2007; Valdes-Sosa et al., 2009) or spiking neuronal modeling (Gerstner et al., 2018; Izhikevich and Edelman, 2008) may better capture the nonlinear nature of neuronal dynamics. Consequently, exploring nonlinearity, its effects on model estimation, and the possibility of adapting backpropagation, SSGD, and SSALS/HGDALS algorithms to accommodate nonlinear generative models are compelling avenues for future research.

To demonstrate the feasibility of our approach, we applied the proposed methodology to solve simultaneously the source localization and FC maps for simulated and real MEG/EEG data. First, in a resting state simulated scenario with favourable SNR conditions, we demonstrated the potential of our method to uncover the hidden dynamics in high-dimensional settings (**Fig. 5**). Then, in more realistic conditions simulating event related data (**Materials and Methods**), we presented the results for a comparison analysis against state-of-the-art inverse solution approaches (**Fig. 6**). In this latter analysis, the proposed SSALS approach showed an average performance attending to spatial accuracy and came closer to the top among the methods with better temporal accuracy. In general, it seems that the methods with significant highest temporal accuracy (MSP and EBB) did achieve so by sacrificing spatial accuracy. Although it is clear in this analysis that none of the methods demonstrated impressive results (see **Supp. Materials Figs 5-12** for a visual inspection of estimated solutions), it should be noted that the SSALS algorithm has a large margin of improvement as we have continued the research to overcome implementation hurdles. The main critical issues are the computational time (around 24 hours) and the high amount of RAM resources (≈0.7 TB) to execute the SSALS algorithm for above high dimensional data. In contrast, the evaluated state-of-the-art methods are much more efficient as they focus only on estimating the ERP components, which otherwise limit their application for estimating FC maps. In addition, the state-of-the-art methods used the whole simulated epoched data, whereas our approach used only a limited number of samples due to RAM limitations, corresponding to the time window (50; 200] ms containing the main simulated FC dynamics (see details in **Materials and Methods**). A similar comparison analysis could have been performed involving state-of-the-art functional connectivity approaches; however, it was beyond our current available resources. In an ongoing research, we are exploring the combination of SSALS with variable selection algorithms to improve the computational limitations and accuracy of estimated spatiotemporal dynamics (refer to Manomaisaowapak et al. (2021) and Yang et al. (2016) for similar but alternative approaches, which may well complement our approach to state-space models).

Moreover, we applied our methodology to analyse real MEG/EEG data from a single subject in the Wakeman and Henson's database (Wakeman and Henson, 2015). The EEG analysis showed activation of the ventral stream of the visual system, with information flowing from occipital to inferior-temporal regions (see **Fig. 7A**). This observation was complemented by the consistent results from another recent investigation from our group, further applying the SSALS algorithm for dynamic FC assessment using only the EEG data from the subjects in this database (Sanchez-Bornot et al., 2023; see also the **Supp. Materials Fig. 13**). Conversely, the MEG analysis identified more active hubs in subcortical areas (**Fig.7B**). These findings, including the significant occipitotemporal activations in both analyses, align well with the existing literature (Dubarry et al., 2014; Dumas et al., 2013; Goodale and Milner, 1992; Miller et al., 2017).

However, there are discernible differences between MEG/EEG outcomes which may be due to various factors. For example, first, the SSALS algorithm may not be fully optimized and therefore the solutions quality can be improved with further research. Second, MEG and EEG signals exhibit different sensitivities to source orientation and depth (Ahlfors et al., 2010). Third, inaccuracies in the head models used for MEG/EEG forward problems, which determine the respective lead field matrices, can significantly influence FC estimation (Cho et al., 2015). This possible negative influence should have different effects for MEG and EEG modalities as their corresponding forward problems are resolved separately, using different assumptions. Additionally, the observed differences may also be explained due to the fact that EEG sensors are attached to the scalp, while MEG sensors do not have a fixed reference with respect to the scalp. In principle, head movements during recording sessions can impact MEG signals and the estimated sources derived from them in non-trivial ways. Further refinements of the MPSS framework and SSALS algorithm can tackle some of these issues and, therefore, deserve further investigation. For instance, more robust spatiotemporal dynamics assessment is possible by extending our approach to perform dynamic FC analysis using the sliding time-window technique (Hindriks et al., 2016) or extending the optimization problem to estimate the whole data with time-varying autoregressive matrices. Critically, the MPSS framework computational implementation must be improved significantly to make possible further developments.

## Data and code availability statement

The main code implementing the algorithms and used to generate simulations and figures will be available on the first author's **Github** repository website (https://github.com/JMSBornot/Multiple-Penalized-State-Space-Models) and will also be accessible in the paper's online version together with the **Supplementary Materials**.

## Acknowledgements

The authors are grateful for access to the Tier 2 High-Performance Computing resources provided by the Northern Ireland High Performance Computing (NI-HPC) facility funded by the UK Engineering and Physical Sciences Research Council (EPSRC), Grant No. EP/T022175/1. DC is supported by a UKRI Turing

AI Fellowship 2021-2025, funded by the EPSRC, Grant No, EP/V025724/1. RCS was partially supported by grant RGPIN-2022-03042 from the Natural Sciences and Engineering Research Council of Canada. JASK's research is supported by the FAU Foundation.

Supplementary material

# Solving large-scale MEG/EEG source localisation and functional connectivity problems simultaneously using state-space models

Jose Sanchez-Bornot[1*], Roberto C. Sotero[2], J. A. Scott Kelso[1,3], Özgür Şimşek[4], and Damien Coyle[1,4*]

[1] Intelligent Systems Research Centre, School of Computing, Engineering and Intelligent Systems, Ulster University, Magee campus, Derry~Londonderry, UK.

[2] Department of Radiology and Hotchkiss Brain Institute, University of Calgary, Calgary, AB, Canada.

[3] Human Brain & Behavior laboratory, Center for Complex Systems & Brain Sciences, Florida Atlantic University, Boca Raton, Florida, USA.

[4] Bath Institute for the Augmented Human, University of Bath, Bath, BA2 7AY, United Kingdom.

[*] Corresponding authors: JSB: jm.sanchez-bornot@ulster.ac.uk; DC: dhc30@bath.ac.uk

# Comparison analysis of Matlab Toolbox's standard and Bayesian implementations of state-space models vs proposed algorithms

Tables 1, 3, 5, and 6 provide the Matlab code for the evaluation of the Monte Carlo simulations based on the functions "ssm.estimate" and "bssm.estimate", which are available in the Econometric toolbox. These functions enable the estimation of state-space models for the standard and Bayesian classical approaches. These scripts, together with the scripts for the other algorithms, are provided in the zip file accompanying the Supplementary Materials, and the first author Github repository: https://github.com/JMSBornot/Multiple-Penalized-State-Space-Models. Complementarily, for more clarity here, the code for the small-scale simulations are provided in Tables 2 and 4. Succinctly, ssm refers to the classical implementation of state-space models in Matlab, while bssm refers to the Bayesian implementation.

The following warning messages appeared during the running of these codes, apparently due to numerical issues.

ssm's warnings:

1) Numerical optimization does not return a convergence flag. Maximum likelihood estimation is less likely to succeed if starting parameter values are far from true values. Try different starting values, other optimization algorithms and settings. Also try to reparameterize the model and rescale the data.

bssm's warning:

1) Numerical optimization does not return a convergence flag. Proposal not necessarily positive definite.

2) Warning: Proposal matrix not positive definite. Bad eigenvalues removed.

For the ssm's issue, we tried different methods for the estimation of the covariance matrix, being the sandwich method the most stable. For both methods we rerun successively the method with different initial random solutions in case of any warning, up to 5 consecutive random initializations. Most of the time, the solution was satisfactory after trying two or three different random initializations. Only for a single Monte Carlo replication the solution was not satisfactory after having tried all the random initializations. This occurred for the noisiest case in the three-variate simulation.

**Table 1:** Matlab R2022a code for a standard solution of a state-space model with "$ssm.estimate$". The solution for different SNR values can be evaluated by changing the value of "$so$" in lines 7-9. The code for function "$mvar1sim1$", for the bivariate (lag=1) simulation, is provided in **Table 2**. Initialization random values for the parameters are provided with "$params0$" (line 36), and up to 5 random initializations are tried if the current solution present numerical problems (lines 34-43). Logically, the lower bound parameter "$lowBound$" (line 21) is set to 0 for the entries corresponding to the covariance variables. Notice that values in "$lowBound$" and "$upBound$" are restricted (lines 21-22) to avoid disparate parameter values, and optimization is performed with fmincon function and interior point parameter values search strategy.

```
1.  % test_ssm_M5T200N2P1_MC
2.  rng('default');
3.
4.  %% and
5.  Nepoch = 100;
6.  ss = 1;
7.  so = 0.1;
8.  % so = 0.5;
9.  % so = 1;
10. M = 5;
11. T = 200;
12. [x, y, A, B] = mvar1sim1(M, T, Nepoch, ss, so);
13. N = size(B,2);
14. P = size(A,3);

15. %% Estimate with Matlab Econometric Toolbox
16. At = [NaN NaN; NaN NaN];
17. Bt = [NaN 0; 0 NaN];
18. Ct = B;
19. Dt = [NaN 0 0 0 0; 0 NaN 0 0 0; 0 0 NaN 0 0; 0 0 0 NaN 0; 0 0 0 0 NaN];
20. Mdl = ssm(At, Bt, Ct, Dt, 'Mean0', 0, 'Cov0', 1*eye(N), 'StateType', zeros(N,1));
21. lowBound = [-10*ones(N*N*P,1); zeros(N+M,1)];
22. uppBound = 10*ones(N*N*P+N+M,1);
23. options = optimoptions(@fmincon, 'ConstraintTolerance', 1e-6, 'Algorithm', 'interior-point', ...
        'MaxFunctionEvaluations', 10000);

24. % CovMethod = 'opg';
25. % CovMethod = 'hessian';
26. CovMethod = 'sandwich';
```

```
27. Ae = zeros(N,N,P,Nepoch);

28. %% Monte Carlo runs
29. tstart = tic;
30. for i = 1:Nepoch
31.     fprintf('\n\n=========== Monte Carlo iteration %d of %d ===========\n\n', i, Nepoch);
32.     frepeat = true;
33.     cont = 0;
34.     while (frepeat && cont<5)
35.         cont = cont + 1;
36.         params0 = [0.1*randn(N*N*P,1); ones(N+M,1)];
37.         lastwarn('','');
38.         [EstMdl,estParams,EstParamCov,logL,Output] = estimate(Mdl, y(:,:,i)', ..
39.             params0, 'lb', lowBound, 'ub', uppBound, 'Options', options, 'CovMethod', CovMethod);
40.         Output.ExitFlag
41.         [warnMsg, warnId] = lastwarn();
42.         frepeat = ~isempty(warnMsg);
43.     end
44.     Ae(:,:,:,i) = EstMdl.A;
45. end
46. time_calc = toc(tstart);
```

**Table 2:** Matlab function "*mvar1sim1*" which generates the bivariate (lag=1) simulation.

```
1.  function [x, y, A, B] = mvar1sim1(M, T, Nepoch, ss, so)
2.  N = 2;
3.  P = 1;
4.  A = [-0.5 0; 0.7 -0.5];
5.  assert(N == size(A,1));
6.  assert(N == size(A,2));
7.  assert(P == size(A,3));
8.
9.  % generate state data: x
10. x = ss*randn(N,Nepoch,T);
11. for t = P+1:T
12.     x(:,:,t) = x(:,:,t) + A*x(:,:,t-1);
13. end
14.
15. % generate observation data: y
16. B = rand(M,N);
17. % B = B/diag(sqrt(sum(B.^2)));
18. y = B*reshape(x,N,[]) + so*randn(M,Nepoch*T);
19. y = reshape(y, [M Nepoch T]);
20.
21. % rearrage so that epochs are in the 3rd dimension
22. x = permute(x, [1 3 2]);
23. y = permute(y, [1 3 2]);
```

**Table 3:** Similar to **Table 1** but for the three-variate (lag=1,2,3) simulation that is generated with the function "*mvar3sim1*", which is presented in **Table 4**. The solution for different SNR values can be evaluated by changing the value of "*so*" in lines 7-9. Initialization random values for the parameters are provided with "*params0*" (line 39), and up to 5 random initializations are tried if the current solution present numerical problems (lines 37-46). Logically, the lower bound parameter "*lowBound*" (line 23) is set to 0 for entries corresponding to the covariance variables. Notice that values in "*lowBound*" and "*upBound*" are restricted (lines 23-24) to avoid disparate parameter values, and optimization is performed with fmincon function and interior point parameter values search strategy.

```
1.  % test_ssm_M5T240N3P3_MC
2.  rng('default');
3.
4.  %% Simulation
5.  Nepoch = 100;
6.  ss = 1;
7.  so = 0.1;
8.  % so = 0.5;
9.  % so = 1;
10. M = 5;
11. Fs = 120;
12. T = 2*Fs;
13. [x, y, A, B] = mvar3sim1(M, Fs, T, Nepoch, ss, so);
14. N = size(B,2);
15. P = size(A,3);
16.
17. %% Estimate with Matlab Econometric Toolbox
18. At = [NaN(N) NaN(N) NaN(N); eye(N) zeros(N) zeros(N); zeros(N) eye(N) zeros(N)];
19. Bt = [diag(NaN(N,1)); zeros(N); zeros(N)];
20. Ct = [B zeros(M,N) zeros(M,N)];
21. Dt = diag(NaN(M,1));
22. Mdl = ssm(At, Bt, Ct, Dt, 'Mean0', 0, 'Cov0', 1*eye(3*N), 'StateType', zeros(3*N,1));
23. lowBound = [-10*ones(N*N*P,1); zeros(N+M,1)];
24. uppBound = 10*ones(N*N*P+N+M,1);
```

```
25. options = optimoptions(@fmincon, 'ConstraintTolerance', 1e-6, 'Algorithm', 'interior-point', ...
26.     'MaxFunctionEvaluations', 10000);

27. % CovMethod = 'opg';
28. % CovMethod = 'hessian';
29. CovMethod = 'sandwich';

30. Ae = zeros(N,N,P,Nepoch);

31. %% Monte Carlo runs
32. tstart = tic;
33. for i = 1:Nepoch
34.     fprintf('\n\n=========== Monte Carlo iteration %d of %d ===========\n\n', i, Nepoch);
35.     frepeat = true;
36.     cont = 0;
37.     while (frepeat && cont<5)
38.         cont = cont + 1;
39.         params0 = [0.1*randn(N*N*P,1); ones(N+M,1)];
40.         lastwarn('','');
41.         [EstMdl,estParams,EstParamCov,logL,Output] = estimate(Mdl, y(:,:,i)', ..
42.             params0, 'lb', lowBound, 'ub', uppBound, 'Options', options, 'CovMethod', CovMethod);
43.         Output.ExitFlag
44.         [warnMsg, warnId] = lastwarn();
45.         frepeat = ~isempty(warnMsg);
46.     end
47.     Ae(:,:,:,i) = reshape(EstMdl.A(1:N,:),[N N P]);
48. end
49. time_calc = toc(tstart);
```

**Table 4:** Matlab function "`mvar3sim1`" which generates the three-variate (lag=1,2,3) simulation.

```
1.  function [x, y, A, B] = mvar3sim1(M, Fs, T, Nepoch, ss, so)
2.  % Fs = 120; % sampling frequency is 120 Hz
3.  % T = 2*Fs; % 2 seconds simulation
4.  r = [0.9 0.7 0.8];
5.  f = [40 10 50];
6.  dt = 1/Fs;
7.  theta = 2*pi*f*dt;
8.
9.  N = 3; % number of nodes
10. p = 3; % true order of the MVAR model
11. A = zeros(N,N,p);
12. A(:,:,1) = diag(2*r.*cos(theta)) + [0 0 0; -0.356 0 0; 0 -0.3098 0];
13. A(:,:,2) = diag(-r.^2) + [0 0 0; 0.7136 0 0; 0 0.5 0];
14. A(:,:,3) = [0 0 0; -0.356 0 0; 0 -0.3098 0];
15. assert(N == size(A,1));
16. assert(N == size(A,2));
17. assert(p == size(A,3));
18.
19. % generate state data: x
20. x = ss*randn(N,Nepoch,T);
21. for t = p+1:T
22.     for k = 1:p
23.         x(:,:,t) = x(:,:,t) + A(:,:,k)*x(:,:,t-k);
24.     end
25. end
26.
27. % generate observation data: y
28. B = rand(M,N);
29. y = B*reshape(x,N,[]) + so*randn(M,Nepoch*T);
30. y = reshape(y, [M Nepoch T]);
31.
32. % rearrage so that epochs are in the 3rd dimension
33. x = permute(x, [1 3 2]);
34. y = permute(y, [1 3 2]);
```

**Table 5:** Matlab code implementing the Monte Carlo analysis for the bivariate simulation, based on the Bayesian solution of a state-space model estimated with the "`bssm.estimate`" function. The solution for different SNR values may be evaluated by changing the value of "`so`" in lines 7-9. Notice that the same model specified above in **Tables 1** and **3** for the classical state-space models is created first in standard form as above (lines 16-20) and then it is converted to the corresponding Bayes form with the function "`ssm2bssm`" (line 21). As can be seen, for the Bayesian estimation, there is no need to use boundary conditions for the parameter estimation. However, the solution may be sensitive to different initial parameter values as it can get stuck in local minima, particularly for low SNR simulations. This issue is ameliorated by trying different random initial solutions inside the while loop (lines 30-37), particularly with the use of "`params0`" (line 32).

```
1.  % test_bssm_M5T200N2P1_MC
2.  rng('default');
3.
4.  %% Simulation
5.  Nepoch = 100;
6.  ss = 1;
```

```
7.  so = 0.1;
8.  % so = 0.5;
9.  % so = 1;
10. M = 5;
11. T = 200;
12. [x, y, A, B] = mvar1sim1(M, T, Nepoch, ss, so);
13. N = size(B,2);
14. P = size(A,3);

15. %% Estimate with Matlab Econometric Toolbox
16. At = [NaN NaN; NaN NaN];
17. Bt = [NaN 0; 0 NaN];
18. Ct = B;
19. Dt = [NaN 0 0 0 0; 0 NaN 0 0 0; 0 0 NaN 0 0; 0 0 0 NaN 0; 0 0 0 0 NaN];
20. Mdl = ssm(At, Bt, Ct, Dt, 'Mean0', 0, 'Cov0', 1*eye(N), 'StateType', zeros(N,1));
21. bayesMdl = ssm2bssm(Mdl);
22. options = optimoptions(@fminunc, 'OptimalityTolerance', 1e-6, 'MaxFunctionEvaluations', 1e4);

23. Ae = zeros(N,N,P,Nepoch);

24. %% Monte Carlo runs
25. tstart = tic;
26. for i = 1:Nepoch
27.     fprintf('\n\n============ Monte Carlo iteration %d of %d ============\n\n', i, Nepoch);
28.     frepeat = true;
29.     cont = 0;
30.     while (frepeat && cont<5)
31.         cont = cont + 1;
32.         params0 = [0.1*randn(N*N*P,1); ones(N+M,1)];
33.         lastwarn('','');
34.         EstPostMdl = estimate(bayesMdl, y(:,:,i)', params0, 'Options', options);
35.         [warnMsg, warnId] = lastwarn();
36.         frepeat = ~isempty(warnMsg);
37.     end
38.     coeff_mean = mean(EstPostMdl.ParamDistribution,2);
39.     Ae(:,:,:,i) = reshape(coeff_mean(1:N*N*P), [N N P]);
40. end
41. time_calc = toc(tstart);
```

**Table 6:** Matlab code implementing the Monte Carlo analysis for the three-variate simulation, based on the Bayesian solution of a state-space model estimated with the "`bssm.estimate`" function. The solution for different SNR values may be evaluated by changing the value of "`so`" in lines 7-9. Notice that the same model specified above in **Tables 1** and **3** for the classical state-space models is created first in standard form as above (lines 18-22) and then it is converted to the corresponding Bayes form with the function "`ssm2bssm`" (line 23). As can be seen, for the Bayesian estimation, there is no need to use boundary conditions for the parameter estimation. However, the solution may be sensitive to different initial parameter values as it can get stuck in local minima, particularly for low SNR simulations. This issue is ameliorated by trying different random initial solutions inside the while loop (lines 30-37), particularly with the use of "`params0`" (line 32).

```
1.  % test_bssm_M5T240N3P3_MC
2.  rng('default');
3.
4.  %% Simulation
5.  Nepoch = 100;
6.  ss = 1;
7.  so = 0.1;
8.  % so = 0.5;
9.  % so = 1;
10. M = 5;
11. Fs = 120;
12. T = 2*Fs;
13. [x, y, A, B] = mvar3sim1(M, Fs, T, Nepoch, ss, so);
14. N = size(B,2);
15. P = size(A,3);
16.
17. %% Estimate with Matlab Econometric Toolbox
18. At = [NaN(N) NaN(N) NaN(N); eye(N) zeros(N) zeros(N); zeros(N) eye(N) zeros(N)];
19. Bt = [diag(NaN(N,1)); zeros(N); zeros(N)];
20. Ct = [B zeros(M,N) zeros(M,N)];
21. Dt = diag(NaN(M,1));
22. Mdl = ssm(At, Bt, Ct, Dt, 'Mean0', 0, 'Cov0', 1*eye(3*N), 'StateType', zeros(3*N,1));
23. bayesMdl = ssm2bssm(Mdl);
24. options = optimoptions(@fminunc, 'OptimalityTolerance', 1e-6, 'MaxFunctionEvaluations', 1e4);

25. % CovMethod = 'opg';
26. % CovMethod = 'hessian';
27. CovMethod = 'sandwich';

28. Ae = zeros(N,N,P,Nepoch);

29. %% Monte Carlo runs
30. tstart = tic;
```

```
31. for i = 1:Nepoch
32.     fprintf('\n\n============ Monte Carlo iteration %d of %d ============\n\n', i, Nepoch);
33.     frepeat = true;
34.     cont = 0;
35.     while (frepeat && cont<5)
36.         cont = cont + 1;
37.         params0 = [0.1*randn(N*N*P,1); ones(N+M,1)];
38.         lastwarn('','');
39.         EstPostMdl = estimate(bayesMdl, y(:,:,i)', params0, 'Options', options);
40.         [warnMsg, warnId] = lastwarn();
41.         frepeat = ~isempty(warnMsg);
42.     end
43.     coeff_mean = mean(EstPostMdl.ParamDistribution,2);
44.     Ae(:,:,:,i) = reshape(coeff_mean(1:N*N*P), [N N P]);
45. end
46. time_calc = toc(tstart);
```

Next, **Tables 7** and **8** illustrate the mean ± error bars results for the 100 Monte Carlo replication of the two small-scale simulations, as demonstrated in the manuscript for both the regularized methods and the Matlab's standard and Bayesian implementations of state-space models. For the proposed algorithms, the cases presented below correspond to the SSGD and SSALS algorithms for the line and plane search regularization approaches.

The estimated solutions are more or less similar among the different methods, although they can be appreciated as slightly better for the state-of-the-art approaches, maybe due to the estimation of covariance matrices or due to the possible estimator's bias introduced by regularized approaches. Additionally, the state-of-the-art methods provided by the Matlab Econometric Toolbox have the clear advantage that, apart from the necessity to evaluate different algorithms options such as an appropriate covariance method and the use of constrained optimization methods (fmincon function in Matlab with "interior point" search) with upper and lower bound constraints, the solution can be found automatically. Whereas, for the provided methods it is necessary to find an appropriate hyperparameter values range. However, once this is done, the computations for the proposed methods were significantly much faster, mainly for the SSALS algorithm.

For example, in this order for the $\sigma_o = 0.1$, $\sigma_o = 0.5$, and $\sigma_o = 1.0$ noise scenarios, for the bivariate Monte Carlo simulations, the computational times were 8.66 sec/replica, 6.48 sec/replica, and 6.87 sec/replica for the standard state-space Matlab Toolbox implementation (ssm.estimate function); and 16.31 sec/replica, 12.52 sec/replica, and 13.30 sec/replica for the Bayesian state-space Matlab Toolbox implementation (bssm.estimate function), respectively. In contrast, for the ==line search==, the computational times for the SSGD algorithm were 22.75 sec/replica, 28.44 sec/replica, and 76.64 sec/replica, and for the SSALS algorithm were ==0.62 sec/replica, 0.73 sec/replica, and 0.95 sec/replica==, respectively for each scenario. For the ==plane search==, the computational times for the SSGD algorithm were 33.43 sec/replica, 36.24 sec/replica, and 46.09 sec/replica, and for the SSALS algorithm were ==1.01 sec/replica, 1.11 sec/replica, and 1.48 sec/replica==, respectively. The explored hyperparameters subspaces have dimension of $28 \times 8$, for the plane search, for $K = 5$ fold cross-validation for all cases, and $40 \times 1$ for the line search. We explored the same subspaces for both the SSGD and SSALS algorithms and for the different noise scenarios, in order to keep the same settings for fairly computational time comparisons.

Similarly, for the three-variate simulation, the computational times were 87.59 sec/replica, 99.62 sec/replica, and 161.70 sec/replica for the ssm.estimate function, while they were 74.93 sec/replica, 93.30 sec/replica and 180.42 sec/replica for bssm.estimate function, respectively for each scenario. In contrast, for the ==plane search==, for the SSGD algorithm, computational times were 29.50 sec/replica, 33.36 sec/replica, and 52.31 sec/replica, and for the SSALS algorithm were ==15.61 sec/replica, 53.78 sec/replica, and 68.53 sec/replica==. Although from the above values, it seems that SSGD algorithm is faster than SSALS for some cases, indeed the SSALS algorithm searched on a bigger subspace, with dimension of $15 \times 16$, which we reduced for the SSGD algorithm to a dimension of $6 \times 5$ because of slower computation. Whereas, for the only solution computed for the hyperparameters ==line search==, using the SSALS algorithm, the computational time was ==3.76 sec/replica==, which was found by searching on a subspace of dimension $10 \times 1$.

In contrast to the proposed algorithms, as mentioned in the manuscript, the state-of the-art methods do not calculate the hidden dynamics. Although, it may be possible to calculate it by following some of the computational tricks presented here (see **Eqs. (18,19)** in the manuscript). Therefore, we only reported the RSE statistics for the SSGD and SSALS algorithms (**Materials and Methods**). For the bivariate simulations, in this order for the $\sigma_o = 0.1$, $\sigma_o = 0.5$, and

$\sigma_o = 1.0$ noise scenarios, for the SSGD algorithm and line search, the RSE statistics were 2.33±0.05% and 1.16±0.02%, 31.51±1.03% and 16.47±0.41%, 88.35±2.74% and 47.77±1.12% for the two latent variables and the different noise scenarios; and 6.61±0.16% and 3.21±0.08%, 25.59±0.43% and 13.65±0.24%, 57.40±1.13% and 32.69±0.63% for the plane search; whereas, for the SSALS algorithm, the RSE statics were 2.33±0.05% and 1.16±0.02%, 31.12±0.96% and 16.29±0.40%, 87.92±2.91% and 47.83±1.28%, for the line search; and 6.47±0.15% and 3.15±0.07%, 25.62±0.42% and 13.67±0.24%, 57.99±1.16% and 32.95±0.65% for the plane search. As observed in the manuscript, the solutions are similar between the SSGD and SSALS algorithms, specially for the regularized approach where solutions are more numerical stable.

**Table 7**: Solutions for the bivariate (lag=1) simulation – same synthetic data as for corresponding analysis discussed in the main text for 100 Monte Carlo replications – using a standard and Bayesian estimation of state-space models as implemented in Matlab R2022a's Econometric Toolbox functions ssm.estimate and bssm.estimate, respectively. From left to right, the solutions are shown for the different simulated noise scenarios, whereas they are shown from top to bottom for the different state-space methods and regularization approaches. The ground-truth autoregressive matrix coefficients is shown in the left-top corner for clarity.

| $\begin{bmatrix} -0.5 & 0 \\ 0.7 & -0.5 \end{bmatrix}$ | $\sigma_o = 0.1$ | $\sigma_o = 0.5$ | $\sigma_o = 1$ |
|---|---|---|---|
| **SSGD: line search ($\lambda$)** | $\begin{bmatrix} -0.51 \pm 0.01 & 0.00 \pm 0.00 \\ 0.74 \pm 0.01 & -0.50 \pm 0.00 \end{bmatrix}$ | $\begin{bmatrix} -0.49 \pm 0.01 & 0.00 \pm 0.00 \\ 0.68 \pm 0.01 & -0.48 \pm 0.01 \end{bmatrix}$ | $\begin{bmatrix} -0.45 \pm 0.01 & 0.03 \pm 0.01 \\ 0.55 \pm 0.02 & -0.44 \pm 0.01 \end{bmatrix}$ |
| **SSGD: plane search ($\lambda, \lambda_2^{(a)}$)** | $\begin{bmatrix} -0.35 \pm 0.01 & 0.03 \pm 0.00 \\ 0.63 \pm 0.01 & -0.52 \pm 0.00 \end{bmatrix}$ | $\begin{bmatrix} -0.35 \pm 0.01 & 0.02 \pm 0.01 \\ 0.61 \pm 0.01 & -0.51 \pm 0.01 \end{bmatrix}$ | $\begin{bmatrix} -0.20 \pm 0.02 & -0.01 \pm 0.01 \\ 0.48 \pm 0.01 & -0.50 \pm 0.01 \end{bmatrix}$ |
| **SSALS: line search ($\lambda$)** | $\begin{bmatrix} -0.51 \pm 0.01 & 0.00 \pm 0.00 \\ 0.74 \pm 0.01 & -0.50 \pm 0.00 \end{bmatrix}$ | $\begin{bmatrix} -0.49 \pm 0.01 & 0.00 \pm 0.01 \\ 0.68 \pm 0.01 & -0.48 \pm 0.01 \end{bmatrix}$ | $\begin{bmatrix} -0.45 \pm 0.02 & 0.03 \pm 0.01 \\ 0.55 \pm 0.02 & -0.44 \pm 0.01 \end{bmatrix}$ |
| **SSALS: plane search ($\lambda, \lambda_2^{(a)}$)** | $\begin{bmatrix} -0.35 \pm 0.01 & 0.03 \pm 0.00 \\ 0.62 \pm 0.01 & -0.52 \pm 0.00 \end{bmatrix}$ | $\begin{bmatrix} -0.35 \pm 0.01 & 0.02 \pm 0.01 \\ 0.60 \pm 0.01 & -0.51 \pm 0.01 \end{bmatrix}$ | $\begin{bmatrix} -0.21 \pm 0.02 & 0.00 \pm 0.01 \\ 0.48 \pm 0.01 & -0.49 \pm 0.01 \end{bmatrix}$ |
| **ssm (Econom. Toolbox)** | $\begin{bmatrix} -0.63 \pm 0.09 & 0.10 \pm 0.07 \\ 0.49 \pm 0.15 & -0.35 \pm 0.10 \end{bmatrix}$ | $\begin{bmatrix} -0.49 \pm 0.01 & -0.01 \pm 0.01 \\ 0.71 \pm 0.01 & -0.49 \pm 0.01 \end{bmatrix}$ | $\begin{bmatrix} -0.48 \pm 0.02 & -0.03 \pm 0.01 \\ 0.78 \pm 0.02 & -0.49 \pm 0.01 \end{bmatrix}$ |
| **bssm (Econom. Toolbox)** | $\begin{bmatrix} -0.50 \pm 0.01 & 0.00 \pm 0.00 \\ 0.70 \pm 0.01 & -0.49 \pm 0.00 \end{bmatrix}$ | $\begin{bmatrix} -0.48 \pm 0.01 & -0.01 \pm 0.01 \\ 0.71 \pm 0.01 & -0.49 \pm 0.01 \end{bmatrix}$ | $\begin{bmatrix} -0.40 \pm 0.02 & -0.06 \pm 0.01 \\ 0.92 \pm 0.05 & -0.51 \pm 0.02 \end{bmatrix}$ |

The same observation cannot be made for the tree-variate simulation, as it appears that solutions can be stuck in local minima, as it is more obvious for the noisier scenarios. The same observation about multiple local minima was observed for the standard and Bayesian Matlab state-space models implementations by using different random initial solutions for the same data. For the three-variate simulation, for the SSGD algorithm and plane search, the RSE statistics were 36.91±0.93%, 25.21±0.38% and 32.71±0.35% for the $\sigma_o = 0.1$ scenario, 40.56±1.03%, 28.48±0.44% and 33.58±0.45% for the $\sigma_o = 0.5$ scenario, and 50.58±1.23%, 38.16±0.67% and 37.77±0.73% for the $\sigma_o = 1.0$ scenario. Whereas, for the SSALS algorithm and plane search, the RSE statistics were 2.27±0.05%, 1.76±0.03% and 1.67±0.03% for the $\sigma_o = 0.1$ scenario, 21.23±0.71%, 13.91±0.41% and 12.64±0.56% for the $\sigma_o = 0.5$ scenario, and 51.17±1.30%, 37.97±0.70% and 37.04±0.80% for the $\sigma_o = 1.0$ scenario. Also, the differences between the SSGD and SSALS solution can be due to the fact that for the SSALS algorithm we used a finer hyperparameters search grid as SSGD solution were much slower. However, the dramatic difference shown in **Table 8** between the SSGD and SSALS solutions for the $\sigma_o = 0.1$ case are mainly due to using different hyperparameter domains. Initially, we set the same interval for $\lambda$ and $\lambda_2^{(a)}$ for both algorithms, in the ranges of [0.1; 0.8] and [6.0; 96.0], respectively. But, after observing that the selected cross-validation values were on the left extreme of both intervals, we adjusted them to the new intervals of [0.002; 0.02] for $\lambda$ and [0.0024; 0.2400] for $\lambda_2^{(a)}$. This change was made only for the SSALS algorithm to highlight the complexity of selected the hyperparameters interval in the comparison between SSGD and SSALS solutions.

Clearly, selecting an unnecessary high value for the hyperparameters will increase the estimators bias. This last observation was also demonstrated by applying the line search regularization approach only for the SSALS algorithm, for the $\sigma_o = 0.1$ case, for the same interval as in the plane search regularization, i.e., $\lambda$ in the range [0.002; 0.02]. Notice that for this case, the "optimal" $\lambda$ for the naïve estimator is $\lambda = \sigma_o^2/\sigma_s^2 = 0.01$, so it made sense to search on an interval that included this value. It is also noticeable by comparing the solutions of the 2$^{nd}$ and 3$^{rd}$ rows in **Table 8**, corresponding to the SSALS solutions for the line and plane search, that by dropping the second regularization parameter (i.e., only selecting $\lambda$), the solutions have less bias.

**Table 8**: Solutions for the three-variate (lag=1,2,3) simulation – same synthetic data as for corresponding analysis discussed in the main text for 100 Monte Carlo replications – using a standard and Bayesian estimation of state-space models, as implemented in Matlab R2022a's Econometric Toolbox functions ssm.estimate and bssm.estimate, respectively. From top to bottom are shown the ground-truth autoregressive coefficients, and the solutions for the three different simulated noise scenarios. For each scenario, the five successive rows show the solutions correspondingly for each of the methods and regularization approaches, in the following order from top to bottom: 1) SSGD algorithm's plane search, 2) SSALS algorithm's line search, 3) SSALS algorithm's plane search, 4) Solutions found by using Matlab ssm.estimate, and 5) Solutions found by Matlab bssm.estimate. Notice that for the 2nd ($\sigma_o = 0.5$) and 3rd scenarios ($\sigma_o = 1.0$), the SSALS algorithm's line search solutions were not calculated (NA – not available), as in these more noisier case, it made sense to calculate only the plane search regularization approach to control better for hight autoregressive coefficient values that may be potentially caused by overfitting.

| Ground-truth autoregressive coefficients | $\mathbf{A} = \left\{ \begin{bmatrix} -0.9000 & 0 & 0 \\ -0.3560 & 1.2124 & 0 \\ 0 & -0.3098 & -1.3856 \end{bmatrix}, \begin{bmatrix} -0.8100 & 0 & 0 \\ 0.7136 & -0.4900 & 0 \\ 0 & 0.5000 & -0.6400 \end{bmatrix}, \begin{bmatrix} 0 & 0 & 0 \\ -0.3560 & 0 & 0 \\ 0 & -0.3098 & 0 \end{bmatrix} \right\}$ |
|---|---|
| \multicolumn{2}{c}{$\sigma_o = 0.1$} |
| | $\left\{ \begin{bmatrix} -0.28 \pm 0.00 & 0.16 \pm 0.00 & -0.19 \pm 0.00 \\ -0.14 \pm 0.00 & 0.55 \pm 0.01 & -0.20 \pm 0.00 \\ 0.23 \pm 0.00 & 0.25 \pm 0.00 & -0.67 \pm 0.01 \end{bmatrix}, \begin{bmatrix} -0.33 \pm 0.00 & -0.03 \pm 0.00 & 0.12 \pm 0.00 \\ 0.20 \pm 0.00 & 0.12 \pm 0.00 & -0.03 \pm 0.00 \\ -0.29 \pm 0.00 & -0.01 \pm 0.00 & 0.28 \pm 0.00 \end{bmatrix}, \begin{bmatrix} 0.37 \pm 0.00 & -0.14 \pm 0.00 & -0.08 \pm 0.00 \\ 0.08 \pm 0.00 & -0.03 \pm 0.01 & -0.20 \pm 0.00 \\ 0.13 \pm 0.00 & 0.12 \pm 0.00 & -0.01 \pm 0.00 \end{bmatrix} \right\}$ |
| | $\left\{ \begin{bmatrix} -0.84 \pm 0.01 & 0.02 \pm 0.01 & -0.05 \pm 0.01 \\ -0.35 \pm 0.01 & 1.13 \pm 0.01 & 0.04 \pm 0.01 \\ -0.07 \pm 0.01 & -0.34 \pm 0.01 & -1.24 \pm 0.01 \end{bmatrix}, \begin{bmatrix} -0.74 \pm 0.01 & -0.03 \pm 0.01 & -0.07 \pm 0.01 \\ 0.68 \pm 0.01 & -0.39 \pm 0.01 & 0.06 \pm 0.01 \\ -0.10 \pm 0.01 & 0.54 \pm 0.01 & -0.42 \pm 0.02 \end{bmatrix}, \begin{bmatrix} 0.04 \pm 0.01 & 0.02 \pm 0.01 & -0.03 \pm 0.01 \\ -0.28 \pm 0.01 & -0.03 \pm 0.01 & 0.02 \pm 0.01 \\ -0.07 \pm 0.02 & -0.33 \pm 0.01 & 0.11 \pm 0.01 \end{bmatrix} \right\}$ |
| | $\left\{ \begin{bmatrix} -0.79 \pm 0.01 & -0.01 \pm 0.01 & 0.03 \pm 0.01 \\ -0.32 \pm 0.01 & 1.07 \pm 0.01 & -0.06 \pm 0.01 \\ 0.03 \pm 0.01 & -0.17 \pm 0.01 & -1.31 \pm 0.01 \end{bmatrix}, \begin{bmatrix} -0.71 \pm 0.01 & -0.01 \pm 0.01 & 0.05 \pm 0.01 \\ 0.73 \pm 0.01 & -0.32 \pm 0.01 & -0.10 \pm 0.02 \\ 0.06 \pm 0.01 & 0.39 \pm 0.01 & -0.50 \pm 0.02 \end{bmatrix}, \begin{bmatrix} 0.07 \pm 0.01 & -0.01 \pm 0.01 & 0.03 \pm 0.01 \\ -0.17 \pm 0.01 & -0.05 \pm 0.01 & -0.06 \pm 0.01 \\ -0.10 \pm 0.01 & -0.29 \pm 0.01 & 0.09 \pm 0.01 \end{bmatrix} \right\}$ |
| | $\left\{ \begin{bmatrix} -0.90 \pm 0.01 & 0.01 \pm 0.01 & 0.01 \pm 0.01 \\ -0.34 \pm 0.01 & 1.21 \pm 0.01 & -0.01 \pm 0.01 \\ -0.01 \pm 0.01 & -0.30 \pm 0.01 & -1.36 \pm 0.01 \end{bmatrix}, \begin{bmatrix} -0.82 \pm 0.01 & -0.02 \pm 0.01 & 0.02 \pm 0.01 \\ 0.72 \pm 0.01 & -0.50 \pm 0.01 & -0.01 \pm 0.01 \\ 0.00 \pm 0.01 & 0.50 \pm 0.02 & -0.60 \pm 0.02 \end{bmatrix}, \begin{bmatrix} -0.03 \pm 0.02 & 0.01 \pm 0.01 & 0.01 \pm 0.01 \\ -0.35 \pm 0.02 & 0.01 \pm 0.01 & 0.00 \pm 0.01 \\ -0.02 \pm 0.02 & -0.32 \pm 0.01 & 0.02 \pm 0.01 \end{bmatrix} \right\}$ |
| | $\left\{ \begin{bmatrix} -0.90 \pm 0.01 & 0.01 \pm 0.01 & 0.01 \pm 0.01 \\ -0.35 \pm 0.01 & 1.21 \pm 0.01 & -0.01 \pm 0.01 \\ -0.01 \pm 0.01 & -0.30 \pm 0.01 & -1.36 \pm 0.01 \end{bmatrix}, \begin{bmatrix} -0.82 \pm 0.01 & -0.02 \pm 0.01 & 0.02 \pm 0.01 \\ 0.72 \pm 0.01 & -0.50 \pm 0.01 & -0.01 \pm 0.01 \\ 0.00 \pm 0.01 & 0.50 \pm 0.02 & -0.60 \pm 0.02 \end{bmatrix}, \begin{bmatrix} -0.03 \pm 0.02 & 0.01 \pm 0.01 & 0.01 \pm 0.01 \\ -0.35 \pm 0.02 & 0.01 \pm 0.01 & 0.00 \pm 0.01 \\ -0.01 \pm 0.02 & -0.32 \pm 0.01 & 0.02 \pm 0.01 \end{bmatrix} \right\}$ |
| \multicolumn{2}{c}{$\sigma_o = 0.5$} |
| | $\left\{ \begin{bmatrix} -0.27 \pm 0.00 & 0.16 \pm 0.00 & -0.18 \pm 0.00 \\ -0.14 \pm 0.01 & 0.55 \pm 0.01 & -0.21 \pm 0.01 \\ 0.24 \pm 0.01 & 0.26 \pm 0.01 & -0.68 \pm 0.01 \end{bmatrix}, \begin{bmatrix} -0.33 \pm 0.01 & -0.03 \pm 0.01 & 0.12 \pm 0.01 \\ 0.21 \pm 0.01 & 0.13 \pm 0.00 & -0.06 \pm 0.01 \\ -0.29 \pm 0.01 & -0.02 \pm 0.01 & 0.29 \pm 0.00 \end{bmatrix}, \begin{bmatrix} 0.38 \pm 0.01 & -0.15 \pm 0.01 & -0.08 \pm 0.01 \\ 0.08 \pm 0.01 & -0.02 \pm 0.01 & -0.21 \pm 0.01 \\ 0.12 \pm 0.01 & 0.11 \pm 0.01 & 0.00 \pm 0.01 \end{bmatrix} \right\}$ |
| | NA |
| | $\left\{ \begin{bmatrix} -0.36 \pm 0.01 & 0.15 \pm 0.01 & -0.09 \pm 0.01 \\ -0.32 \pm 0.02 & 0.67 \pm 0.01 & -0.21 \pm 0.02 \\ 0.02 \pm 0.01 & 0.12 \pm 0.01 & -0.94 \pm 0.01 \end{bmatrix}, \begin{bmatrix} -0.34 \pm 0.01 & -0.18 \pm 0.01 & 0.05 \pm 0.01 \\ 0.52 \pm 0.02 & 0.09 \pm 0.01 & -0.22 \pm 0.03 \\ -0.02 \pm 0.01 & 0.30 \pm 0.02 & 0.08 \pm 0.01 \end{bmatrix}, \begin{bmatrix} 0.28 \pm 0.01 & 0.01 \pm 0.01 & -0.01 \pm 0.01 \\ 0.04 \pm 0.02 & -0.09 \pm 0.01 & -0.09 \pm 0.02 \\ -0.30 \pm 0.02 & -0.29 \pm 0.02 & 0.41 \pm 0.01 \end{bmatrix} \right\}$ |
| | $\left\{ \begin{bmatrix} -0.75 \pm 0.03 & 0.07 \pm 0.03 & -0.08 \pm 0.02 \\ -0.29 \pm 0.03 & 1.07 \pm 0.04 & 0.16 \pm 0.03 \\ -0.18 \pm 0.06 & -0.42 \pm 0.06 & -1.28 \pm 0.04 \end{bmatrix}, \begin{bmatrix} -0.66 \pm 0.04 & -0.14 \pm 0.05 & -0.11 \pm 0.03 \\ 0.68 \pm 0.04 & -0.29 \pm 0.05 & 0.25 \pm 0.05 \\ -0.22 \pm 0.08 & 0.67 \pm 0.08 & -0.46 \pm 0.06 \end{bmatrix}, \begin{bmatrix} 0.03 \pm 0.06 & 0.09 \pm 0.04 & -0.06 \pm 0.02 \\ -0.19 \pm 0.06 & -0.08 \pm 0.03 & 0.11 \pm 0.03 \\ -0.08 \pm 0.10 & -0.41 \pm 0.06 & 0.11 \pm 0.04 \end{bmatrix} \right\}$ |
| | $\left\{ \begin{bmatrix} -0.81 \pm 0.03 & 0.06 \pm 0.04 & -0.03 \pm 0.02 \\ -0.29 \pm 0.03 & 1.09 \pm 0.04 & 0.12 \pm 0.03 \\ -0.25 \pm 0.04 & -0.39 \pm 0.06 & -1.23 \pm 0.04 \end{bmatrix}, \begin{bmatrix} -0.74 \pm 0.04 & -0.09 \pm 0.06 & -0.02 \pm 0.04 \\ 0.70 \pm 0.04 & -0.35 \pm 0.06 & 0.17 \pm 0.05 \\ -0.31 \pm 0.06 & 0.63 \pm 0.09 & -0.38 \pm 0.06 \end{bmatrix}, \begin{bmatrix} -0.01 \pm 0.06 & 0.05 \pm 0.04 & -0.01 \pm 0.02 \\ -0.23 \pm 0.06 & -0.03 \pm 0.04 & 0.07 \pm 0.03 \\ -0.20 \pm 0.09 & -0.39 \pm 0.05 & 0.15 \pm 0.03 \end{bmatrix} \right\}$ |
| \multicolumn{2}{c}{$\sigma_o = 1.0$} |
| | $\left\{ \begin{bmatrix} -0.25 \pm 0.01 & 0.18 \pm 0.01 & -0.19 \pm 0.01 \\ -0.18 \pm 0.02 & 0.49 \pm 0.02 & -0.17 \pm 0.02 \\ 0.21 \pm 0.02 & 0.28 \pm 0.01 & -0.63 \pm 0.01 \end{bmatrix}, \begin{bmatrix} -0.29 \pm 0.01 & -0.01 \pm 0.01 & 0.11 \pm 0.01 \\ 0.17 \pm 0.01 & 0.14 \pm 0.01 & -0.03 \pm 0.02 \\ -0.30 \pm 0.01 & -0.03 \pm 0.01 & 0.31 \pm 0.01 \end{bmatrix}, \begin{bmatrix} 0.38 \pm 0.01 & -0.15 \pm 0.01 & -0.08 \pm 0.01 \\ 0.04 \pm 0.02 & 0.00 \pm 0.01 & -0.19 \pm 0.02 \\ 0.11 \pm 0.01 & 0.07 \pm 0.02 & -0.01 \pm 0.02 \end{bmatrix} \right\}$ |
| | NA |
| | $\left\{ \begin{bmatrix} -0.27 \pm 0.01 & 0.18 \pm 0.01 & -0.18 \pm 0.01 \\ -0.20 \pm 0.02 & 0.54 \pm 0.02 & -0.20 \pm 0.02 \\ 0.22 \pm 0.01 & 0.29 \pm 0.01 & -0.67 \pm 0.01 \end{bmatrix}, \begin{bmatrix} -0.33 \pm 0.01 & -0.02 \pm 0.01 & 0.13 \pm 0.01 \\ 0.18 \pm 0.02 & 0.14 \pm 0.01 & -0.05 \pm 0.02 \\ -0.30 \pm 0.01 & -0.02 \pm 0.01 & 0.31 \pm 0.01 \end{bmatrix}, \begin{bmatrix} 0.39 \pm 0.01 & -0.14 \pm 0.01 & -0.10 \pm 0.01 \\ 0.05 \pm 0.02 & -0.01 \pm 0.01 & -0.21 \pm 0.02 \\ 0.10 \pm 0.02 & 0.07 \pm 0.02 & -0.01 \pm 0.02 \end{bmatrix} \right\}$ |
| | $\left\{ \begin{bmatrix} -0.56 \pm 0.05 & 0.17 \pm 0.05 & -0.10 \pm 0.04 \\ -0.30 \pm 0.08 & 0.87 \pm 0.06 & 0.35 \pm 0.05 \\ -0.26 \pm 0.10 & -0.67 \pm 0.10 & -1.04 \pm 0.07 \end{bmatrix}, \begin{bmatrix} -0.46 \pm 0.05 & -0.19 \pm 0.08 & -0.07 \pm 0.06 \\ 0.54 \pm 0.07 & -0.02 \pm 0.08 & 0.51 \pm 0.08 \\ -0.37 \pm 0.11 & 0.96 \pm 0.15 & -0.21 \pm 0.10 \end{bmatrix}, \begin{bmatrix} 0.14 \pm 0.08 & 0.13 \pm 0.05 & -0.02 \pm 0.04 \\ -0.07 \pm 0.09 & -0.17 \pm 0.06 & 0.24 \pm 0.05 \\ 0.06 \pm 0.12 & -0.53 \pm 0.09 & 0.22 \pm 0.07 \end{bmatrix} \right\}$ |
| | $\left\{ \begin{bmatrix} -0.60 \pm 0.05 & 0.11 \pm 0.07 & -0.05 \pm 0.04 \\ -0.33 \pm 0.08 & 0.93 \pm 0.06 & 0.22 \pm 0.05 \\ -0.12 \pm 0.14 & -0.51 \pm 0.09 & -1.09 \pm 0.05 \end{bmatrix}, \begin{bmatrix} -0.51 \pm 0.06 & -0.13 \pm 0.10 & -0.06 \pm 0.07 \\ 0.56 \pm 0.06 & -0.14 \pm 0.09 & 0.33 \pm 0.07 \\ -0.30 \pm 0.09 & 0.74 \pm 0.17 & -0.19 \pm 0.08 \end{bmatrix}, \begin{bmatrix} 0.17 \pm 0.07 & 0.09 \pm 0.06 & -0.01 \pm 0.05 \\ -0.16 \pm 0.09 & -0.09 \pm 0.06 & 0.10 \pm 0.04 \\ -0.03 \pm 0.12 & -0.39 \pm 0.12 & 0.25 \pm 0.06 \end{bmatrix} \right\}$ |

# Validation of AROC and AWROC measures for sparse and smooth random simulations

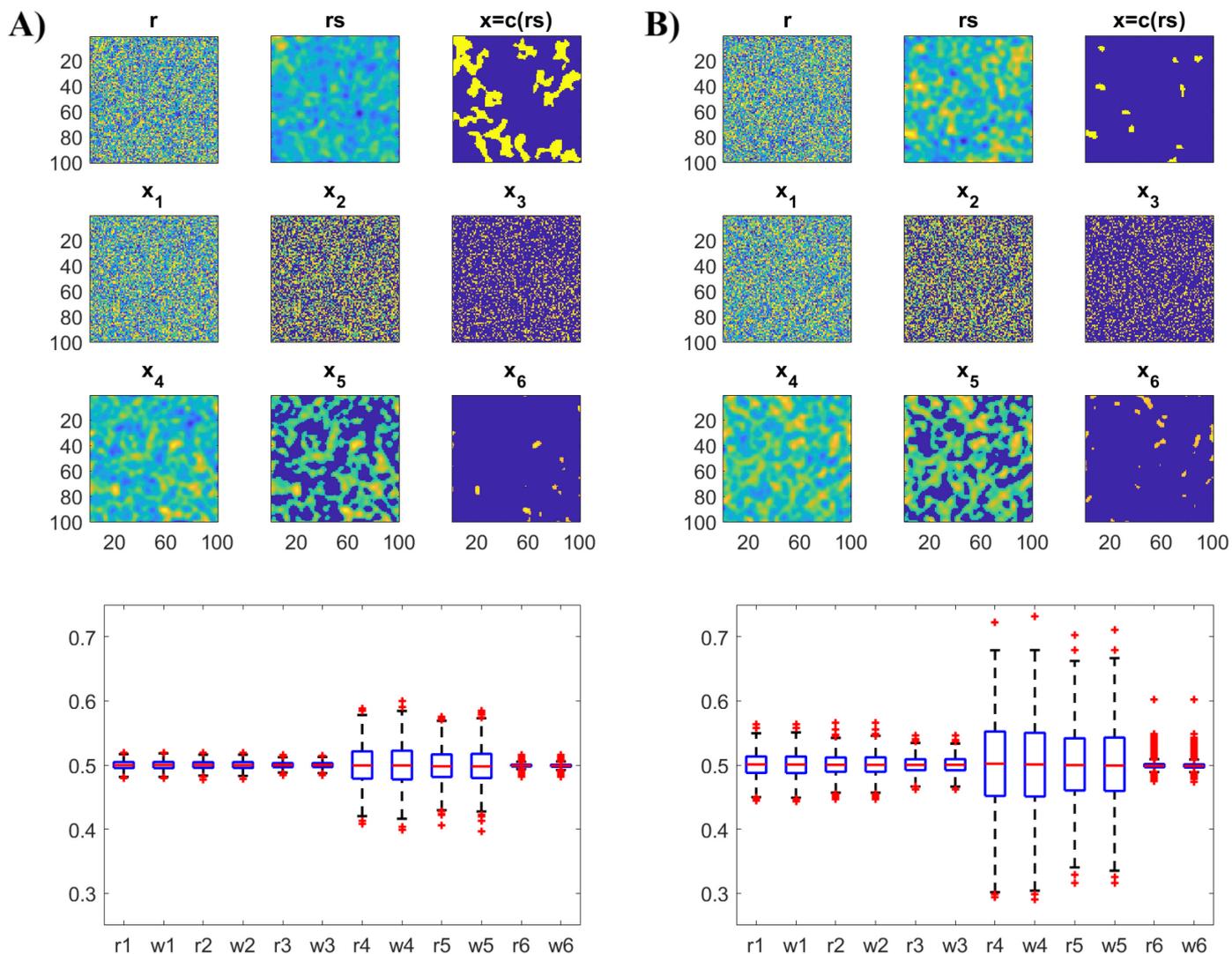

**Figure 1:** Validation of new area under weighted ROC (AWROC) curve statistic using Monte Carlo simulations. Simulated ground-truth data is generated as follows: 1) a random image is generated by i.i.d. sampling from the uniform random distribution in the range [0; 1], named as "**r**" in the plots' title; 2) the random image is smoothed using a 2D image Gaussian filter (Matlab's imgaussfilt function), names as "**rs**" in the title; 3) after linear transforming the smoothed image into the range [0; 1], the latter was thresholded to create less and more sparse masks (threshold values equals to 0.5 or 0.8 for images shown under title "**x=c(rs)**" in **A** and **B**, respectively), and finally the 10 larger connected components were extracted (Matlab's function bwconncomp) to create sparse images, as named "**x=c(rs)**" in the plots. The masked images **x** are the ground-truth data for the two simulated less and more sparse scenarios as illustrated in **A** and **B**. Similarly, random solutions were generated inside the Monte Carlo iterations using the random uniform distribution and its smoothed image filter outcome, as mentioned above, represented by $x_1$ and $x_4$ in the plots. While $x_2$ and $x_3$ are the thresholded images obtained from $x_1$ (same threshold values as above), similarly, $x_5$ and $x_6$ are obtained from $x_4$. Finally, from 10,000 Monte Carlo replications of this procedure, we obtained the values for the calculation of AROC and AWROC statistics (**Materials and Methods**) for each of the solutions $x_1,...,x_6$, when comparing against the ground truth **x**, separately for the less and more sparse cases shown in **A** and **B**. The AROC and AWROC calculated values are represented by the boxplots under labels $r_1,...,r_6$, and $w_1,...,w_6$, respectively, in the graphic plots in the bottom row (here in the boxplot labels, "r" stands for ROC and "w" stands for AWROC; see **Materials and Methods** for more information).

# Solving large-scale MPSS models for synthetic resting state data

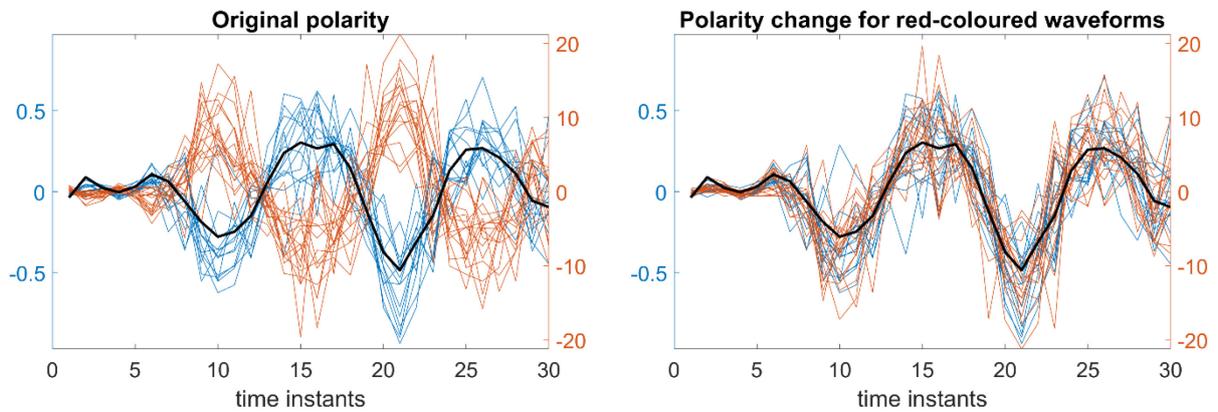

**Figure 2:** Manual correction of the polarity for the estimated waveforms of the most salient dipoles. This operation is illustrated for the estimated sources corresponding to the fourth ground-truth source, the same subplot shown in **Fig. 5C**, fourth row, and seventh column. **Left side:** original waveforms as estimated by HGDALS. **Right side:** after manually correcting the polarity for the waveform (red colour). In both cases, the black curve represents the ground-truth waveform for the simulated source.

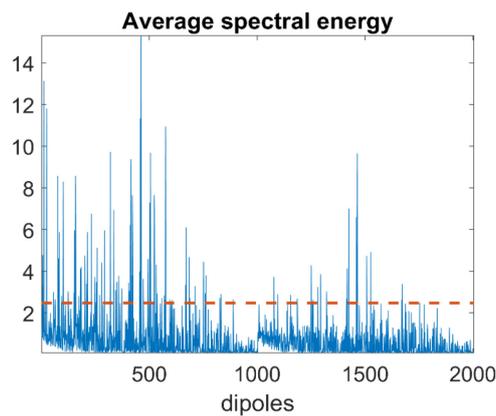

**Figure 3:** Average spectral energy calculated from the estimated time series $\{x_t^{(e)}\}$, $e = 1, \ldots, E$, by calculating their Fourier transform and averaging from the second to sixth spectral component across epochs. The dipoles in the plot's left (right) half are located in the left (right) hemisphere. The threshold (red discontinuous) line indicates the lower limit corresponding to selected 100 most salient dipoles.

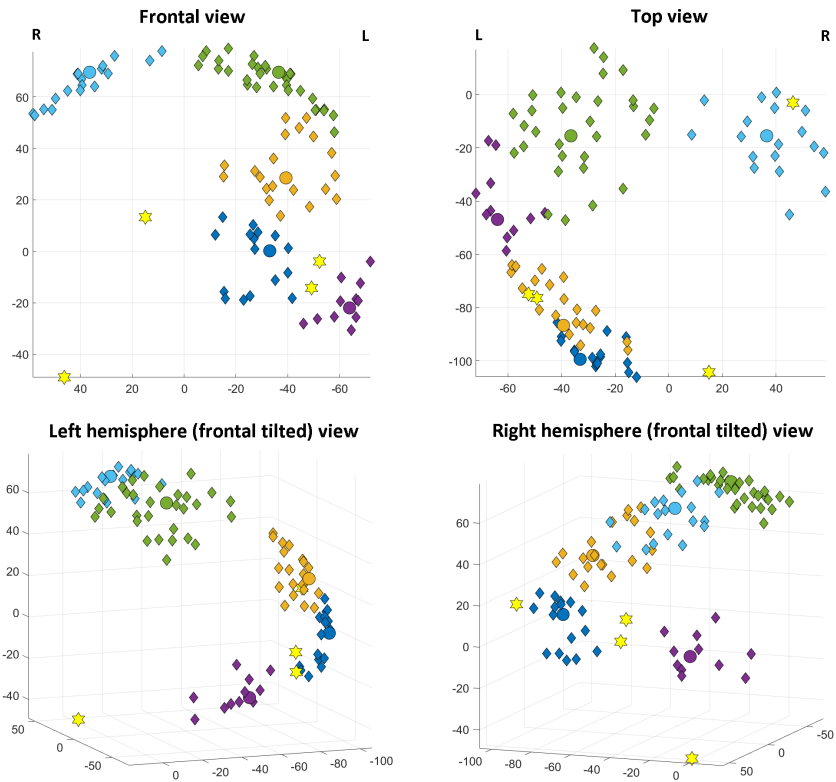

**Figure 4:** Visualization of estimated dipoles clustered nearby simulated sources, as obtained from **Fig. 5B** by removing the brain cortical surface and displaying only the ground-truth (coloured sphere points) and estimated (coloured diamond points) dipoles, and outliers (coloured star points). Multiple views of the same sources highlight the clustering characteristics.

## Solving large-scale MPSS models for synthetic event-related data

Here, the temporal and spatial components of the solutions achieved with the SSALS and six state-of-the-art source localization methods are shown as complementary information to visually appreciate the quality of the methods. For each case, at the left side of the figure, the ground truth is shown clearly illustrating the true components for the simulated scenarios: either with two or five simulated dipole of extension patch of either 6 or 15 cm$^2$.

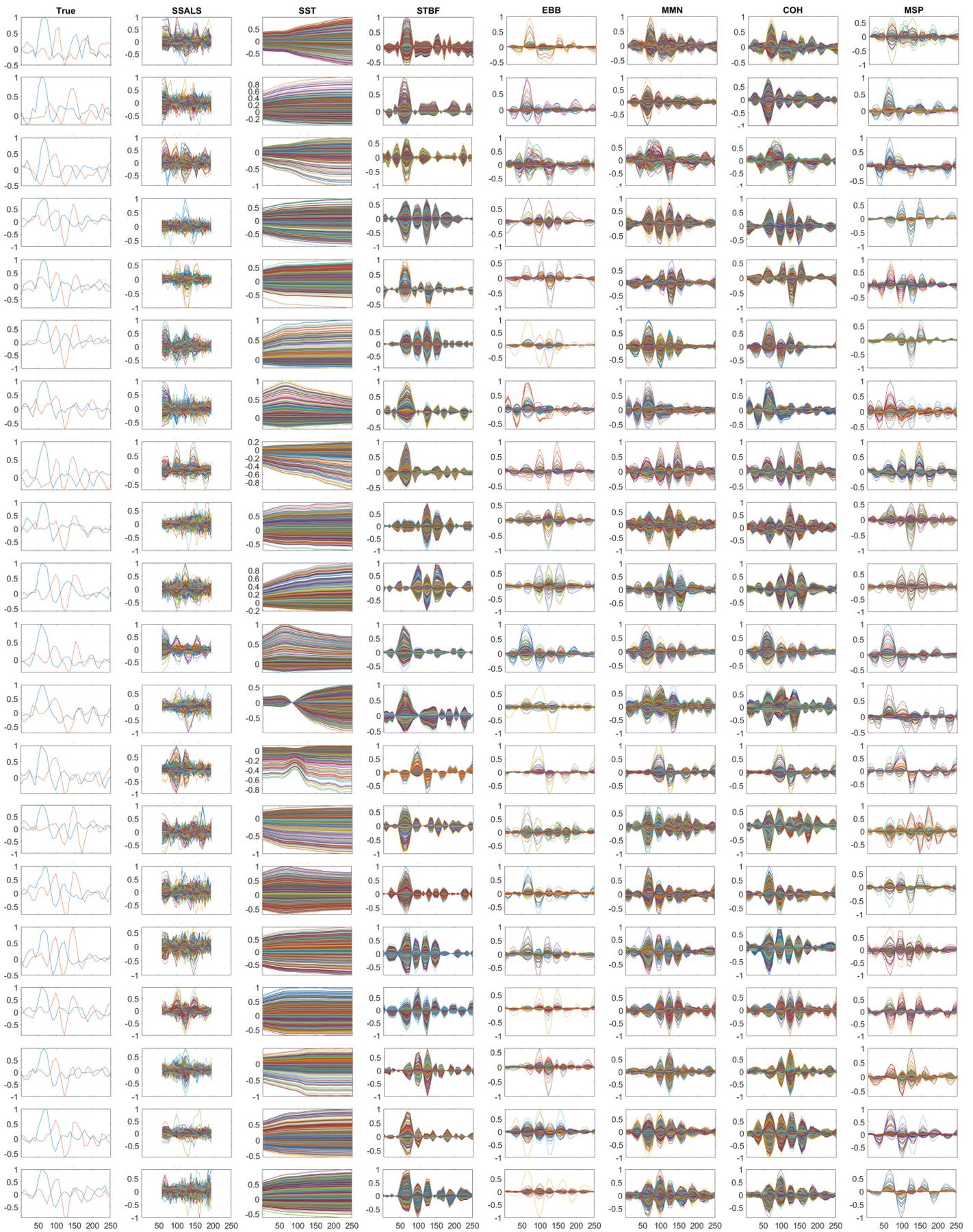

**Figure 5:** Inverse solution plotted to assess the temporal accuracy by visual inspection: simulation case for 2 ROIs with 6cm$^2$ extension.

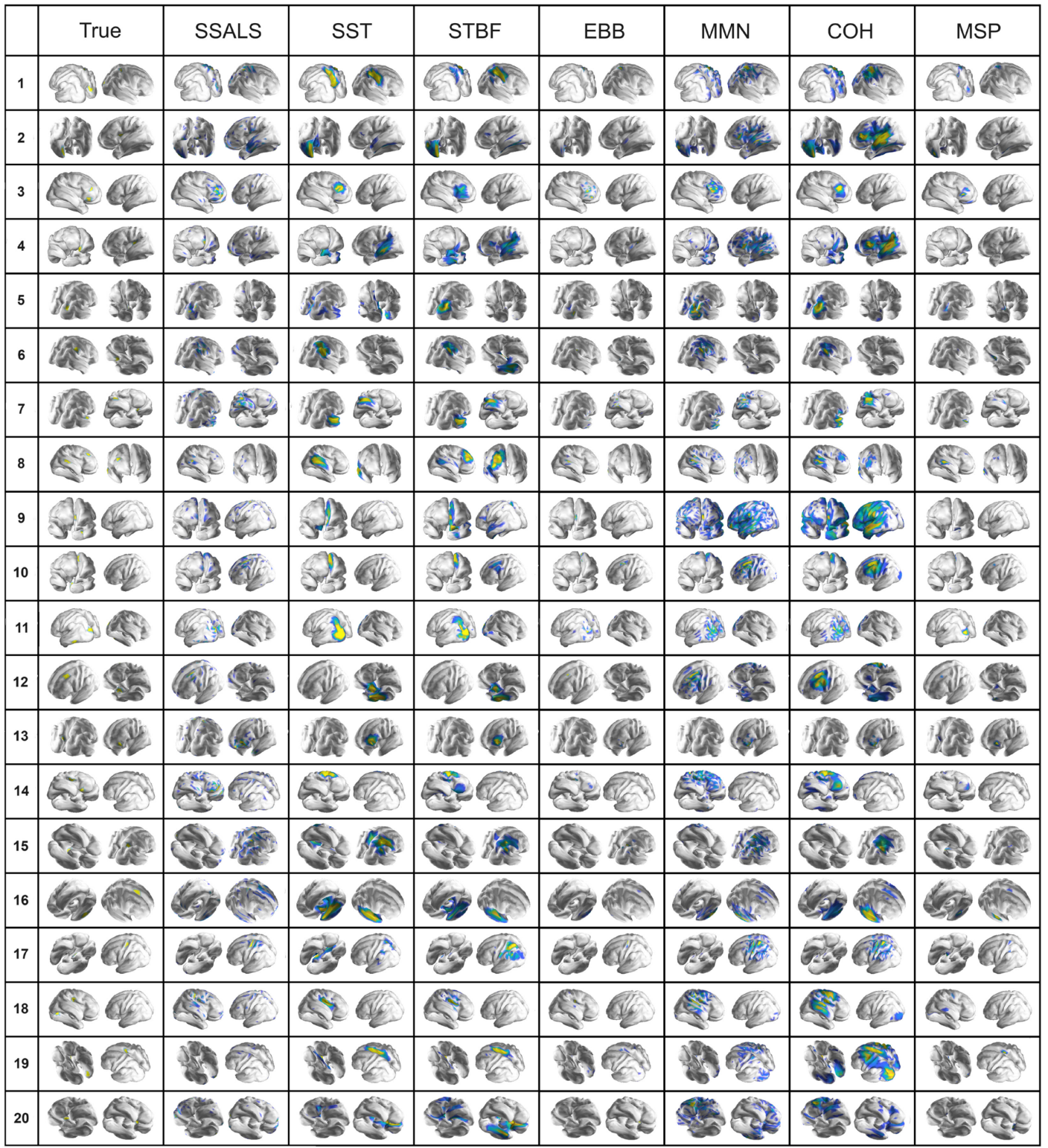

**Figure 6:** Inverse solution plotted to assess the spatial accuracy by visual inspection: simulation case for 2 ROIs with 6 cm$^2$ extension.

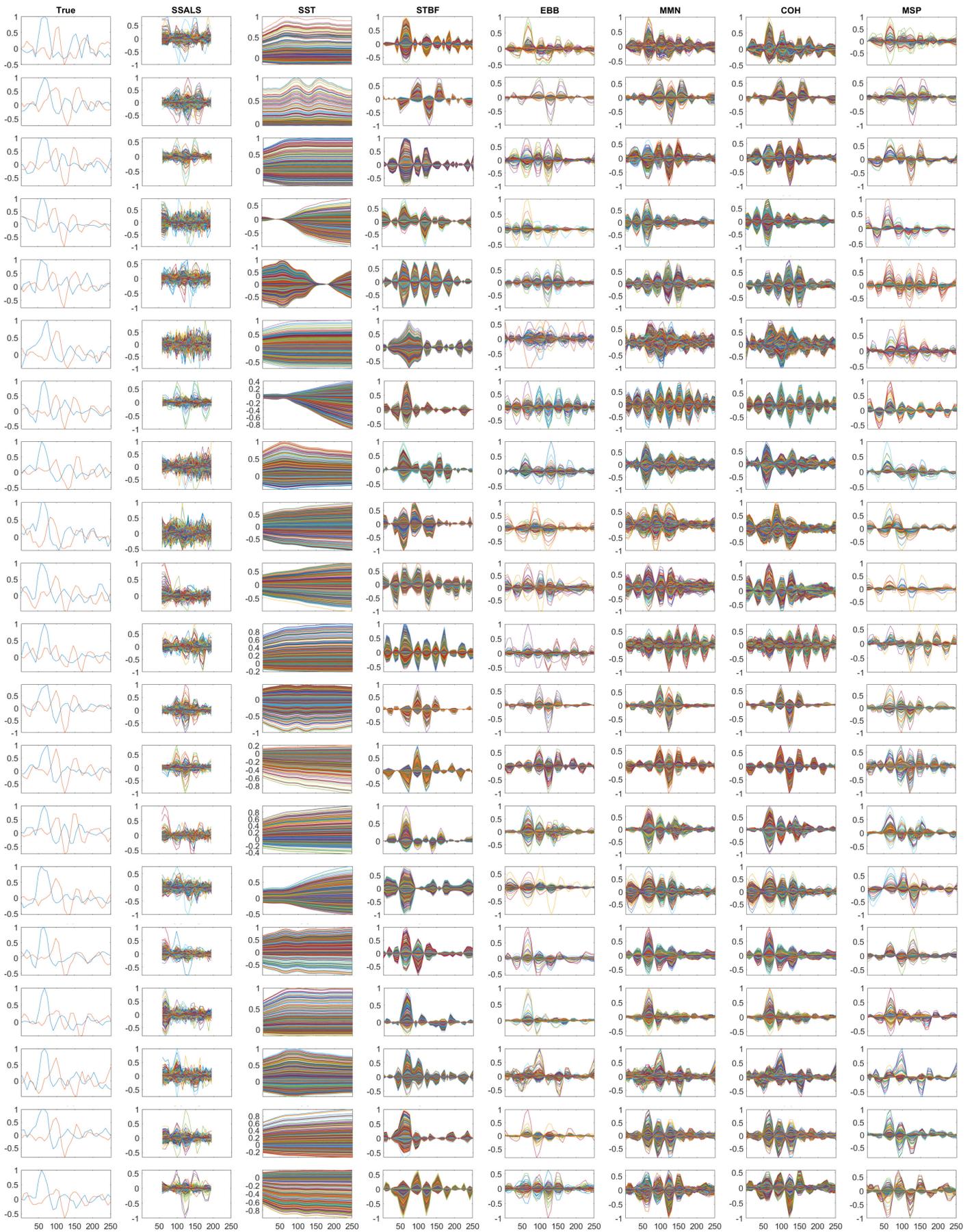

**Figure 7:** Inverse solution plotted to assess the temporal accuracy by visual inspection: simulation case for 2 ROIs with 15 cm² extension.

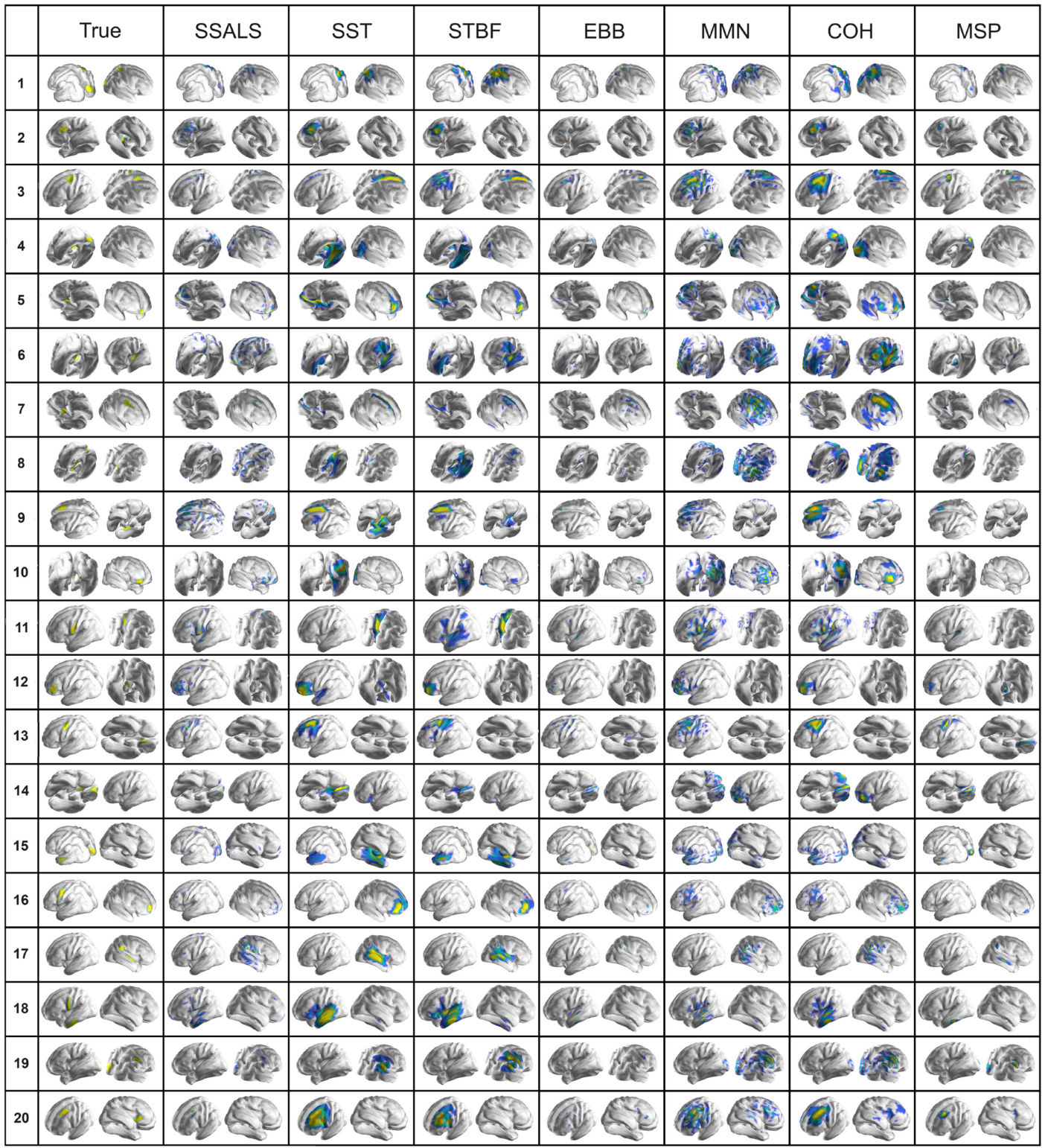

**Figure 8:** Inverse solution plotted to assess the spatial accuracy by visual inspection: simulation case for 2 ROIs with 15 cm² extension.

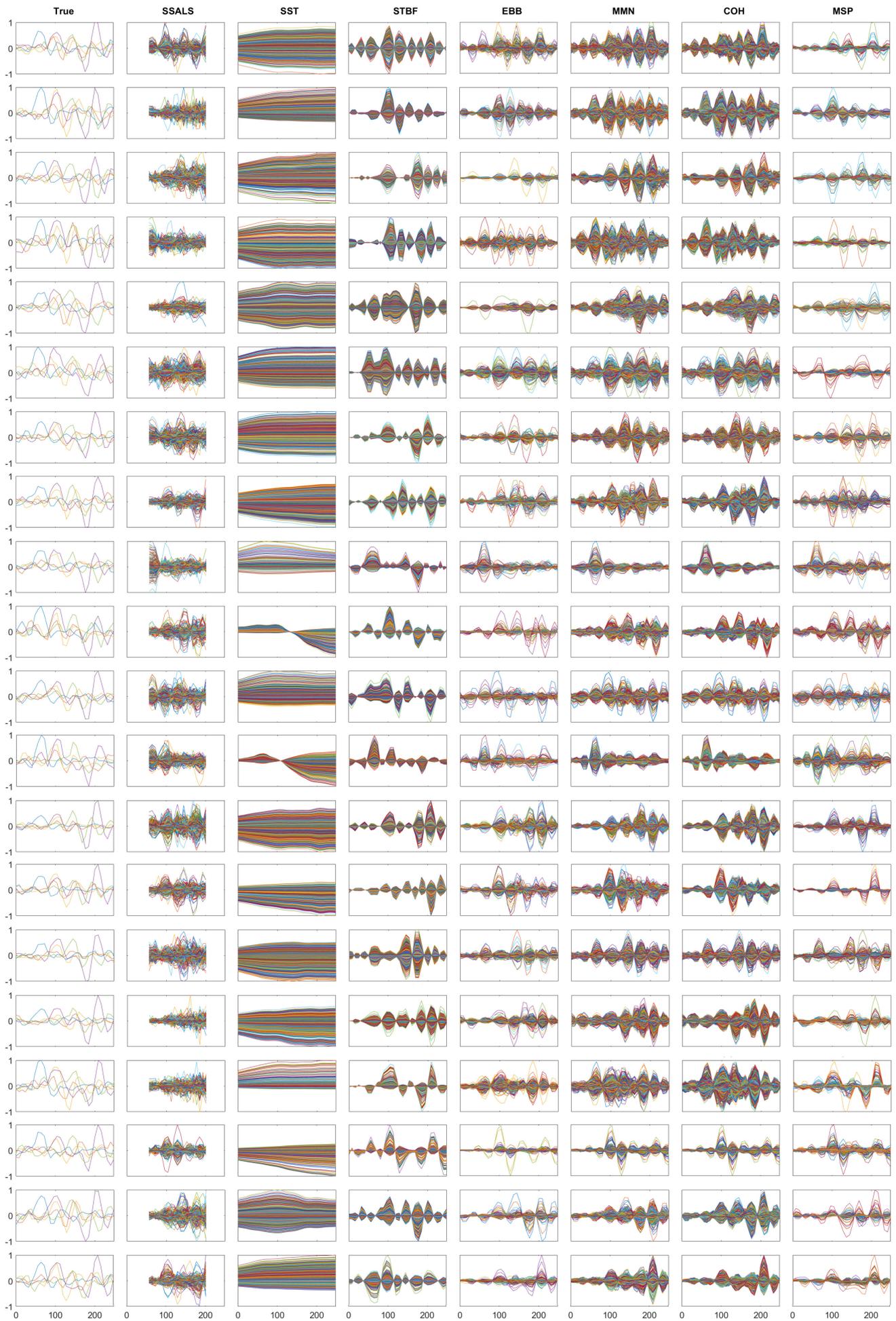

**Figure 9:** Inverse solution plotted to assess the temporal accuracy by visual inspection: simulation case for 5 ROIs with 6 cm$^2$ extension.

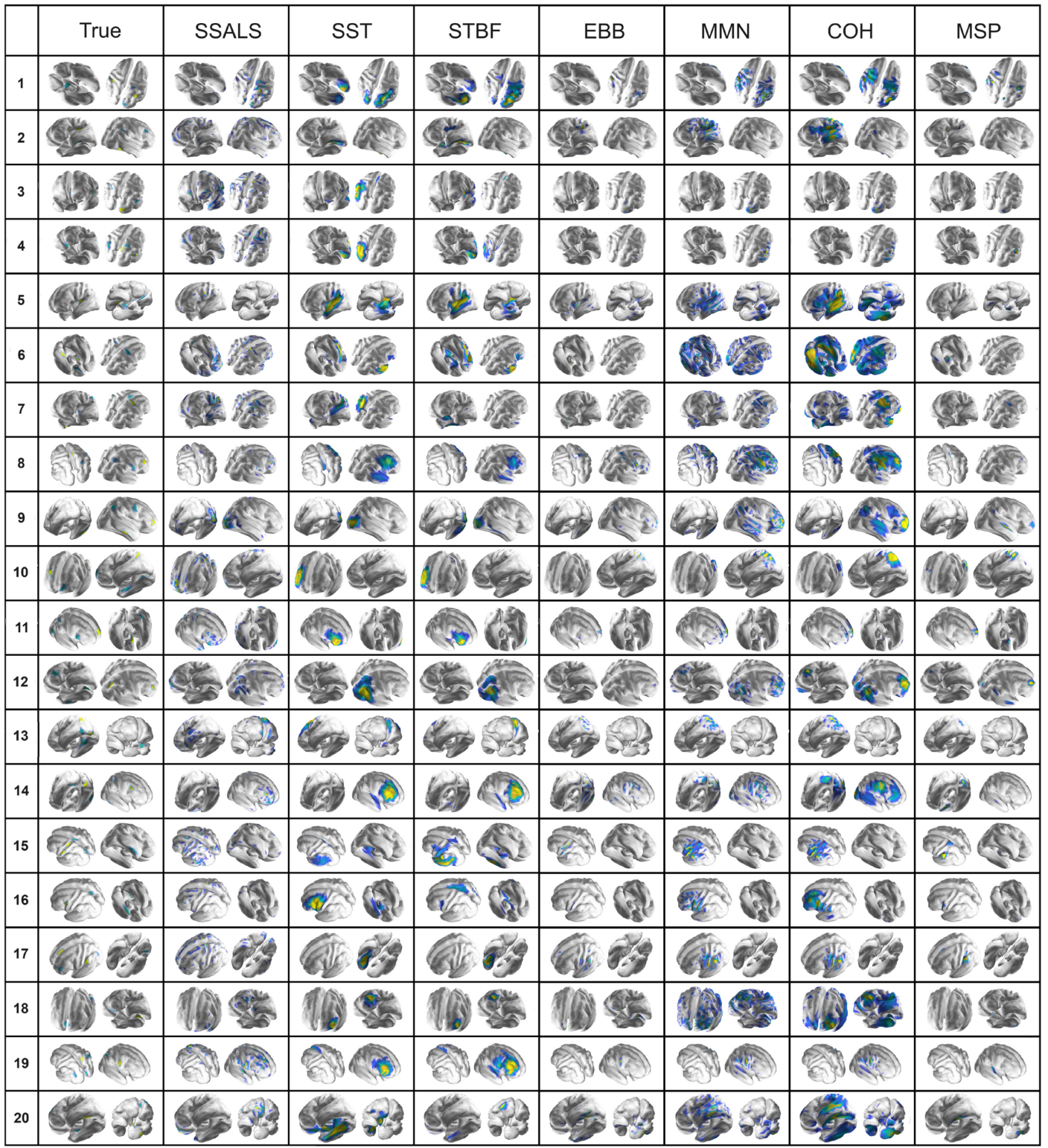

**Figure 10:** Inverse solution plotted to assess the spatial accuracy by visual inspection: simulation case for 5 ROIs with 6 cm$^2$ extension.

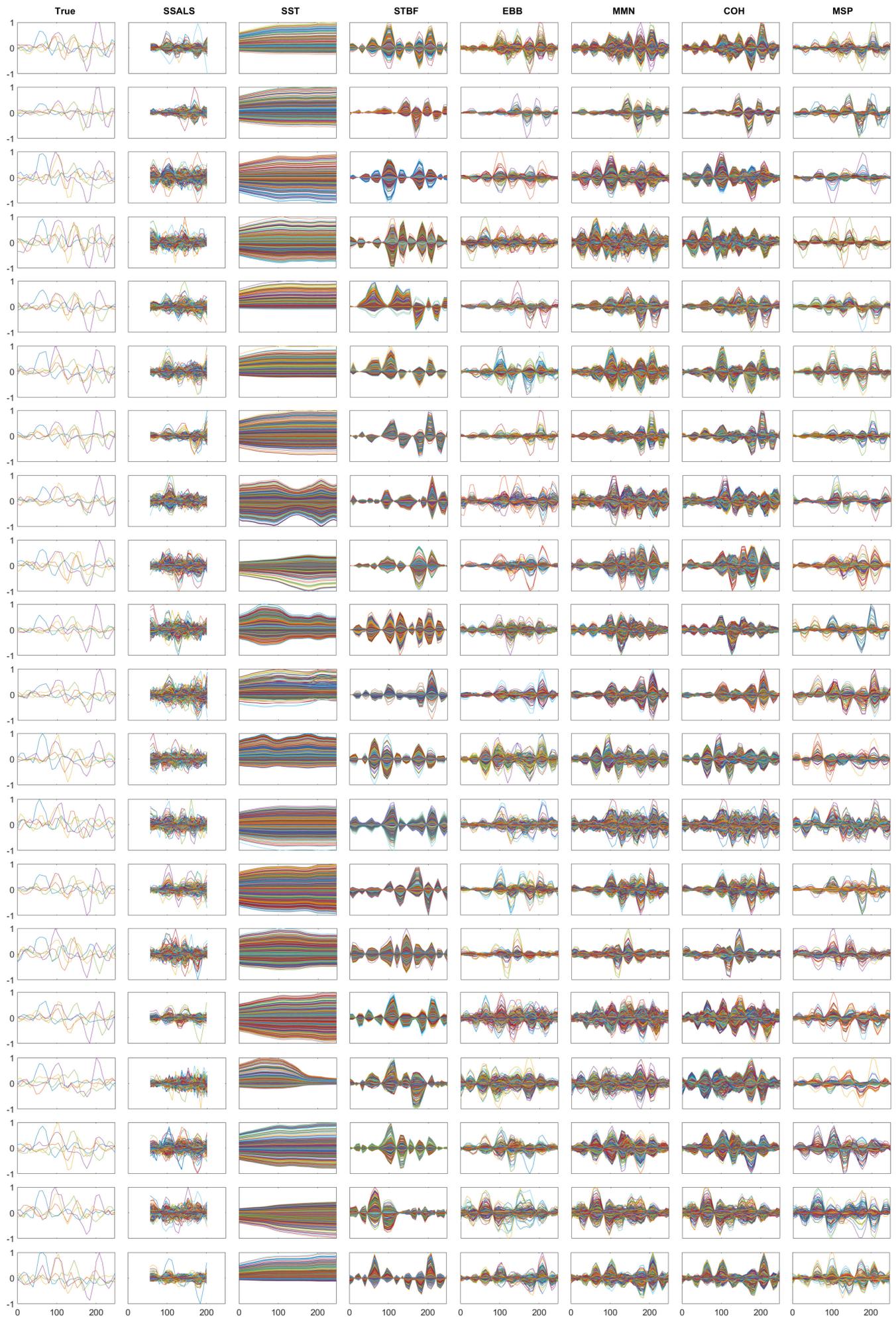

**Figure 11:** Inverse solution plotted to assess the temporal accuracy by visual inspection: simulation case for 5 ROIs with 15 cm² extension.

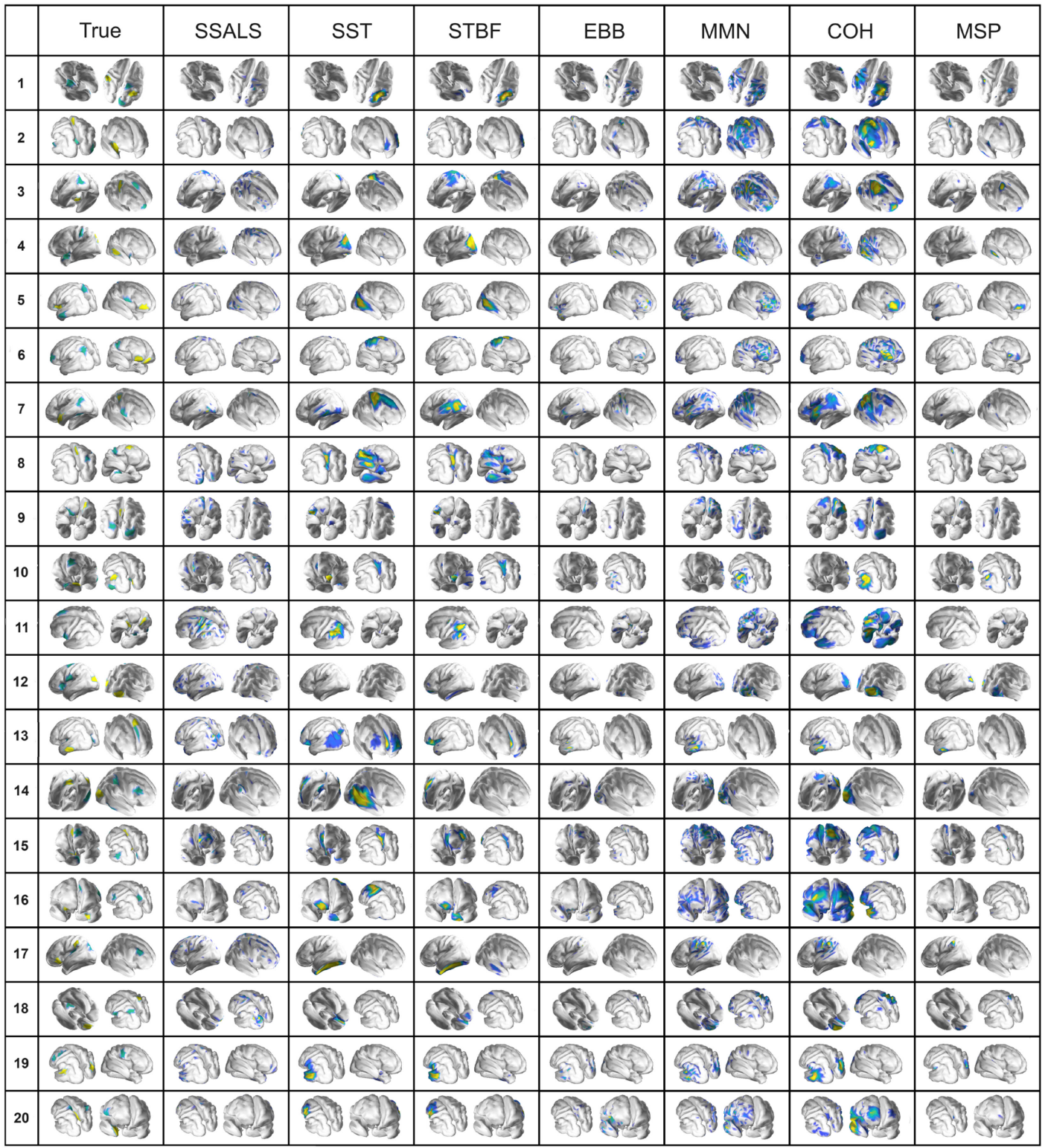

**Figure 12:** Inverse solution plotted to assess the spatial accuracy by visual inspection: simulation case for 5 ROIs with 15 cm² extension.

# Solving large-scale MPSS models for real event-related data

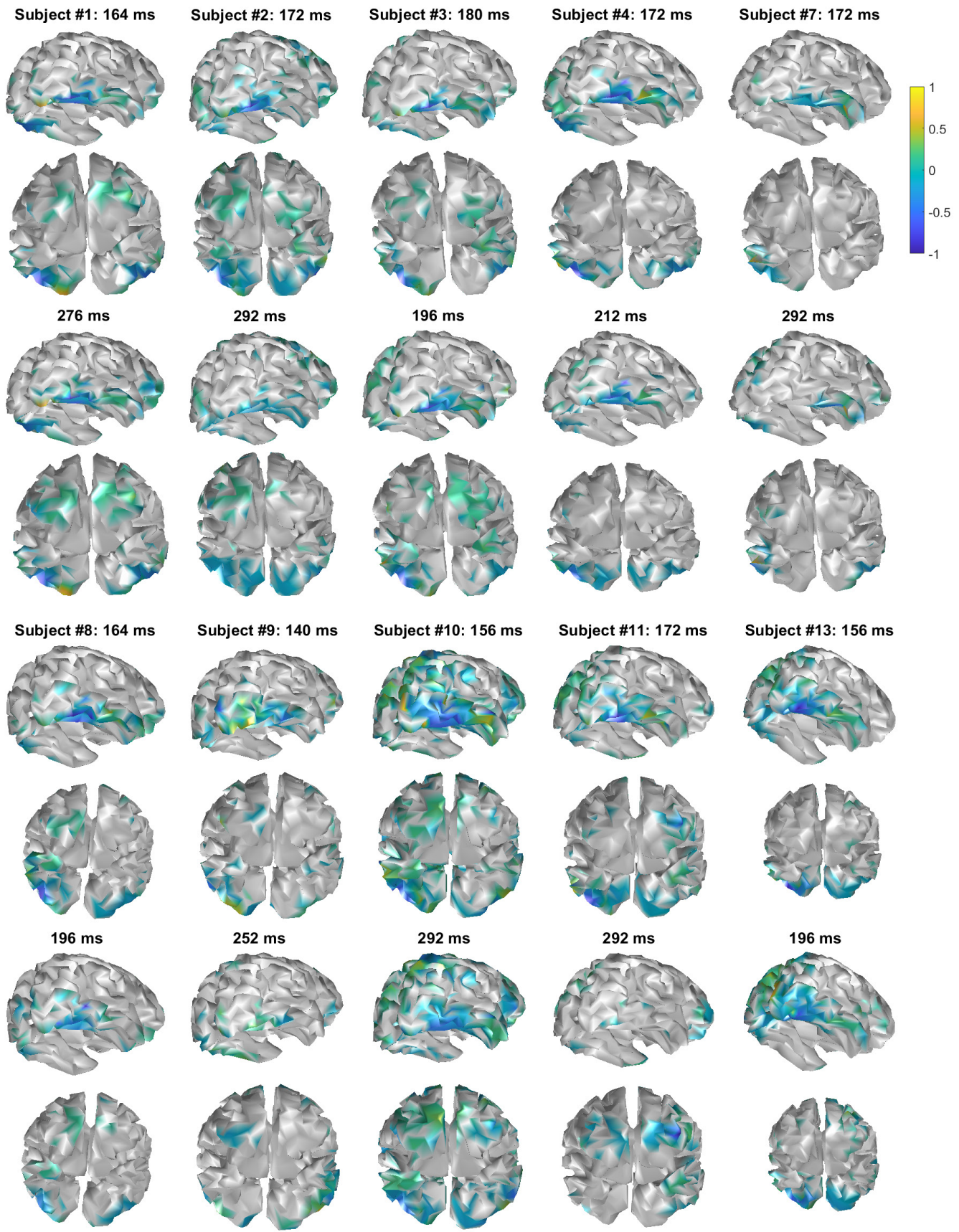

**Figure 13:** Source localization based on SSALS solutions for 10 subjects from the Wakeman and Rik Henson database using only the EEG signals. The solution is shown for the activity peak as estimated for two successive time windows, approximately from 100 to 200 ms, and 200 to 300 ms. The first studied participants (subjects #1,2,3,4,7) are shown in the upper half plots, while the bottom half show the results for the remaining data analysis (subjects #8,9,10,11,13). The remaining subjects in the Wakeman and Henson database were not shown as failures occurred with the calculation of the lead field matrices and co-registration with a reduced-size template. Two views are shown for each participant solution at corresponding time peaks. The views are from a left-posterior-inferior tilted angle (top map) and a frontal-inferior tilted angle (bottom map). Consistently, the results show the activation of the occipitotemporal ventral visual stream and frontal lobe for most of the participants data and across the successive time windows.